\pdfoutput=1

\documentclass[12pt]{article}
\usepackage{jheppub}
\pdfoutput=1
\usepackage{epstopdf}

\usepackage{mathrsfs}
\usepackage{psfrag}
\usepackage{color}
\usepackage{hyperref}

\usepackage{slashed}
\usepackage{simplewick}
\usepackage{cancel}
\usepackage[utf8]{inputenc}

\usepackage[table]{xcolor}

\usepackage{amsmath,bbm,array,amsfonts,graphicx,wrapfig,arydshln,lscape,float,multirow,longtable,rotating,makecell}
\usepackage{url}

\hypersetup{pdftitle={},pdfcreator={},linkcolor=[rgb]{0.15,0.35,0.75},colorlinks=true,citecolor=[rgb]{0.675,0,0.2},urlcolor=[rgb]{0.15,0.35,0.65}}
\let\olditemize\itemize\renewcommand{\itemize}{\vspace{-2pt}\olditemize\setlength{\itemsep}{1pt}\setlength{\parskip}{0pt}\setlength{\parsep}{-0pt}}
\let\oldenumerate\enumerate\renewcommand{\enumerate}{\vspace{-4pt}\oldenumerate\setlength{\itemsep}{1pt}\setlength{\parskip}{0pt}\setlength{\parsep}{0pt}}

\newcommand{\be}{\begin{equation}}
\newcommand{\ee}{\end{equation}}
\newcommand{\bea}{\begin{eqnarray}}
\newcommand{\eea}{\end{eqnarray}}

\def\eqn#1{eq.~(\ref{#1})}

\def\spa#1.#2{\left\langle#1\,#2\right\rangle}
\def\spb#1.#2{\left[#1\,#2\right]}
\newcommand{\e}{\epsilon}

\renewcommand{\phi}{\varphi}
\renewcommand{\bar}{\overline}
\newcommand{\eq}[1]{\vspace{-3.5pt}\begin{equation}\hspace{2pt}#1\hspace{-0pt}\vspace{-3.5pt}\end{equation}}

\newcommand{\ab}[1]{\langle #1 \rangle}
\newcommand{\sqb}[1]{[ #1 ]}

\newcommand{\sab}[1]{s_{#1}}

\newcommand{\lam}[1]{\lambda_{#1} }
\newcommand{\lamt}[1]{\widetilde \lambda_{#1} }

\newcommand{\mi}{\raisebox{0.75pt}{\scalebox{0.75}{$\hspace{-1pt}\,-\,\hspace{-0.75pt}$}}}
\newcommand{\pl}{\raisebox{0.75pt}{\scalebox{0.75}{$\hspace{-1pt}\,+\,\hspace{-0.75pt}$}}}

\newcommand{\eqnDiag}[1]{ \vcenter{\hbox{#1}} }
\def\lra{\leftrightarrow}
\def\dlog{d\log}

\def\O{\mathcal{O}}
\def\N{\mathcal{N}}
\def\M{\mathcal{M}}
\def\I{\mathcal{I}}
\def\J{\mathcal{J}}
\def\Q{\mathcal{Q}}

\def\tr{\text{tr}}

\def\eps{\epsilon}

\newcommand{\NeqFour}{\mathcal{N}\!=\!4\text{ SYM}}
\newcommand{\NeqEight}{\mathcal{N}\!=\!8\text{ SUGRA}}

\definecolor{airforceblue}{rgb}{0.36, 0.54, 0.66}
\definecolor{bananayellow}{rgb}{1.0, 0.88, 0.21}
\definecolor{bittersweet}{rgb}{1.0, 0.44, 0.37}
\definecolor{blue(ncs)}{rgb}{0.0, 0.53, 0.74}
\definecolor{bole}{rgb}{0.47, 0.27, 0.23}
\definecolor{brass}{rgb}{0.71, 0.65, 0.26}
\definecolor{bronze}{rgb}{0.8, 0.5, 0.2}
\definecolor{brgreen}{rgb}{0.0, 0.26, 0.15}
\definecolor{burgundy}{rgb}{0.5, 0.0, 0.13}
\definecolor{cherry}{rgb}{1.0, 0.72, 0.77}
\definecolor{cocao}{rgb}{0.82, 0.41, 0.12}
\definecolor{citrine}{rgb}{0.99, 0.82, 0.07}

\newcommand{\twhite}[1]{\textcolor{white}{#1}}

\makeatletter\DeclareRobustCommand*{\bfseries}{\not@math@alphabet\bfseries\mathbf\fontseries\bfdefault\selectfont\boldmath}\makeatother

\title{The two-loop five-point amplitude \\  in $\N=8$ supergravity}

\author[1]{Samuel~Abreu,}
\affiliation[1]{Center for Cosmology, Particle Physics and Phenomenology (CP3), \\
Universit\'e Catholique de Louvain, 1348 Louvain-La-Neuve, Belgium}
\emailAdd{samuel.abreu@uclouvain.be}

\author[2,3]{Lance~J.~Dixon,}
\affiliation[2]{SLAC National Accelerator Laboratory, Stanford University, Stanford, CA 94039, USA}
\affiliation[3]{Institut f\"ur Physik und IRIS Adlershof, Humboldt-Universit\"at zu Berlin, \\
Zum Gro\ss en Windkanal 6, D-12489 Berlin, Germany}
\emailAdd{lance@slac.stanford.edu}

\author[2]{Enrico Herrmann,}
\emailAdd{eh10@stanford.edu}

\author[4]{Ben~Page,}
\affiliation[4]{Institut de Physique Th\'eorique, CEA, CNRS, Universit\'e Paris-Saclay, \\
F-91191 Gif-sur-Yvette cedex, France}
\emailAdd{bpage@ipht.fr}

\author[5]{Mao~Zeng}
\affiliation[5]{Institut f\"ur Theoretische Physik, Eidgen\"ossische Technische Hochschule Z\"urich, \\
Wolfgang-Pauli-Strasse 27, 8093 Z\"urich, Switzerland}
\emailAdd{mzeng@phys.ethz.ch}

\preprint{CP3-19-06, IPhT-19/003, SLAC--PUB--17377, HU-EP-19/01}

\abstract{
We compute the symbol of the two-loop five-point amplitude in 
$\N=8$ supergravity. We write an ansatz for the amplitude whose
rational prefactors are based on not only 4-dimensional leading
singularities, but also 
$d$-dimensional ones, as the former are insufficient.
Our novel $d$-dimensional unitarity-based approach to the 
systematic construction of an amplitude's rational structures is
likely to  have broader applications, for example to analogous
QCD calculations.
We fix parameters in the ansatz by performing numerical
integration-by-parts reduction of the known integrand.
We find that the two-loop five-point $\N=8$ supergravity 
amplitude is uniformly transcendental.
We then verify the soft and collinear limits of the amplitude.
There is considerable similarity with the corresponding
amplitude for $\N=4$ super-Yang-Mills theory: all the
rational prefactors are double copies of the Yang-Mills ones
and the transcendental functions overlap to a large degree.
As a byproduct, we find new relations between
color-ordered loop amplitudes in $\N=4$ super-Yang-Mills theory.
}

\begin{document}
\maketitle

\vspace{-0pt}
\newpage
\section{Introduction}
\label{sec:intro}\vspace{-8pt}
%

Scattering amplitudes in gauge and gravity theories with high degrees of
supersymmetry are known to exhibit a wide variety of simplifications in
their analytic form that are obscured in traditional Feynman-diagram
computations. A posteriori, these structures have often been found to be
linked to hidden symmetries, such as dual conformal
symmetry~\cite{DualConformalMagic,Bern:2006ew,Alday:2007hr,Drummond:2008vq}
in planar maximally-supersymmetric gauge theory. These results have
also impacted calculations in theories with lower
degrees of
supersymmetry, as techniques born to organize the supersymmetric cases,
such as the symbol map~\cite{Goncharov:2010jf, Duhr:2011zq,Duhr:2012fh}
and generalized unitarity~\cite{Bern:1994zx}, have proven
indispensable
in computations of phenomenological relevance.  As a result,
supersymmetric amplitudes have been used as a laboratory, both to extend
our general understanding of quantum field theories and to develop new
computational tools to meet the precision goals for current and future
collider experiments.  Crucial to making progress on these dual fronts
has been the availability of `theoretical data' - explicit expressions
for scattering amplitudes. 

In the past decade, great leaps have been made in the understanding of
\emph{integrands} of scattering amplitudes.  For $\N\!=\!4$ super-Yang--Mills 
theory ($\NeqFour$) in the planar limit there exist
recursive all-multiplicity formulae for amplitude integrands to
any loop order (in principle)~\cite{ArkaniHamed:2010kv}. Local integrand
representations have also been
derived~\cite{ArkaniHamed:2010gh,Bourjaily:2017wjl} by making full use of
generalized unitarity
\cite{Bern:1994zx,Bern:1994cg,Britto:2004nc,MaximalCuts,Bourjaily:2017wjl}.
In parallel, there has also been enormous progress in `geometrizing'
scattering amplitudes by relating them to mathematical
objects like the Grassmannian~\cite{postnikov,ArkaniHamed:2012nw} and
the amplituhedron~\cite{Arkani-Hamed:2013jha}. 

In theories of gravitation the construction of integrands is
dramatically eased by the color-kinematics duality and
double-copy procedure of Bern, Carrasco and
Johansson (BCJ)~\cite{BCJ},
where gravity integrands are represented as `squares' of their much simpler
gauge-theory counterparts. Even though this construction has been proven
to work for tree-level
amplitudes~\cite{BjerrumBohr:2009rd,Stieberger:2009hq,BCJSquare},
a loop-level proof remains elusive.
Nonetheless, on a case-by-case basis, the existence of
BCJ-satisfying representations~\cite{BCJLoop}
 has been established up to the four-loop
order for four-particle amplitudes~\cite{ColorKinematics}. At higher
multiplicities, the integrand of the two-loop five-point amplitude in
the maximally supersymmetric theory of gravity, $\mathcal{N}=8$ supergravity
(SUGRA), has been known in a compact form for a number of
years~\cite{Carrasco:2011mn, Mafra:2015mja} and
still constitutes the state of the art in this direction.  Starting at
five loops, novel ideas~\cite{Bern:2017yxu,Bern:2017ucb}
were required to sidestep the difficulty of finding a BCJ form for
the integrand.
In light of this progress, it is hard to overstate the importance of the
double-copy procedure.  It has led to an explosion of gravity
integrand calculations and has fostered an
improved understanding of the ultraviolet character of $\NeqEight$ as
well as other theories of quantum gravity. For
the latest progress see refs.~\cite{Bern:2018jmv,BCJreviewToAppear}
and references therein. 

At the level of amplitudes, rather than integrands, whilst considerable
progress has been made in the planar sector of $\NeqFour$ 
(where bootstrap methods~\cite{Dixon:2011pw} have allowed the 
computation of six-point five-loop~\cite{Caron-Huot:2016owq} 
and seven-point four-loop~\cite{Dixon:2016nkn,Drummond:2018caf} amplitudes),
much less is known beyond the planar limit.  Supersymmetric theories of
gravitational interactions are inherently nonplanar.  For
$\NeqEight$, the maximally helicity violating (MHV) one-loop
amplitudes
were computed over 20 years ago~\cite{Bern:1998sv} and many other
one-loop computations have been performed since then.
At two loops, however, the state of the art has been the four-point
amplitude in
$\NeqEight$~\cite{Bern:1998ug,SchnitzerN8UniformTrans,QueenMaryN8UniformTrans}
as well as in $\mathcal{N}\geq4$
supergravity~\cite{BoucherVeronneau:2011qv},\footnote{See the
noted added at the end of the introduction.} 
with partial two-loop results available for the four- and
five-point all-plus amplitudes in Einstein
gravity~\cite{Bern:2015xsa,Bern:2017puu,Dunbar:2017qxb}.

In the absence of a bootstrap program 
for nonplanar amplitudes,
the main obstacle to obtaining higher multiplicity results in  
 nonplanar 
sectors has been
the difficulty of constructing the relevant integration-by-parts (IBP) identities
\cite{IBP1,IBP2}, required for both the reduction of the
integrand and the calculation of the master integrals.
However, this field has
seen major developments in recent years, in particular with its
reformulation in terms of unitarity cuts and computational
algebraic geometry~\cite{Gluza:2010ws,Ita:2015tya,Larsen:2015ped,Boehm:2018fpv,Abreu:2017hqn, Kosower:2018obg}, as
well as with the usage of finite-field methods
\cite{vonManteuffel:2014ixa, Peraro:2016wsq, Maierhoefer:2017hyi, 
Abreu:2017hqn, Smirnov:2019qkx}. A combination of
these improvements has unlocked the pathway to computing more complex higher
multiplicity amplitudes at two loops in a variety of theories.
Employing the method of differential 
equations~\cite{Kotikov:1990kg,Bern:1992em,Gehrmann:1999as}
in a canonical basis~\cite{Henn:2013pwa}, by now all master integrals 
relevant for two-loop five-point massless amplitudes are known,
both in the planar~\cite{Gehrmann:2000zt,Gehrmann:2015bfy,%
Papadopoulos:2015jft,Gehrmann:2018yef}
and nonplanar~\cite{Gehrmann:2001ck,Chicherin:2017dob,Abreu:2018rcw,%
Abreu:2018aqd,Chicherin:2018mue,Chicherin:2018old}
sectors (at least at the level of the symbol~\cite{Goncharov:2010jf, Duhr:2011zq,Duhr:2012fh}). Furthermore, the complete set of leading-color (planar)
five-point two-loop planar amplitudes in QCD is now known
numerically~\cite{Badger:2017jhb,Badger:2018gip,Abreu:2017hqn,Abreu:2018jgq}
and the two-loop five-gluon scattering 
amplitudes in pure Yang-Mills are known 
analytically~\cite{Gehrmann:2015bfy,Badger:2018enw,Abreu:2018zmy}.
Very recently, these methods have led to the first analytic results for the 
symbol of the two-loop five-point $\NeqFour$ amplitude
including nonplanar
contributions~\cite{Abreu:2018aqd,Chicherin:2018yne}. 
This amplitude is simpler to compute than the one we study in
this paper because its integrand only involves numerators with
one power of loop momentum, while in $\NeqEight$ the numerators
have two powers of loop momentum~\cite{Carrasco:2011mn}.

In this work, we combine these advances in integration technology 
with integrand-level \emph{leading singularity}
techniques~\cite{Cachazo:2008vp} in order to 
compute the symbol of the two-loop five-point scattering amplitude 
in $\NeqEight$. Whilst for $\NeqFour$, leading singularities for 
MHV amplitudes are completely understood from the
Grassmannian~\cite{Arkani-Hamed:2014bca}, the situation in
$\NeqEight$ is less developed. Nonetheless, efficient techniques exist to
compute analytically the 4-dimensional leading singularities
on a case-by-case
basis~\cite{Bern:2018jmv,Herrmann:2016qea,Heslop:2016plj}.
These well-defined on-shell quantities encode non-trivial
properties of the theory and are therefore
interesting to study in their own right, see
e.g.~refs.~\cite{ArkaniHamed:2012nw,Herrmann:2018dja}.
As will be relevant for this paper, these functions are not
linearly independent but satisfy a number of residue theorems,
which were used recently to establish the absence of poles at infinity
in the two-loop five-point integrand for
$\NeqEight$~\cite{Bourjaily:2018omh}. 

For $\NeqFour$, the four-dimensional
leading singularities are MHV tree amplitudes,
or Parke-Taylor factors~\cite{Parke:1986gb}.
This fact was crucial for efficiently computing the symbol of 
the two-loop five-point 
amplitude~\cite{Abreu:2018aqd,Chicherin:2018yne}.
In this paper, the leading singularities of $\NeqEight$, not just in four
dimensions but also in $d=4-2\e$ dimensions, will systematically guide us to
construct an ansatz for the amplitude's symbol.
Employing the symbols of the master integrals from ref.~\cite{Abreu:2018aqd}
and numerical IBP reductions of the BCJ integrand~\cite{Carrasco:2011mn}
in a finite field, we can fix all parameters in the ansatz
and determine the symbol uniquely.
As predicted from the integrand's logarithmic singularity
structure~\cite{Bourjaily:2018omh},
our integrated result has uniform transcendentality
\cite{BDS,Dixon:2011pw,ArkaniHamed:2012nw,LipatovTranscendentality},
just like the four-point amplitude
\cite{SchnitzerN8UniformTrans,QueenMaryN8UniformTrans,BoucherVeronneau:2011qv}
and its four- and five-point $\NeqFour$
counterparts~\cite{SchnitzerN8UniformTrans,Abreu:2018aqd,Chicherin:2018yne}.
Furthermore, the result satisfies a number of interesting structural
properties.  For example, the function space is surprisingly 
simple
and closely related to that of the corresponding amplitude in
$\NeqFour$, and after an appropriate infrared subtraction
the contributions of $d$-dimensional leading singularities drop out.

The structure of the paper is as follows. We begin, in
section~\ref{sec:Integrand}, by describing the known integrand 
of the two-loop five-point scattering amplitude in $\NeqEight$.
From this integrand, we construct in section~\ref{sec:leadingSing}
a set of $4$- and $d$-dimensional leading singularities.
Next, in section~\ref{sec:amplitude_decomposition}, we
discuss our method for computing the symbol of the
amplitude. Then, in section~\ref{sec:validation} we discuss
various consistency checks satisfied by our result. In
section~\ref{sec:Structure} we discuss interesting features of the
amplitude.  Finally, we conclude in section~\ref{sec:outlook}.
We provide an appendix detailing our conventions for the kinematics
and symbol letters.
We also include a number of ancillary files, described below,
containing computer-readable expressions that are too lengthy to print.

\medskip

\textbf{Note added:} In the final stages of this work, the
preprint~\cite{Chicherin:2019xeg} appeared which also
investigated the two-loop five-point amplitude in $\N=8$
supergravity. The two computed amplitudes are in complete agreement.

\newpage
\vspace{-0pt}
\section{The $\N=8$ supergravity integrand}
\label{sec:Integrand}\vspace{-8pt}
%

In this paper we compute the two-loop five-point amplitude
in $\N=8$ supergravity. We first briefly discuss our
conventions and introduce some useful notation. 
We define normalized 
$L$-loop $n$-point amplitudes $M^{(L)}_n$ as
\begin{equation}
    \M_n^{(L)}(1,2,\ldots,n)
    =\left(\frac{\kappa}{2}\right)^{n+2(L-1)} \delta^{(16)}(\Q)
    \left(\frac{ e^{-\e\gamma_E} }{(4\pi)^{2-\e}}\right)^L \,
    M_n^{(L)}(1,2,\ldots,n)\,,
\end{equation}
where $\kappa^2 = 32\pi G_N$ is the gravitational coupling, and,
since we are concerned with MHV scattering amplitudes in the
maximally supersymmetric $\N=8$ theory, we also strip off the 
super-momentum conserving delta-function $\delta^{(16)}(\Q)$, 
which relates the scattering amplitudes with only graviton 
external states to all other scattering amplitudes for states in 
the same super-multiplet.  (All 256 states in $\NeqEight$ are
in the same super-multiplet.)
Defined in this
way, the amplitudes are totally Bose-symmetric in \emph{all} 
labels. The normalized four- and five-point tree amplitudes are 
given by 
\cite{Berends:1988zp}
\begin{align}\label{eq:trees}
M_4^{(0)} &= \frac{\sqb{12}}{\ab{34} \, N(4)} \,,
\qquad
M_5^{(0)}  = \frac{\tr_5}{N(5)} \,,
\qquad
\text{where }\quad N(n) \equiv \prod_{i=1}^{n-1} \prod_{j=i+1}^n
\spa{i}.{j} \,,
\end{align}
where we introduced the parity-odd 
$\varepsilon$-tensor contraction $\tr_5$ defined as
\begin{align}\begin{split}
\tr_5 &\equiv \varepsilon(1,2,3,4) \equiv 4 i \varepsilon_{\mu
\nu \rho \sigma} k_1^{\mu } k_2^{\nu} k_3^{\rho} k_4^{\sigma} 
	= \tr (\gamma^5 \slashed k_1 \slashed k_2 \slashed k_3
	\slashed k_4) \\
	&=\sqb{12}\ab{23}\sqb{34}\ab{41} - \ab{12}\sqb{23}\ab{34}
	\sqb{41}\,.
\label{eq:tr5Def}
\end{split}\end{align}

\begin{figure}[h]
\centering
\includegraphics[scale=.50]{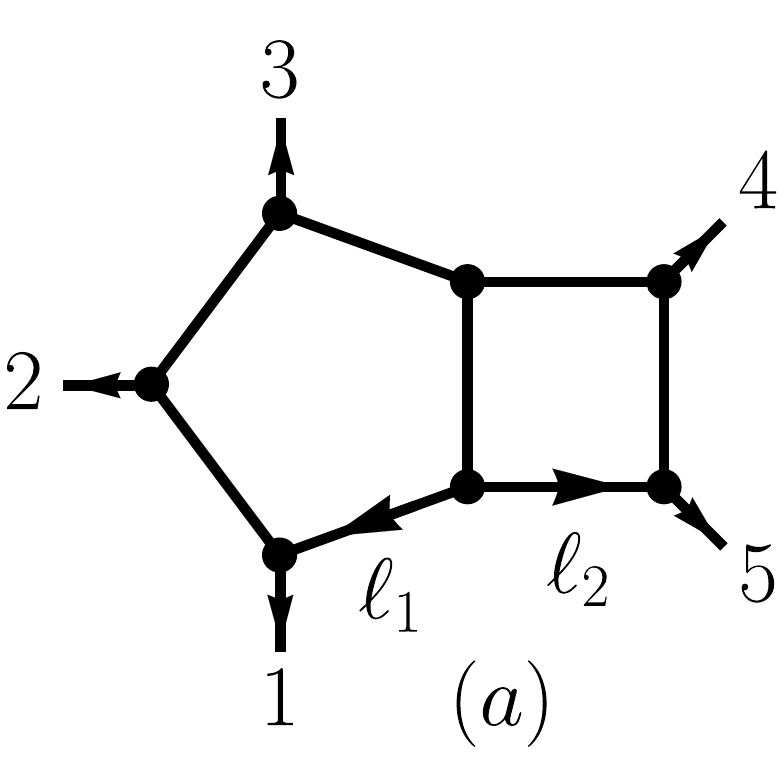} 
\hspace{.5cm}
\includegraphics[scale=.50]{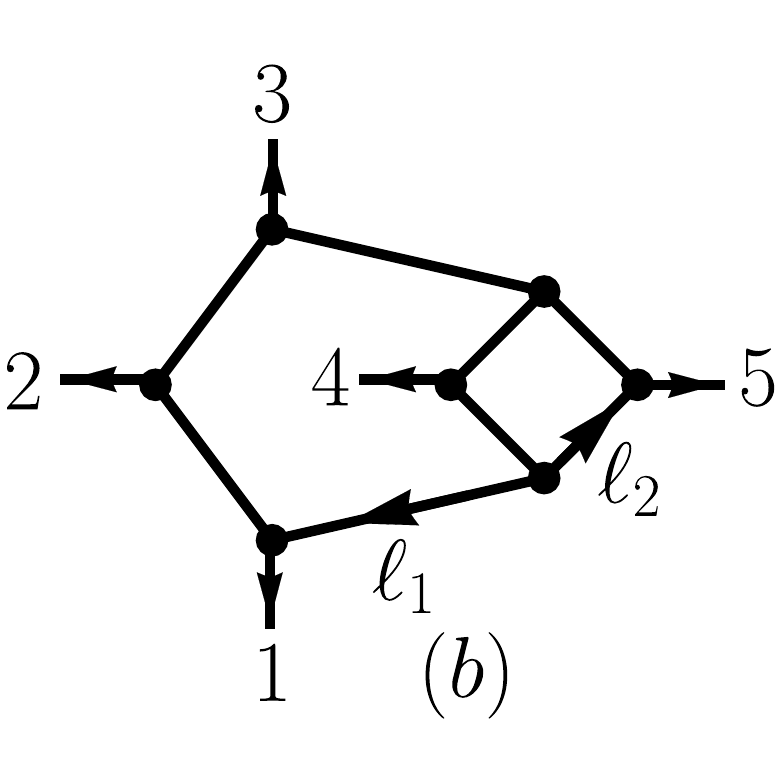}
\hspace{.7cm}
\includegraphics[scale=.50]{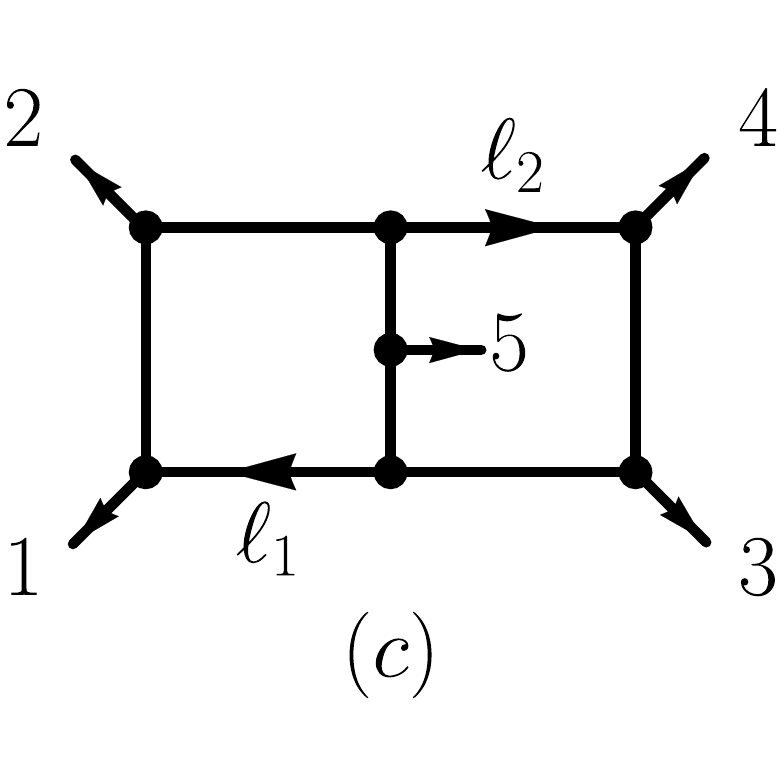} 
\newline
\vskip -.8cm
\includegraphics[scale=.50]{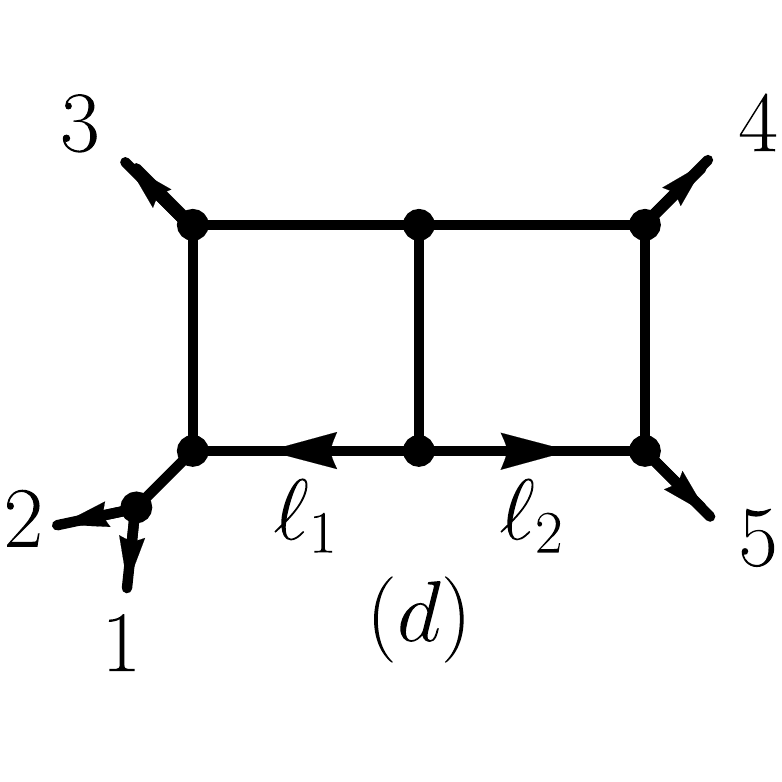}
\hspace{.5cm}
\includegraphics[scale=.50]{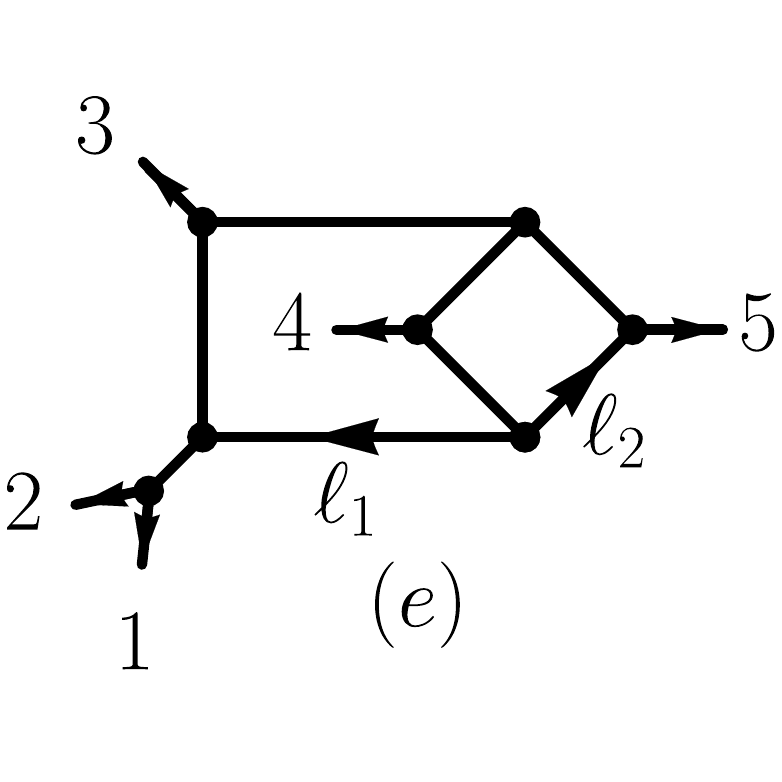}
\hspace{.5cm}
\includegraphics[scale=.50]{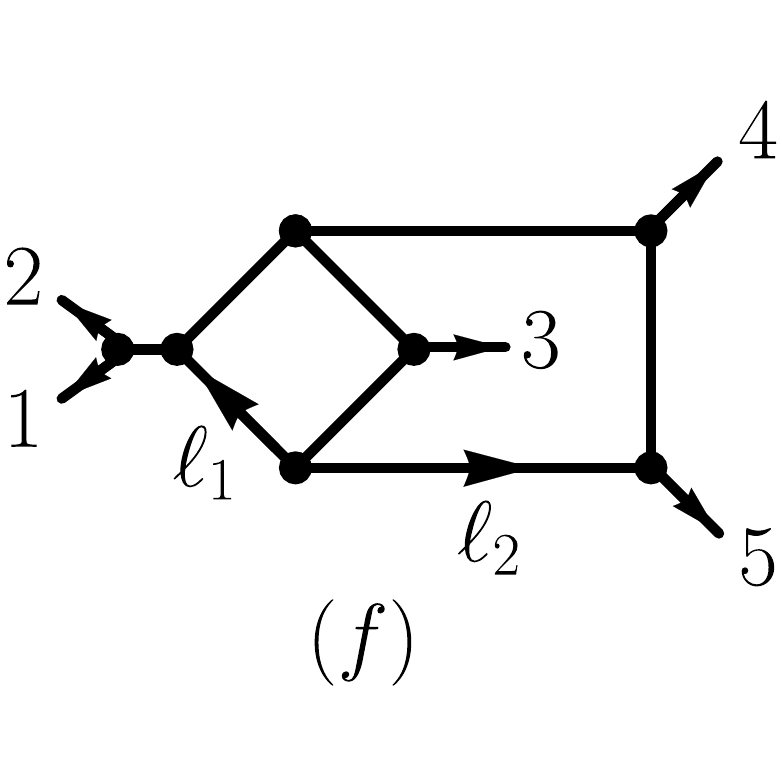}
\vspace{-20pt}
\caption{\label{fig:bcj_integrands}Diagram topologies entering the local representation of the 
two-loop five-point integrand of $\N=8$ supergravity~\cite{Carrasco:2011mn}. 
Each diagram has an associated kinematic numerator which we give in the main text.}
\end{figure}

For the two-loop five-point $\NeqEight$ amplitude, our starting
point is the integrand 
of~ref.~\cite{Carrasco:2011mn} which is valid in $d=4\mi2\epsilon$ 
space-time dimensions and is given in terms of the six topologies 
in Fig.~\ref{fig:bcj_integrands}. It was obtained using the BCJ
double-copy procedure~\cite{BCJ,BCJLoop,BCJSquare}.
Here, we adopt the conventions of~ref.~\cite{Carrasco:2011mn} 
and define the supergravity amplitude by
\begin{align}
M^{(2)}_5 = 
	\sum_{S_5} 	\left(\frac{ I^{(a)}}{2} \pl \frac{I^{(b)}}{4}
	\pl \frac{ I^{(c)}}{4} \pl 
	\frac{I^{(d)} }{2} \pl \frac{ I^{(e)}}{4} \pl \frac{I^{(f)}}{4} \right)\,. 
\hspace{-.3cm}
\label{eq:amp_bcj_rep}
\end{align}
The sum is over all $5!$ permutations of external legs and the 
rational numbers correspond to diagram symmetry factors. 
In \eqn{eq:amp_bcj_rep}, the integrals $I^{(x)}$ are
normalized as follows:
\begin{align}
 I^{(x)} = e^{2\epsilon\gamma_E}\int \frac{d^d\ell_1}
 {i\pi^{d/2}} \frac{d^d\ell_2}
 {i\pi^{d/2}} \frac{ \big[N^{(x)}
 (1,2,3,4,5;\ell_1,\ell_2)\big]^2}
 {\rho_1\ldots \rho_8}\,,
 \label{eq:defIx}
\end{align}
where the $\rho_i$ are inverse propagators 
(diagrams $(d)$, $(e)$ and $(f)$ include a loop-momentum independent 
$1/\sab{ij}$ propagator so that all integrals have the same 
mass dimension)
and the $N^{(x)}$ are the color-kinematics 
duality satisfying Yang--Mills numerators. For completeness, 
we provide the $\NeqFour$ BCJ numerators~\cite{Carrasco:2011mn} here,
\begin{align}
\label{eq:bcj_numerators}
\hskip -.6cm
\begin{split}
 N^{(a,b)}	& = \frac{1}{4} 
 \Big[ \gamma_{12} (2 s_{45} \mi s_{12} \pl \tau_{2\ell_1} \mi  \tau_{1\ell_1})
 \pl \gamma_{23} (s_{45}  \pl 2 s_{12} \mi \tau_{2\ell_1} \pl \tau_{3\ell_1}) 	\\
& \hskip .6cm \pl 2 \gamma_{45} (\tau_{5\ell_1} \mi \tau_
{4\ell_1})
 \pl \gamma_{13} (s_{12}  \pl s_{45} \mi \tau_{1\ell_1} \pl 
 \tau_{3\ell_1}) \Big]\,, \\[+5pt]
N^{(c)}	& = \frac{1}{4} \Big[\!
\gamma_{15} (\tau_{5\ell_1} \mi \tau_{1\ell_1}) 
 \pl \gamma_{25} (s_{12} \mi \tau_{2\ell_1} \pl \tau_{5\ell_1}) 
 \pl \gamma_{12} (s_{34} \pl \tau_{2\ell_1} \mi \tau_{1\ell_1} 
 \pl 2[s_{15} \pl  \tau_{1\ell_2} \mi  \tau_{2\ell_2}])\\
&  \hskip .7cm
\pl \gamma_{45}  (\tau_{4\ell_2} \mi \tau_{5\ell_2}) 
  \mi \gamma_{35} (s_{34} \mi \tau_{3\ell_2} \pl \tau_{5\ell_2}) 
\pl \gamma_{34} (s_{12} \pl \tau_{3\ell_2} \mi \tau_{4\ell_2} 
 \pl 2 [s_{45} \pl  \tau_{4\ell_1} \mi  \tau_{3\ell_1}] ) \!\Big], \\
 N^{(d,e,f)}\! &=  \gamma_{12} s_{45} \mi \frac{1}{4} \Big[2
 \gamma_{12} \pl  \gamma_{13} \mi \gamma_{23}\Big] s_{12} \,,
\end{split}\hskip -.8cm
\end{align}
where we follow the notation of ref.~\cite{Carrasco:2011mn}
and define
\begin{align}
s_{ij}=(k_i+k_j)^2 = 2 k_i \cdot k_j\,, \quad
\tau_{i\ell_j} = 2 k_i \cdot \ell_j \,,
\end{align}
and the various permutations of the function
\begin{align}
\label{eq:gamma_ij_def_BCJ}
 \gamma_{12}	\equiv \gamma_{12345} 
 			\equiv i \frac{\sqb{12}^2 \sqb{34}\sqb{45}\sqb{35}}
								{\sqb{12}\ab{23}\sqb{34}\ab{41}-\ab{12}\sqb{23}\ab{34}\sqb{41}} 
 			= i \frac{\sqb{12}^2 \sqb{34}\sqb{45}\sqb{35}}{\tr_5}\,.
\end{align}
The $\gamma_{ijklm}$ are totally symmetric in the last three
labels. Therefore, every $\gamma$-function can be uniquely
specified by its first two indices, in which it is
antisymmetric, $\gamma_{ij}=-\gamma_{ji}$. 
Five-point massless amplitudes depend on five independent
Mandelstam invariants, which can be chosen to be $s_{12}, 
s_{23}, s_{34}, s_{45}$ and $s_{51}$, and on the parity-odd
$\tr_5$ defined in eq.~\eqref{eq:tr5Def}.

A drawback of the BCJ representation in 
\eqn{eq:bcj_numerators} is the introduction of spurious
poles that cancel in the final amplitude.
For instance, from \eqn{eq:gamma_ij_def_BCJ} we see that
the various $\gamma_{ij}$-terms introduce poles at $\tr_5=0$,
which are known to be spurious in $\NeqFour$.
In ref.~\cite{Abreu:2018aqd}, detailed knowledge of the
Yang--Mills
leading singularities was valuable for efficiently computing
the two-loop five-point $\NeqFour$ amplitude.
This warrants the study of supergravity leading singularities in
order to follow the same approach in $\N=8$. More precisely, we
are going to use this information to identify a minimal set of
(linearly independent) rational coefficients relevant to
the two-loop five-point amplitude in $\NeqEight$.

\vspace{-0pt}
\section{Leading singularities}
\label{sec:leadingSing}\vspace{-8pt}

All \emph{known} amplitudes in $\NeqFour$ and $\NeqEight$ share 
the common feature of being functions of uniform transcendental (UT) weight 
\cite{SchnitzerN8UniformTrans,QueenMaryN8UniformTrans,
Bern:2014kca,Caron-Huot:2016owq,Bourjaily:2018omh}.
Whether this property persists at higher numbers of loops or legs
is an outstanding open question which the 
present work touches on. Following common 
`integrand lore'~\cite{ArkaniHamed:2012nw,Log} that logarithmic singularities 
imply uniform transcendentality of amplitudes, one expects that four point 
amplitudes in $\NeqEight$ remain uniformly transcendental through three loops. 
Starting at four loops, however, there are known pieces in the 
integrand~\cite{Bern:2014kca} that have non-logarithmic poles at 
infinity, which are expected to cause a transcendentality drop. 
Whether such contributions cancel in the final amplitudes---similar in 
spirit to enhanced cancellations of UV divergences 
(see e.g.~ref.~\cite{N5FourLoop})---remains an interesting open
problem. 
Staying at two loops but increasing the number of external legs shows 
a similar behavior. Starting at seven particles, non-logarithmic singularities 
appear in individual terms~\cite{Bourjaily:2018omh}, again signaling the 
potential for a transcendentality drop.
Nonetheless, for the two-loop five-particle amplitude under consideration here, 
these complications are absent and we therefore expect a uniform transcendental result. 

Furthermore, from general considerations 
\cite{Weinberg:1965nx,Akhoury:2011kq}, it can be
shown that there are no virtual collinear divergences
in a gravitational scattering amplitude. 
In the absence of UV divergences,
at each loop order one only finds (potentially overlapping) soft
divergences, leading to one pole in $\e$ per loop.
Concretely, this means that the two-loop five-point amplitude 
in $\NeqEight$, cf.~\eqn{eq:amp_bcj_rep}, can be
schematically written as
\begin{align}
\label{eq:schematicAmp}
M^{(2)}_5=\sum^4_{k=2} \frac{1}{\eps^{4-k}} \sum_{j}\,  
r_j \,f^{(k)}_j +\O(\eps)\,.
\end{align}
Here, the $f^{(k)}_j$ are \emph{pure functions} 
given by $\mathbb{Q}$-linear combinations of polylogarithmic
functions of weight $k$.\footnote{It is well known that all
master integrals for two-loop five-point massless amplitudes can
be written in terms of polylogarithms, as can be seen for
instance from their recent explicit calculation at symbol level
\cite{Abreu:2018aqd,Chicherin:2018old}.} 
We used the fact that from the analysis of the four-dimensional
integrand in ref.~\cite{Bourjaily:2018omh} it is clear that 
there are
only logarithmic poles, implying a maximal uniform weight
result according to common expectations 
\cite{ArkaniHamed:2012nw}. That is, if we assign
weight $-1$ to $\epsilon$, every term in \eqn{eq:schematicAmp}
is expected to be of weight $4$.
The~$r_j$ are in general ($d$-independent) algebraic functions 
of the kinematic data.
Using a convenient parametrization of  
massless five-point kinematics, such as the one obtained 
from momentum-twistor variables~\cite{Hodges}
in ref.~\cite{Badger:2013gxa} 
(cf.~appendix~\ref{app_subsec:twistor_parametrization} for details), 
we can guarantee that the $r_j$ are rational functions.
These rational functions are (linear combinations of) the
leading singularities we shall be discussing in this section.\footnote{In the context 
of correlation functions, the connection between leading singularities 
and rational functions was pointed out in ref.~\cite{Drummond:2013nda}.}

\subsection{Leading singularities in four dimensions}

As we mentioned in the introduction, a Grassmannian
representation for on-shell diagrams in 
$\NeqFour$~\cite{ArkaniHamed:2012nw} has been 
exploited to show that all leading singularities (maximal
codimension residues of the loop integrand, 
see e.g.~ref.~\cite{Cachazo:2008vp}) are given by certain linear 
combinations of Parke-Taylor 
factors~\cite{Arkani-Hamed:2014bca}.
In $\NeqFour$, all these leading singularity analyses were based on
inherently 4-dimensional arguments.
While the understanding of leading singularities in $\NeqEight$ 
is much less developed, it is nevertheless reasonable to
assume that at least a subset of the rational 
functions~$r_i$ in \eqn{eq:schematicAmp} are also linear 
combinations of 4-dimensional $\NeqEight$ leading
singularities. We will start by investigating these types of
rational functions.

We note that there now exists a very elegant and efficient way
for computing these leading singularities 
in gravity via the
Grassmannian duality~\cite{Heslop:2016plj,Herrmann:2016qea}. 
For gravity on-shell diagrams (on-shell functions that are given 
solely as products of three-point amplitudes) there is an efficient 
alternative method.  Because the BCJ double-copy is trivial at 
the level of three-point amplitudes, we can compute a gravity 
on-shell diagram as the square of the respective Yang-Mills one, 
multiplied by a Jacobian factor originating from the fact that propagators 
do not get squared in the double-copy procedure.
For readers more familiar with the BCJ representation in terms of cubic 
graphs, this double-copy structure of on-shell diagrams is equivalent to 
the statement that maximal cuts of cubic graphs always double-copy. 
The simplest two-loop five-point example is the planar on-shell function, 
\begin{align}
\raisebox{-40pt}{\includegraphics[scale=.40]{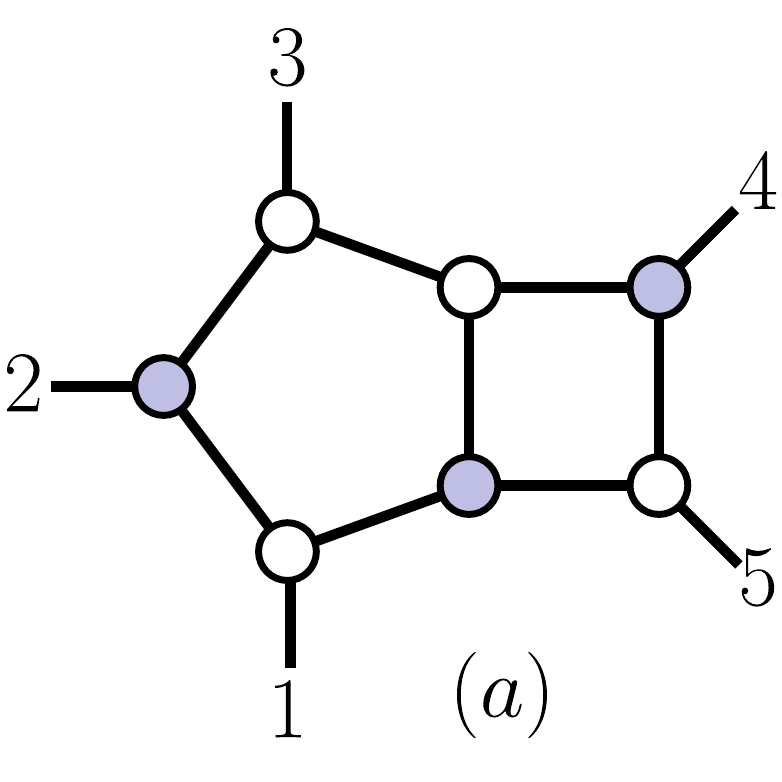}}   
\hskip .5cm
\begin{array}{cc}
 \text{LS}_{\text{SYM}}        & = \frac{1}{\ab{12}\ab{23}\ab{34}\ab{45}\ab{51}} \,, \\[6pt]
 \text{LS}_{\text{SUGRA}}   &\hskip .3cm = \frac{\sqb{12}\sqb{23}\sqb{45}^2}{\ab{12}\ab{23}\ab{34}\ab{45}\ab{51}\ab{13}}\,, \\
\end{array}
\end{align}
which we compute both in $\NeqFour$ and in $\NeqEight$ 
(suppressing coupling constants and super-momentum conserving 
delta functions).
Evaluating the residue where all inverse propagators $\rho_i$
are put on-shell, $\rho_i=0$, 
introduces a Jacobian
$\J$, and completely localizes the eight degrees of freedom of
the two 4-dimensional loop momenta $\ell_j$.
The on-shell Jacobian is
\begin{align}
\J= 
\det \frac{\partial \rho_i}{\partial \ell_j}\bigg\vert_{\rho_i=0}
= \frac{\ab{12}
\ab{23}\ab{34}\ab{45}\ab{51}\sqb{12}\sqb{23}\sqb{45}^2}{\ab{13}}\,.
\label{eq:onShellJac}
\end{align}
It is now easy to see that the gauge and gravity leading singularities are related in the prescribed way
\begin{align}
 \text{LS}_{\text{SUGRA}}  = \text{LS}^2_{\text{SYM}} \times \J\,.
 \label{eq:doubleCopyLS}
\end{align}
For two-loop five-point scattering, the relevant $\NeqEight$ leading singularities are all permutations of the following basic structures:
\begin{align}
\label{eq:os_funcs_2loop_5pt_0}
\begin{split}
\eqnDiag{\includegraphics[scale=.45]{./figures/planar_penta_box_int_bcj_label_os}} \twhite{d}& = 
	\frac{\sqb{12}\sqb{23}\sqb{45}^2}{\ab{12}\ab{23}\ab{34}\ab{45}\ab{51}\ab{13}} 
\\
\eqnDiag{\includegraphics[scale=.47]{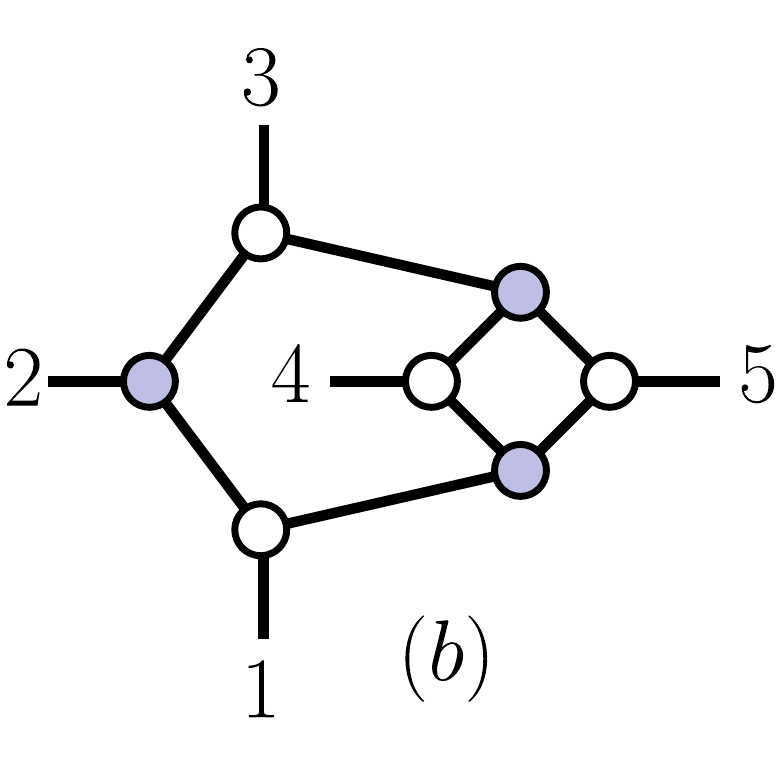}}  &= 
	\frac{\sqb{12}\sqb{23}\sqb{45}^2}{\ab{12}\ab{23}\ab{14}\ab{34}\ab{35}\ab{51}} 
\end{split}
\end{align}
\begin{align}
\label{eq:os_funcs_2loop_5pt}
\begin{split}
\raisebox{-54pt}{\includegraphics[scale=.45]
{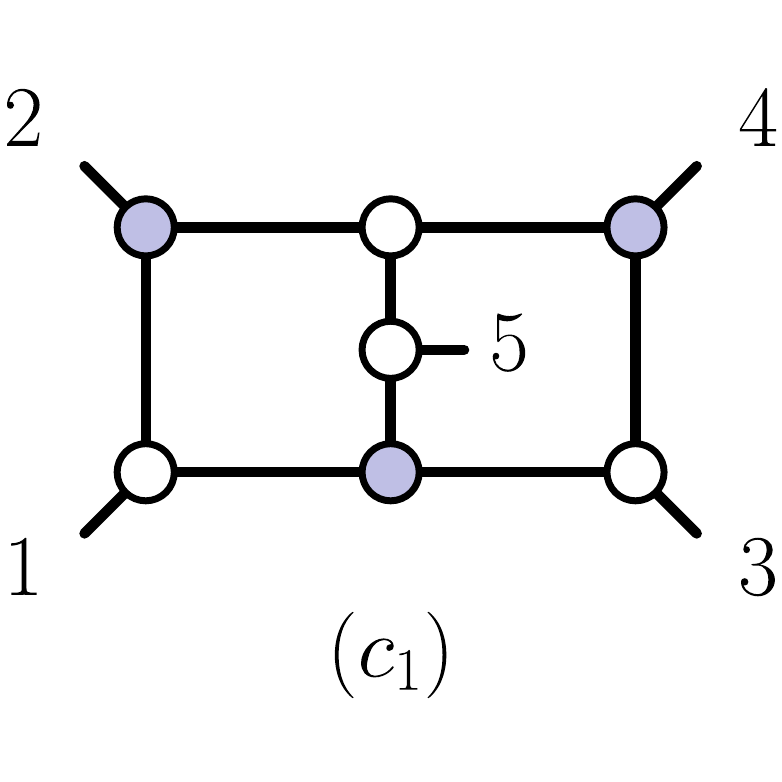}}& =
	\frac{\sqb{24}\sqb{34}\sqb{12}^2}{\ab{13}\ab{25}\ab{34}\ab{35}\ab{45}\ab{51}} 
	+ (1\leftrightarrow3, 2\leftrightarrow4)
\\[-19pt]
\raisebox{-54pt}{\includegraphics[scale=.45]
{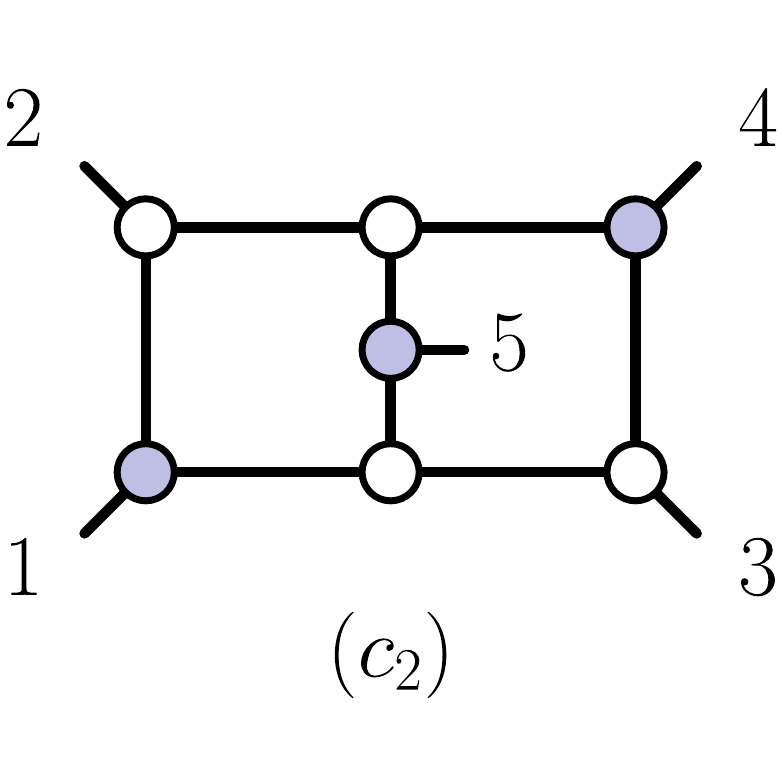}}& = 
	\frac{\sqb{12}\sqb{34}\sqb{45}\sqb{51}}{\ab{12}\ab{13}\ab{24}\ab{25}\ab{34}\ab{35}}\\[-15pt]
\end{split}
\end{align}
These on-shell diagrams are not all independent but satisfy 
a number of linear relations due to residue theorems, 
see e.g.~ref.~\cite{Bourjaily:2018omh}. Taking all 120
permutations of the on-shell functions in eqs.~(\ref{eq:os_funcs_2loop_5pt_0}) and (\ref{eq:os_funcs_2loop_5pt}),
we find 40 linearly independent terms.
They can be chosen, for example, from the set of 60
inequivalent permutations of the
on-shell diagrams $(c_2)$.
If all rational factors $r_i$ in \eqn{eq:schematicAmp}
could be identified with 4-dimensional on-shell diagrams, we would
conclude that the space spanned by the $r_i$ 
is 40-dimensional, in the
same way that the six independent five-point
Parke-Taylor factors were found from
4-dimensional on-shell diagrams 
in $\NeqFour$~\cite{Arkani-Hamed:2014bca}.

To verify whether the set of 40 independent leading singularities
is really adequate for the
decomposition in \eqn{eq:schematicAmp}, it is sufficient 
to numerically reduce the amplitude in 
\eqn{eq:amp_bcj_rep} via IBP
relations onto a basis of master integrals, e.g.~the one
introduced in ref.~\cite{Abreu:2018aqd}.
Since the $r_i$ are rational functions, the efficiency of the
reduction can be improved by using finite-field 
techniques. We will describe the reduction procedure in more detail in 
section~\ref{sec:NumericalReduction}. For now we simply note that by reducing
the amplitude on sufficiently many kinematic points (more than 45),
we find that the space spanned by the coefficient functions $r_i$ is
actually $45$-dimensional.
This observation is confirmed by analyzing the amplitude on a 
so-called \emph{univariate slice}, 
which, following the procedure introduced in 
ref.~\cite{Abreu:2018zmy}, can be used to completely determine
the denominators of the $r_i$. Indeed, we find that
there are new coefficients with poles at $\tr_5=0$, which are
inconsistent with the results obtained from the 4-dimensional
leading singularities.

\subsection{Leading singularities in $d$ dimensions}

In order to find the missing rational structures we relax the
condition of working strictly in 4 dimensions, and compute
leading singularities in $d$ dimensions. 
This extension is natural given that the amplitude is not well
defined in exactly 4 dimensions, and it is expected that
pieces that vanish in strictly $d=4$ potentially become 
important in the context of dimensional regularization.
To further motivate the need for $d$-dimensional leading
singularities, we note that they are already necessary for
one-loop five-point amplitudes beyond $\epsilon^0$. Indeed, 
while the scalar pentagon in 4 dimensions is trivially
reducible to boxes, the leading singularity of the massless
scalar pentagon integral in $d=6-2\epsilon$ dimensions, which
contributes to the amplitude at order 
$\epsilon$~\cite{Bern:1998sv},
is precisely given by $1/\tr_5$, 
see e.g.~ref.~\cite{Abreu:2017ptx}.

\begin{figure}
  \centering
  \includegraphics[width=0.4\textwidth]{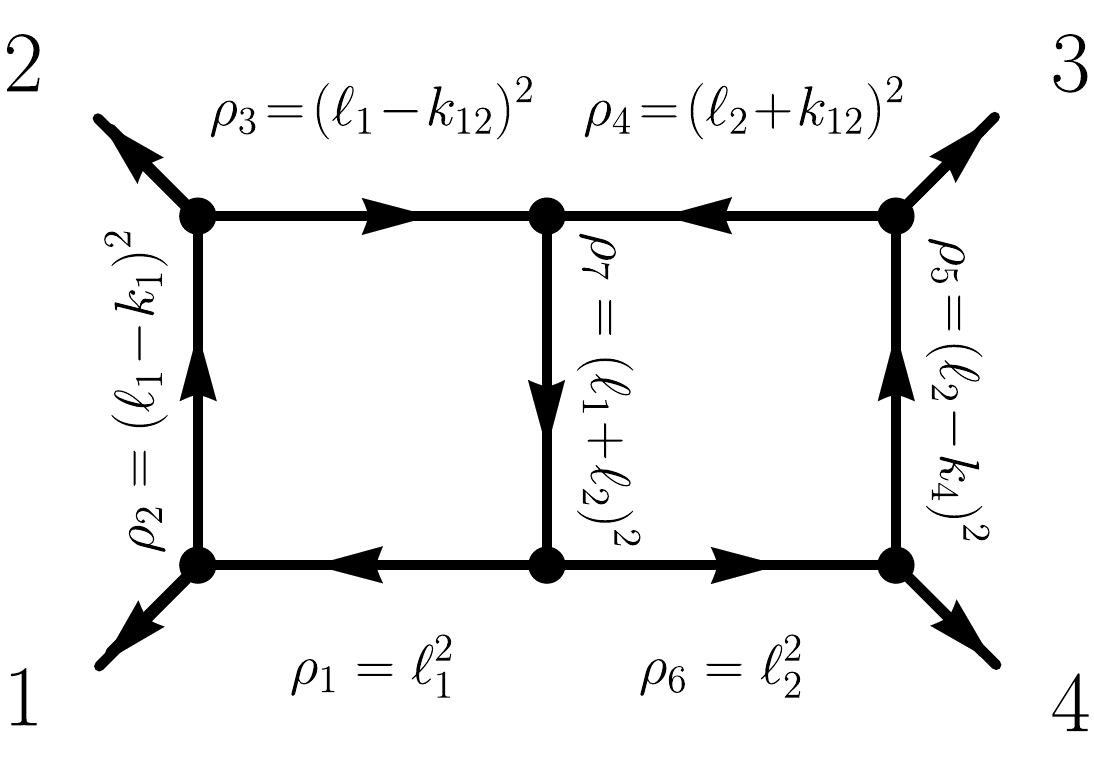}
  \caption{The double-box diagram.}
  \label{fig:doublebox}
\end{figure}

In order to compute the $d$-dimensional leading singularities,
we use the
Baikov representation~\cite{Baikov:1996rk, Baikov:1996iu,
Grozin:2011mt, Frellesvig:2017aai} for the topologies in the
$\NeqEight$ integrand given in \eqn{eq:amp_bcj_rep}.
To explain our approach to these calculations in a simple
setting, we first consider the all-massless
planar double-box integral in fig.~\ref{fig:doublebox} 
with numerator $\N$ and perform an analysis similar to that of
ref.~\cite{Chicherin:2018old}.
The kinematic variables for the double box are $s=(k_1 + k_2)^2$
and $t=(k_2+k_3)^2$. The inverse propagators 
$\rho_1, \, \rho_2, \,\ldots \, , \rho_7$ are labelled in fig.~\ref{fig:doublebox}, 
and we complete them by the irreducible numerators
\begin{equation}
\rho_8 = (\ell_1 + k_4)^2, \quad \rho_9 = (\ell_2 + k_1)^2 \, . 
\end{equation}
By integrating out ``angular" variables, we rewrite the loop 
integral in terms of the Baikov variables 
$\rho_1, \, \rho_2, \, \dots , \rho_9$, introducing a Jacobian
from the change of variables. Omitting 
constant normalization factors, the integral is
\begin{equation}
\label{eq:dboxN}
I_{\textrm{dbox}} [\mathcal N] = \int d\rho_1 d\rho_2 \dots d\rho_9 \frac {\mathcal N} {\rho_1 \rho_2 \dots \rho_7} \frac{ G(k_1, k_2, k_3)^\epsilon } {G(\ell_1, \ell_2, k_1, k_2, k_3)^{1+\epsilon}},
\end{equation}
where we use $G(q_1,q_2,\dots,q_r)$ to denote
the Gram determinant of the set of vectors $\{q_1,\ldots,q_r\}$,
which is given by $\det (2q_i \cdot q_j),\,1\leq i,j \leq r$.
Since there is a linear map between the Baikov $\rho$ variables
and scalar
products involving the loop momenta $\ell_i$, 
the Gram determinants are polynomial in the Baikov variables 
$\rho_1, \rho_2,\dots, \rho_9$ and the dot products of external
momenta. The \emph{Baikov polynomial} $P(\rho_i)$ is defined as
\begin{equation}
P(\rho_i) \equiv G(\ell_1, \ell_2, k_1, k_2, k_3) \, .
\end{equation}
The leading singularities correspond to evaluating codimension nine residues 
where all nine $\rho_i$ variables are fixed. Correspondingly, this fixes
nine degrees of freedom for the loop-momenta $\ell_i$. In
strictly $d=4$ dimensions, the system would be over constrained 
as the space of loop-momenta only has eight degrees of
freedom.
At leading order in the Laurent-expansion in $\epsilon$, we can
thus compute the $d$-dimensional leading singularities of the
double box in \eqn{eq:dboxN} by evaluating the global
residues of the nine-form $\Omega$ defined by\footnote{
We stress here the difference between maximal cuts and leading 
singularities, as discussed in 
e.g.~refs.~\cite{Kosower:2011ty,Abreu:2017ptx}. 
The former are a property of the \emph{integral} which can be
interpreted as some iterated discontinuity.
Computing them requires specifying an integration contour
and residues are not taken at the Jacobian poles.
The latter are a
property of the \emph{integrand}, and correspond to
some residue at a global pole, with no interpretation as
discontinuities in general. Evaluating the global residues
requires setting $\epsilon=0$ in eq.~\eqref{eq:dboxN} to remove
branch-cut ambiguities.}
\begin{equation}
\int \Omega\equiv
\int d\rho_1 d\rho_2 \dots d\rho_9\ \frac {\mathcal N(\rho_i)} 
{\rho_1 \rho_2 \dots \rho_7 \, P(\rho_i)} \,.
\end{equation}
To proceed, we first take residues at 
$\rho_1 =\rho_2 =\dots = \rho_7 = 0$, i.e.~we impose the 
maximal-cut conditions, upon which the Baikov polynomial
only depends on the irreducible numerators and external
kinematics,
\begin{equation}
P_{\textrm{max-cut}} = 2s\, \rho_8 \rho_9 \left[(s\pl\rho_8)\rho_9 \pl s(\rho_8 \mi t)\right].
\end{equation}
On the maximal cut, we obtain a two-form in the two variables $\rho_8$ and $\rho_9$,
\begin{equation}
\Omega_{\text{max-cut}} = \frac{d\rho_8 d\rho_9 \, \mathcal N} {
2s\,\rho_8 \rho_9 
\left[(s\pl\rho_8)\rho_9 \pl s(\rho_8 \mi t) \right]}\,.
\label{eq:r8r9integral}
\end{equation}
We can now take further residues of $\Omega_{\text{max-cut}}$ at $\rho_8 = \rho_8^0$ and then at 
$\rho_9 = \rho_9^0$, for all possible choices of $\rho_8^0$ and $\rho_9^0$. 
More precisely, we calculate
\begin{align}
\underset{\rho_9= \rho_9^0}{\text{Res}}\left[
\underset{\rho_8= \rho_8^0}{\text{Res}}\ \frac{\mathcal N} { 
2s\, \rho_8 \rho_9 \left[(s\pl\rho_8)\rho_9 \pl s(\rho_8 \mi t) \right] }
\right] \, .
\label{eq:r8r9residue}
\end{align}
For illustration purposes, consider the scalar double-box
integral with $\mathcal N = 2 s^2 t$.
It is easy to see that \eqn{eq:r8r9residue} evaluates to
$\pm 1$ for any of the four different choices of singularities,
\begin{align}
\begin{split}
\bullet\, &  \rho_8^0=0, \hskip 2.05cm  \rho_9^0 = 0, \\
\bullet\, &  \rho_8^0=0, \hskip 2.05cm   \rho_9^0 = t,  \\[-2pt]
\bullet\, &  \rho_8^0= \frac{s(t\mi\rho_9)}{(s\pl\rho_9)}, \qquad  \rho_9^0 = 0, \\
\bullet\, &  \rho_8^0= \frac{s(t\mi\rho_9)}{(s\pl\rho_9)},  \qquad  \rho_9^0 = t.
\end{split}
\end{align}
In other words, the integral $I_{\textrm{dbox}}[2 s^2 t]$ has \emph{unit leading
singularities} in $d$ dimensions.
In fact, this property can be made manifest by a change of variables to recast the two-form $\Omega_{\text{max-cut}}$ 
into a ``dlog-form",
\begin{equation}
\Omega_{\text{max-cut}}
 = \frac{s^2 t \, d\rho_8 d\rho_9} { s\,\rho_8 \rho_9 \left[(s\pl\rho_8)\rho_9 \pl s(\rho_8 \mi t)\right] } 
	      = d\log \frac{\rho_8 \mi t}{\rho_8} \wedge  d\log \frac{\rho_9}{(s\pl\rho_8)\rho_9 \pl s(\rho_8 \mi t)}\,.
\end{equation}
We stress again that the above formalism is inherently
$d$-dimensional, with integration variables and integration
measures differing from the 4-dimensional case. In particular,
the leading singularities computed are sensitive to components 
of the loop momenta beyond 4 dimensions. For example, 
consider the numerator 
$\mathcal N = P(\rho_i) = G(\ell_1, \ell_2,k_1, k_2, k_3)$, 
which vanishes identically for 4-dimensional loop momenta due 
to anti-symmetrization over more than 4 momenta in the Gram 
determinant. Such a numerator is ``undetectable" by 
$4$-dimensional leading singularities, but will contribute to 
double poles at $\rho_8 = \infty, \rho_9 = \infty$ in 
\eqn{eq:r8r9integral} when considering $d$-dimensional 
residues.

Let us now return to two-loop five-point topologies.
To find the full space of rational prefactors $r_i$ in the 
$\NeqEight$ amplitude~(\ref{eq:schematicAmp}), which, as we have established, 
has five extra elements beyond the $40$-dimensional space of 
4-dimensional leading singularities, we first compute the
$d$-dimensional leading singularities of the planar top-level
diagram $(a)$ in fig.~\ref{fig:bcj_integrands}.
In this case, the original Baikov representation is not
the most convenient. Instead, we follow the method of 
ref.~\cite{Chicherin:2018old} to compute leading singularities 
using the loop-by-loop Baikov representation of 
ref.~\cite{Frellesvig:2017aai}. We define the Baikov variables
for the planar pentabox, consisting of eight inverse propagators $\rho_1, \rho_2, \dots, \rho_8$,
followed by three irreducible numerators $\rho_9, \rho_{10}, \rho_{11}$,
\begin{align}
&\rho_1 = \ell_1^2, \quad \rho_2 = (\ell_1 - k_1)^2, \quad \rho_3 =(\ell_1 - k_1 - k_2)^2,
\quad \rho_4 = (\ell_1 + k_4 + k_5)^2, \nonumber \\
&\rho_5 = (\ell_2 - k_4 - k_5)^2, \quad \rho_6 = (\ell_2 - k_5)^2, \quad \rho_7 = \ell_2^2,
\quad \rho_8 = (\ell_1 + \ell_2)^2, \nonumber \\
&\rho_9 = (\ell_2 - k_3)^2, \quad \rho_{10} = (\ell_2 - k_1)^2, \quad \rho_{11} = (\ell_1 + k_5)^2 \, .
\label{eq:pentaboxBaikovVars}
\end{align}

We first consider the pentagon sub-loop on the left of
diagram $(a)$, with loop momentum $\ell_1$ and outgoing external 
momenta $k_1,~k_2,~k_3,~k_4 + k_5 - \ell_2,~\ell_2$. The numerator
in the BCJ integrand is $[N^{(a)}]^2$, as defined in eqs.~\eqref{eq:defIx}
and \eqref{eq:bcj_numerators}. Performing standard one-loop tensor reduction 
for this sub-loop, we eliminate all
$\ell_1$ dependence in $[N^{(a)}]^2$
and produce an expression $\tilde N$ which is nonlocal in $\ell_2$.
This step removes all dependence on $\rho_{11}$ in the integrand.
The remaining numerators can all be expressed in terms of the
irreducible numerators $\rho_{9}$ and $\rho_{10}$ of eq.~\eqref{eq:pentaboxBaikovVars},
as well as the inverse propagators which are set to zero on the maximal cut.

As discussed above for the two-loop double box, we 
then change the integration variables of the pentagon sub-loop
from $\ell_1^\mu$ to the five inverse propagators of the pentagon, which
are among the Baikov variables in eq.~\eqref{eq:pentaboxBaikovVars}.
Up to constant factors, we have
\begin{equation}
\int\! d^d \ell_{1} \propto \int\! \frac{d\rho_1 d\rho_2 d\rho_3 d\rho_4 d\rho_8 \, G(\ell_2, k_1, k_2, k_3)^{1/2 + \epsilon}} {G(\ell_1, \ell_2, k_1, k_2, k_3)^{1 + \epsilon}},
\end{equation}
where we again used the Gram determinant notation introduced after eq.~\eqref{eq:dboxN}.

Finally, we also change the
integration variables $\ell_2^\mu$ of the remaining triangle
sub-loop to the three inverse propagators, 
$\rho_5, \rho_6, \rho_7$ and the two $\ell_2$-dependent
irreducible numerators $\rho_9, \rho_{10}$,
\begin{equation}
\int\! d^d \ell_{2} \propto \int\! \frac{d\rho_5 d\rho_6 d\rho_7 d\rho_{9} d\rho_{10} \, G(k_1, k_3, k_4, k_5)^{1/2 + \epsilon}} {G(\ell_2, k_1, k_3, k_4, k_5)^{1 + \epsilon}} \, .
\end{equation}
The $(d-5)$ remaining ``angular" variables of the $\ell_2$ integration have been trivially integrated over, because after $\ell_1$-integration, the pentagon sub-loop produces an expression which depends only on $\rho_5, \rho_6, \dots, \rho_{10}$.
Now the differential form associated to the
pentabox contribution to the amplitude is written as (up to
constant factors)
\begin{equation}
\Omega_{\text{penta-box}} \sim
\frac{ \left(\prod_{i=1}^{10} d \rho_i \right) \tilde N \, G(\ell_2, k_1, k_2, k_3)^{1/2 + \epsilon} \, G(k_1, k_3, k_4, k_5)^{1/2 + \epsilon}} {G(\ell_1, \ell_2, k_1, k_2, k_3)^{1 + \epsilon} \, G(\ell_2, k_1, k_3, k_4, k_5)^{1 + \epsilon}},
\end{equation}
where all the Gram determinants are expressed in terms of the
Baikov variables $\rho_1$ through $\rho_{10}$. 
Recall that $\tilde N$ is obtained
from the original BCJ numerator $[N^{(a)}]^2$ via
tensor reduction for the $\ell_1$ sub-loop, and is a rational function of
the Baikov variables. As in the double-box example,
we neglect $\epsilon$ in the exponents, and obtain leading
singularities by successively computing residues in the 10 Baikov variables.

To complete the example and explicitly compute one of the 
leading singularities, we
cut the 8 propagators $\rho_1$ through $\rho_8$, then
take the residue of $\rho_{10} = (\ell_2 - k_1)^2$ at~0, and
finally take the residue of $\rho_9 = (\ell_2 - k_3)^2$ at 
$s_{45} - s_{12}$. 
The leading singularity obtained in this way is, up to a
constant,
\begin{equation}
	\text{LS}^{\text{penta-box}}_{\text{SUGRA}} \sim
	\frac{s_{12} [12][23][34][45][51]}
	{\tr_5\,\langle 12 \rangle \langle 23 \rangle \langle 34 
	\rangle \langle 45 \rangle \langle 51 \rangle} \, .
	\label{eq:sugraLeadingSingD}
\end{equation}
This expression turns out to be enough to identify the remaining
five rational functions needed for the decomposition in 
\eqn{eq:schematicAmp}, which means we do not need to
study the leading singularities of diagrams $(b)$ and $(c)$ in 
fig.~\ref{fig:bcj_integrands}.
Indeed, the above expression and its images under permutations 
of external legs produce exactly the five extra rational
prefactors in the amplitude which were not captured by the 
4-dimensional leading singularities discussed in the previous
subsection. We note that this rational function has a single
pole at $\tr_5=0$, which is consistent with the behavior
expected from analyzing the amplitude on a univariate slice.
Furthermore, since all the eight propagators are cut in the 
above calculation,
the $d$-dimensional leading singularity we computed for $\mathcal N=8$
SUGRA is again a double copy of the $\mathcal N=4$ SYM
counterpart,
due to the trivial double-copy property of the three point amplitudes
in arbitrary dimensions.\footnote{In this case, the double copy relation
eq.~\eqref{eq:doubleCopyLS} involves a different Jacobian from eq.~\eqref{eq:onShellJac},
computed from the Baikov representation. This new Jacobian is the source of
$\tr_5$ in the denominator of eq.~\eqref{eq:sugraLeadingSingD}.}

In summary, we find that for $\NeqEight$ the 4-dimensional
leading singularities are not sufficient to determine all
rational functions and a genuine $d$-dimensional analysis is
required. Relevant for the remainder of this work, we choose 
the following leading singularities (and permutations thereof)
\begin{align}
\label{eq:gravity_LS_basic_4d}
 \text{$d=4$:}\quad  &  \frac{\sqb{12}\sqb{34}\sqb{45}\sqb{51}}{\ab{12}\ab{13}\ab{24}\ab{25}\ab{34}\ab{35}} 
 		 &&+ 39 \text{ perms.} \\
 \text{general }d:\quad  & \frac{s_{12} [12][23][34][45][51]} {\tr_5 \ab{12}\ab{23}\ab{34}\ab{45}\ab{51}}
 		&&  + 4 \text{ perms.}
\label{eq:gravity_LS_basic_d}
\end{align}
as the basis of 45 rational coefficients $r_i$ required to expand the two-loop 
five-point amplitude in $\NeqEight$ in \eqn{eq:schematicAmp}. 
The explicit choice of all $r_i$ is given 
in the ancillary file \texttt{ri$\_$to$\_$brackets.txt}. 

One might have already expected the necessity for 
considering $d$-dimensional cuts given that the
amplitude is not defined in strictly 4 dimensions. This
observation highlights once more the very special properties of
$\NeqFour$, where the 4-dimensional leading singularities were
sufficient. However, the fact that we are 
able to construct \emph{all} rational coefficients of the 
amplitude from a cut analysis is very encouraging, and has a
large potential for applications outside maximally 
supersymmetric theories. In fact, we envision that a similar
analysis can help organize QCD computations in a clean and
systematic manner.

\section{Construction of the amplitude} 
\label{sec:amplitude_decomposition}

In the previous section we discussed the fact that two-loop
five-point $\NeqEight$ amplitudes are of uniform transcendental
weight, i.e., at each order in $\epsilon$ they can be written as
kinematically-dependent linear combinations of pure
transcendental functions, see eq.~\eqref{eq:schematicAmp}. Here,
we will start by further characterizing the pure 
functions~$f_j^{(k)}$. 
They are $\mathbb{Q}$-linear combinations 
of polylogarithms of weight $k$, which can be written
as iterated integrals over so-called ``$\dlog$-forms''. That is,
they can be written as
\begin{equation}
    f_j^{{(k)}} = \sum_{\alpha_1, \ldots, \alpha_k} c^j_{\alpha_1, \ldots, \alpha_k}
    \int \dlog W_{\alpha_1} \cdots \dlog W_{\alpha_k},
    \label{eq:IteratedIntegral}
\end{equation}
where the weight corresponds to the number of integration
kernels and the $c^j_{\alpha_1,\ldots \alpha_k}$ are rational numbers. 
In equation \eqref{eq:IteratedIntegral} there is an
implicit integration contour, but a large amount of the 
analytic properties of the functions is contained in the
$k$-fold $\dlog$ integrand, which is a differential form 
on the space of external kinematics. As such, in the remainder 
of this paper we will work at the level of the so-called
\emph{symbol}~\cite{Goncharov:2010jf, Duhr:2011zq,Duhr:2012fh},
denoted $S\left[f_j^{(k)}\right]$, and given by:
\begin{equation}
    S\left[f_j^{(k)}\right] = 
    \sum_{\alpha_1, \ldots, \alpha_k}
    c^j_{\alpha_1, \ldots, \alpha_k} [W_{\alpha_1}, \ldots, W_{\alpha_k}].
    \label{eq:SymbolMap}
\end{equation}
Here, we use square brackets to indicate a formal tensor 
product of the symbol letters~$W_\alpha$. Although we will often
omit the map $S$, from now on we consider all transcendental
functions at the symbol level.

In equations \eqref{eq:IteratedIntegral} and 
\eqref{eq:SymbolMap}, the $W_\alpha$ are algebraic functions of
the external kinematics. The full set is referred
to as an \emph{alphabet}, and each $W_\alpha$ as a 
\emph{letter}. For massless five-point scattering at
two loops, the symbol alphabet is given by a set of 31 
letters~\cite{Chicherin:2017dob} which we summarize in appendix 
\ref{kinematicsappendix} for convenience. 
Most letters correspond to permutations of the four-point
one-mass two-loop alphabet, and only 6 letters are truly
five-point. They can be graded according to their parity, i.e.,
their transformation under complex conjugation 
$\ab{\cdot} \lra \sqb{\cdot}$ or, equivalently, under
$ \tr_5\to-\tr_5$ with $\tr_5$ as defined in \eqn{eq:tr5Def}.
Five letters are parity-odd  ($\alpha \in \{26,\!...,30\}$), 
and can be expressed as ratios of spinor-brackets, 
see eq.~\eqref{eq:simple_alphabet} in appendix
\ref{kinematicsappendix}.
The parity-even letter ($\alpha\!=\!31$) is $\tr_5$.
All letters with $\alpha \in \{1,\!...,25\}$ are even under
parity because they do not depend on $\tr_5$.
With this grading, the amplitude is
naturally split into parity-even and parity-odd parts.
At symbol level, the parity grading can be
found from the number of parity-odd letters, 
$W_{26},...,W_{30}$, in a given symbol tensor.

Returning to the $\NeqEight$ two-loop five-point amplitude, it
can then be decomposed as 
\begin{equation}
    M_5^{(2)} = \sum_{k=2}^4 \frac{1}{\epsilon^{4-k}} 
    M_{5,k}^{(2)}
    + O(\epsilon),
    \label{eq:AmpEpsilonExpansion}
\end{equation}
where
\begin{equation}
    S[M_{5,k}^{(2)}] = \sum^{31}_{\alpha_1=1}\!\! \cdots\!\!
    \sum^{31}_{\alpha_{k}=1}\sum^{45}_{j=1}
    c^j_{\alpha_1, \ldots, \alpha_{k}}\  r_j \times 
    \left[W_{\alpha_1}, \ldots, W_{\alpha_{k}}\right], \quad
    k=2,3,4 \,.
\label{eq:AmplitudeSymbol}
\end{equation}
The coefficients $r_j$ are the 45 rational functions identified
in the previous section and the
$c^j_{\alpha_1, \ldots, \alpha_{k}}\in \mathbb{Q}$ 
are rational numbers. 
Computing the symbol of the amplitude amounts to computing these
rational numbers.

\subsection{Pure basis of master integrals} 
\label{sec:PureBasis}

The first step in computing the symbol of the $\NeqEight$
amplitude is the calculation of the symbol of a complete set of
master integrals, on which we can then project the 
representation
in eq.~\eqref{eq:amp_bcj_rep} using IBP relations. In this
section, we review the approach we recently used to perform this
calculation~\cite{Abreu:2018aqd}.

A powerful method for computing master integrals is through
differential equations, especially when written in 
canonical form~\cite{Henn:2013pwa}. If we denote a set of
master integrals by~$\{\I_a\}$, 
then their differential equation with respect to the external
kinematic variables $x_i$
is said to be canonical if it has the form
\begin{align}
\partial_{x_i}  \I_a \equiv \frac{\partial  \I_a}{\partial x_i} = \epsilon \sum_{\alpha}
\frac{\partial \log W_\alpha} {\partial x_i}  M_\alpha^{ab} \,  
\I_b \,,
\label{eq:diffCanonical}
\end{align}
where the index $\alpha$ runs over the letters of the alphabet
and the indices $a$ and $b$ run over all master integrals in the
set $\{\I_a\}$. 
Importantly, the dimensional regulator $\epsilon$
factorizes and the matrix $M_\alpha^{ab}$ consists solely of
rational numbers.
Conjecturally, there is a one-to-one correspondence between the
basis of master integrals being pure and their differential
equation being in canonical form. 

Even when a pure basis is
known, the conventional way to construct the differential
equations suffers from the computational bottleneck of IBP 
reduction when the number of kinematic invariants and masses is
large. Indeed, the large number of variables makes
the size of analytic expressions swell up to an often
unmanageable size.
In ref.~\cite{Abreu:2018rcw}, a new method of
constructing the differential equations was presented that 
builds on the prior knowledge of the symbol alphabet and of a
basis of pure master integrals. The matrix $M_\alpha^{ab}$ in 
eq.~\eqref{eq:diffCanonical} is then determined by
performing IBP reduction on a small number of \emph{numerical}
phase-space points, avoiding large intermediate analytic
expressions in the IBP reduction.
For the amplitude we are concerned with, the
symbol alphabet is known~\cite{Chicherin:2017dob} and, in order
to apply the procedure of ref.~\cite{Abreu:2018rcw}, we simply
need to discuss how we identified the pure bases for topologies
$(a)$, $(b)$ and $(c)$ in fig.~\ref{fig:bcj_integrands}.

The pure bases of master integrals for the planar pentabox and
nonplanar hexabox, i.e.~diagrams $(a)$ and $(b)$ in 
fig.~\ref{fig:bcj_integrands}, and their sub-topology integrals
are known in the literature~\cite{Gehrmann:2000zt,Gehrmann:2001ck,%
Gehrmann:2015bfy,Papadopoulos:2015jft,Bern:2015ple,Gehrmann:2018yef,%
Abreu:2018rcw,Chicherin:2018mue}.
Here we review how we identified the nine pure integrals for 
the nonplanar double pentagon 
\cite{Abreu:2018aqd}.\footnote{An alternative basis is 
given in ref.~\cite{Chicherin:2018old}.} 
To find a parity-even pure integral, we start from
the four-dimensional pure integral with numerator $N_1^{{\mathrm{(a)}}}$
identified in ref.~\cite{Bern:2015ple} and
rewritten with the labels of fig.~\ref{fig:bcj_integrands},
\begin{align}
N_1^{( \mathrm { a } )} &=\ab{15}\ab{24}\left[ 
	[24] [15] \bigg( \ell_{7} + \frac {[43]} {[24]} \lambda_{3} \,\widetilde {\lambda}_{2} \bigg)^2 
		      \bigg( \ell_{6} - \frac {(k_1 \pl k_2)\! \cdot\! \widetilde {\lambda}_{5} \,\widetilde {\lambda}_{1} } {[15]} \bigg)^2 \right. \nonumber \\
&\left. \qquad\qquad\hspace{.3cm}
 - [14] [25] \bigg(\ell_ {7} + \frac {[43]} {[14]} \lambda_{3} \,\widetilde {\lambda}_{1} \bigg)^2
		 \bigg(\ell_{6} - \frac {(k_1 \pl k_2)\! \cdot\! \widetilde {\lambda }_{5} \,\widetilde {\lambda}_{2}} {[25]} \bigg)^2 
\right],
\end{align}
where we refer the reader to appendix \ref{kinematicsappendix}
for the definition of the $\lambda_i$ and 
$\widetilde{\lambda}_i$.
The notation $\ell_6$, $\ell_7$ for the loop momenta is from
ref.~\cite{Bern:2015ple}, and is related to our labels by
\begin{equation}
\ell_6 = \ell_1, \quad \ell_7 = k_3 + k_4 - \ell_2 \, .
\end{equation}
This integral has a hidden symmetry 
\cite{Bern:2018oao, Bern:2017gdk, Chicherin:2018wes} which is a 
nonplanar generalization of dual conformal symmetry for planar 
diagrams. 
In ref.~\cite{Bern:2018oao}, the numerator $N_1^{{\mathrm{(a)}}}$ is
rewritten in terms of spinor traces to make the symmetry 
manifest,
\begin{align}
N_1^{( \mathrm { a } )} &=
-\tr \left[ \frac{1-\gamma^5} {2} \slashed k_5 \slashed k_1 \slashed k_2 \slashed k_4 (\slashed k_4 \mi \slashed \ell_2) 
   (\slashed \ell_1 \mi \slashed \ell_2 \pl \slashed k_3 \pl \slashed k_4) \slashed \ell_1 (\slashed k_3 \pl \slashed k_4) \right] \nonumber \\
&\quad  - \ell_1^2 \ell_2^2 \, \tr \left[ \frac{1-\gamma^5} {2} \slashed k_5 \slashed k_1 \slashed k_2 \slashed k_4 \right]\, .
\label{eq:pureNumeratorTrace}
\end{align}
Removing the projector $(1-\gamma^5)/2$ from the two traces, we obtain twice the parity-even part,
\begin{equation}
2N_1^{( \mathrm { a } )}\big\vert_{\textrm{even}}
\!=\!
-\tr \left[ \slashed k_5 \slashed k_1 \slashed k_2 \slashed k_4 (\slashed k_4 \mi \slashed \ell_2) 
   (\slashed \ell_1 \mi \slashed \ell_2 \pl \slashed k_3 \pl \slashed k_4) \slashed \ell_1 (\slashed k_3 \pl \slashed k_4) \right] 
   - \ell_1^2 \ell_2^2 \tr \left[ \slashed k_5 \slashed k_1 \slashed k_2 \slashed k_4 \right]\, .
\label{eq:pureNumeratorTraceEven}
\end{equation}
By elementary Dirac-matrix manipulations, the above traces 
evaluate to an expression in terms of Lorentz dot products 
involving both internal and external momenta, without 
any explicit $d$ dependence. This numerator gives a
$d$-dimensional pure integral. Using the
$Z_2\times Z_2$ symmetry of the nonplanar double-pentagon
diagram, including a horizontal and a vertical flip, we obtain
two more similar pure integrals.

\begin{figure}[]
\centering
\includegraphics[scale=.35]{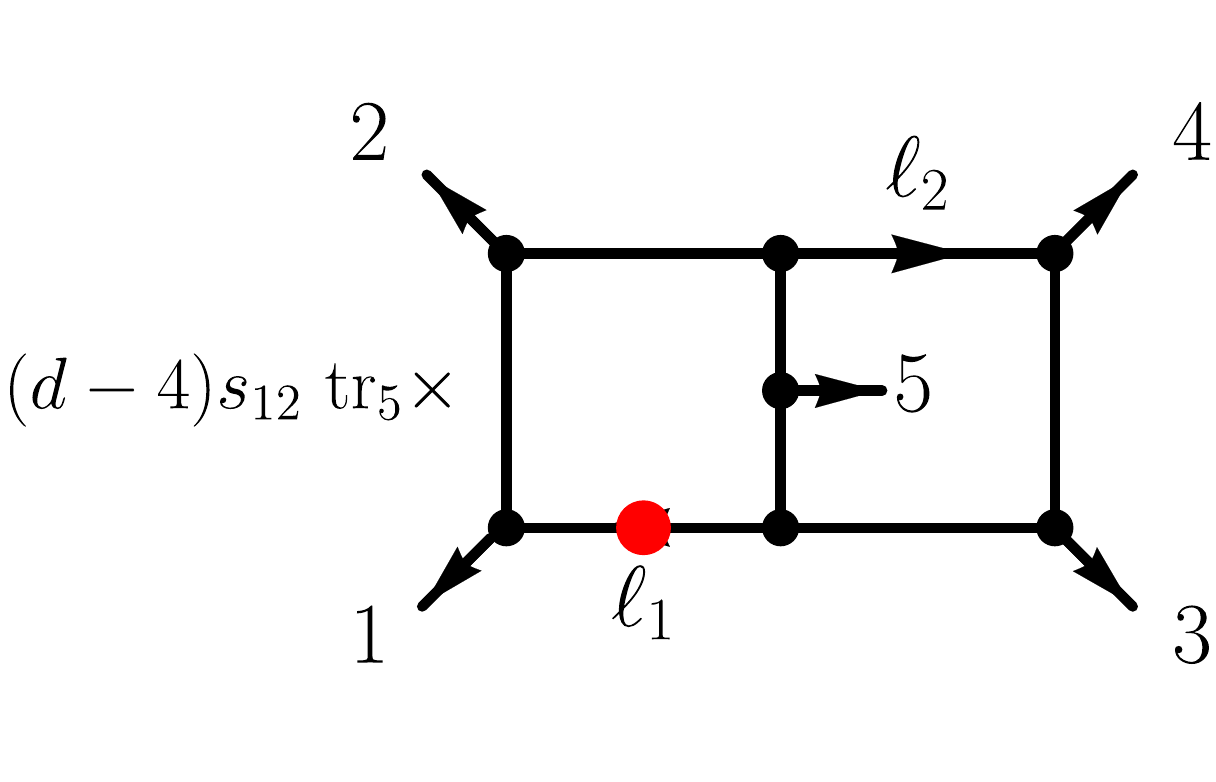} 
\hspace{.5cm}
\includegraphics[scale=.35]{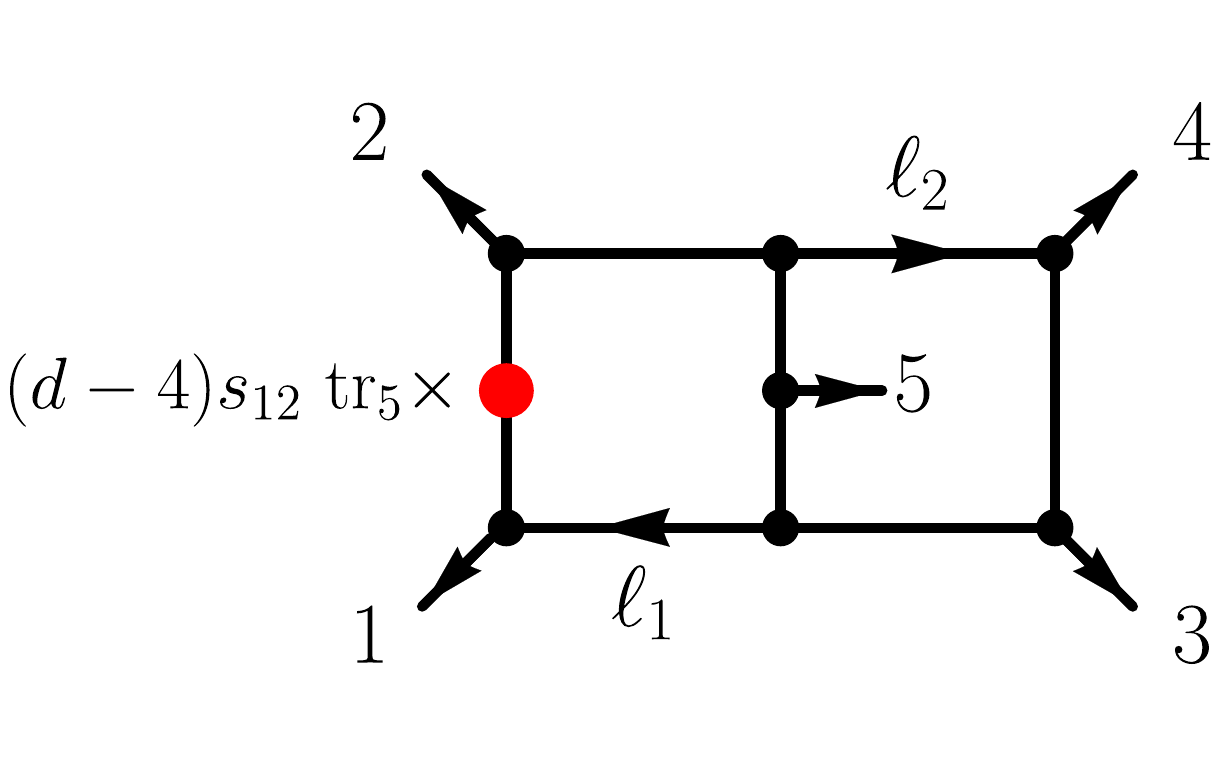}
\hspace{.5cm}
\includegraphics[scale=.35]{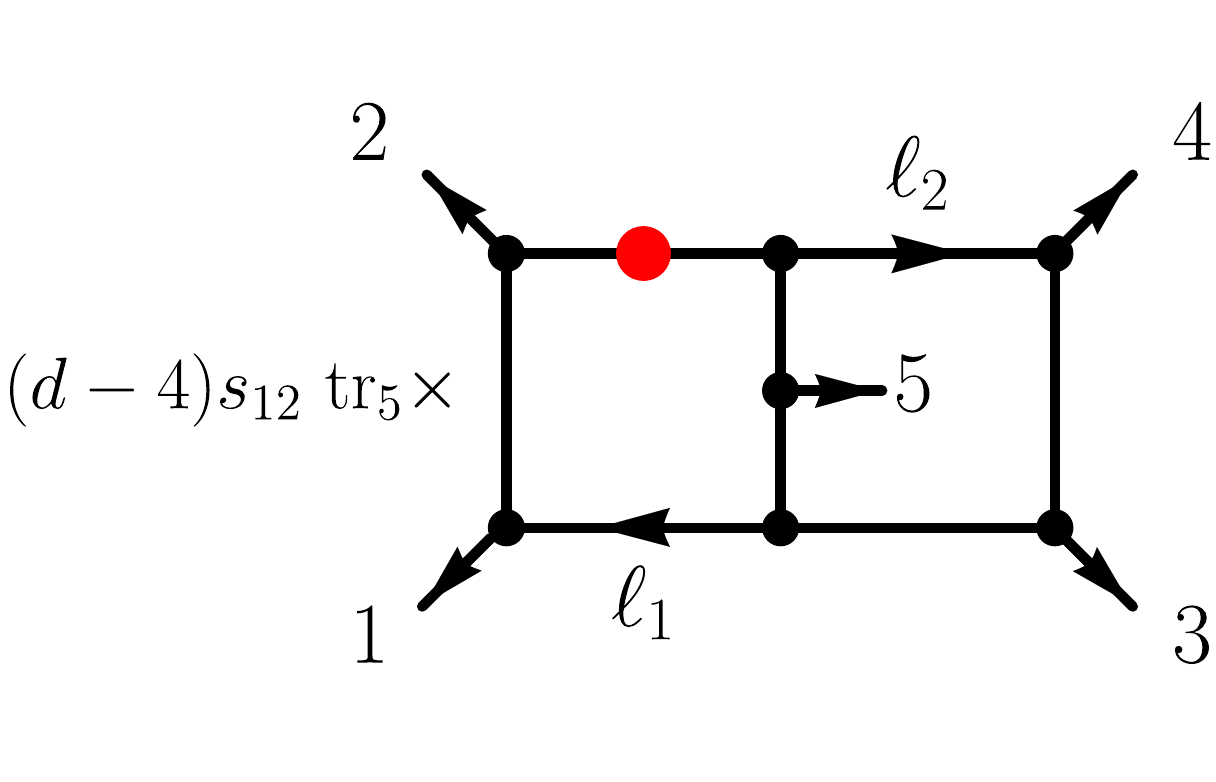} 
\newline
\vskip -.8cm
\hspace{-0.3cm}
\includegraphics[scale=.35]{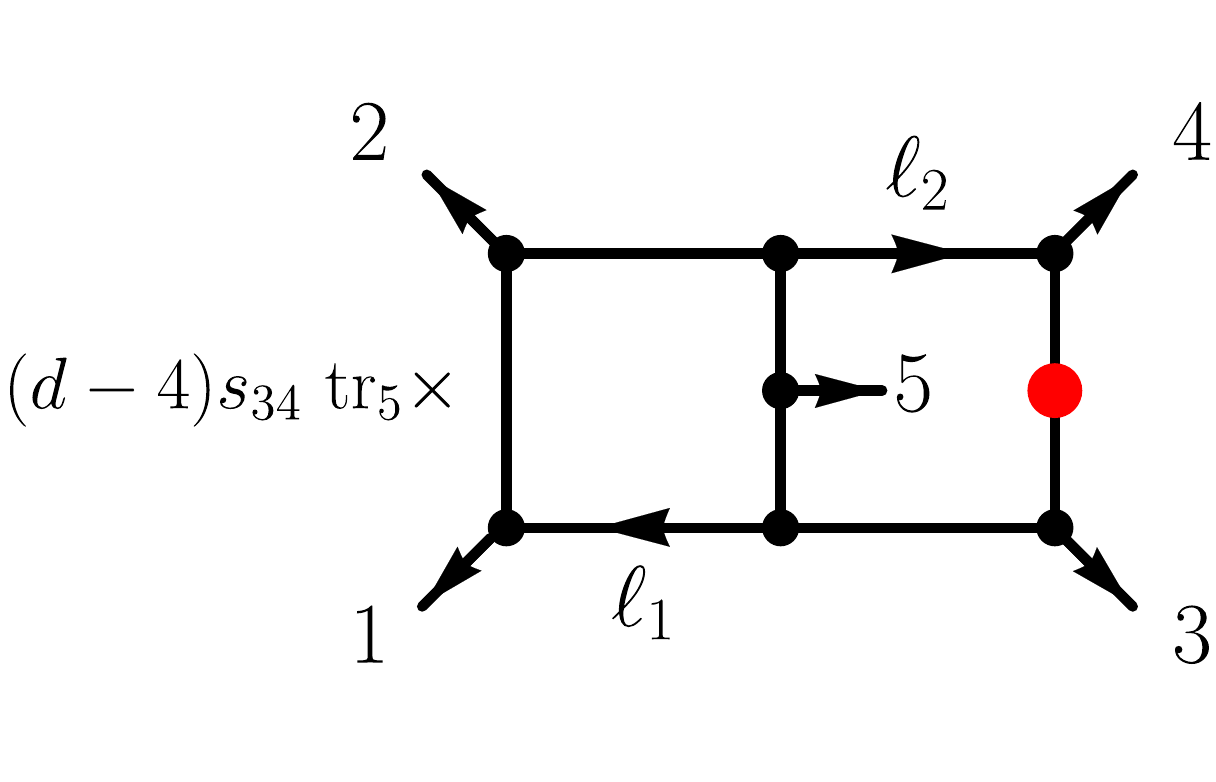}
\includegraphics[scale=.39]{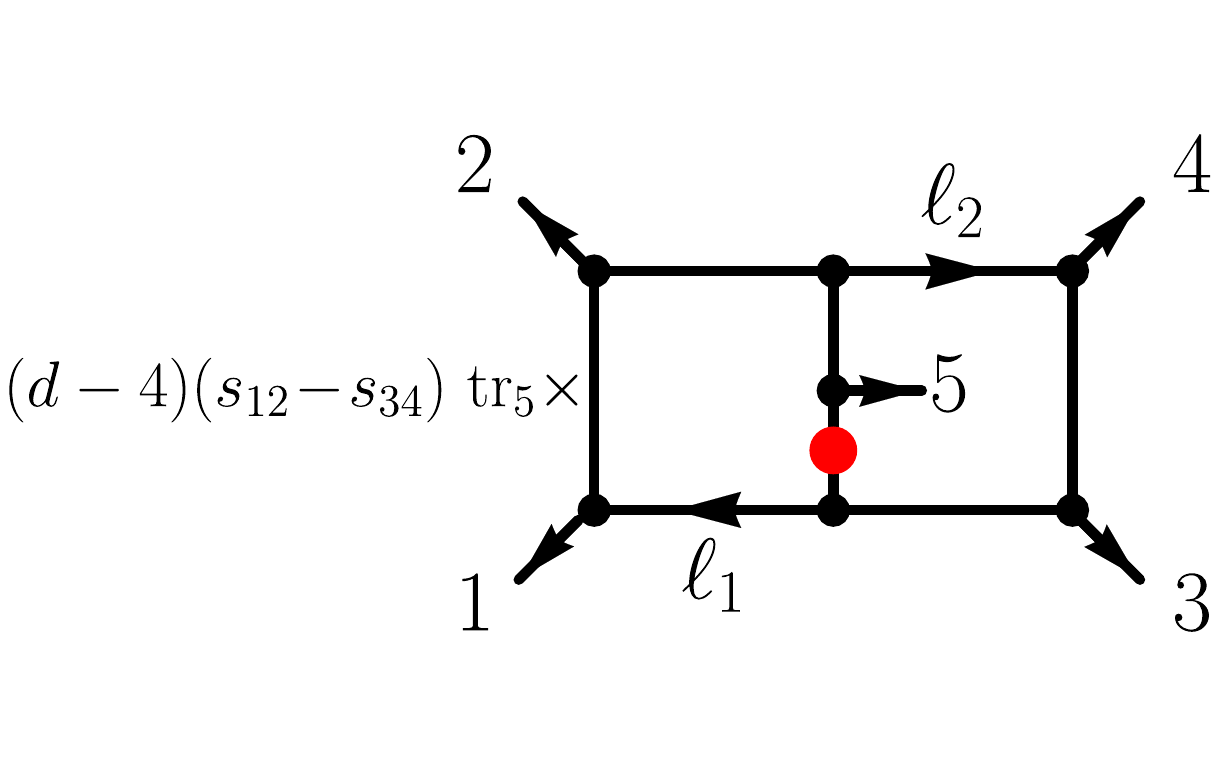}
\hspace{.1cm}
\includegraphics[scale=.39]{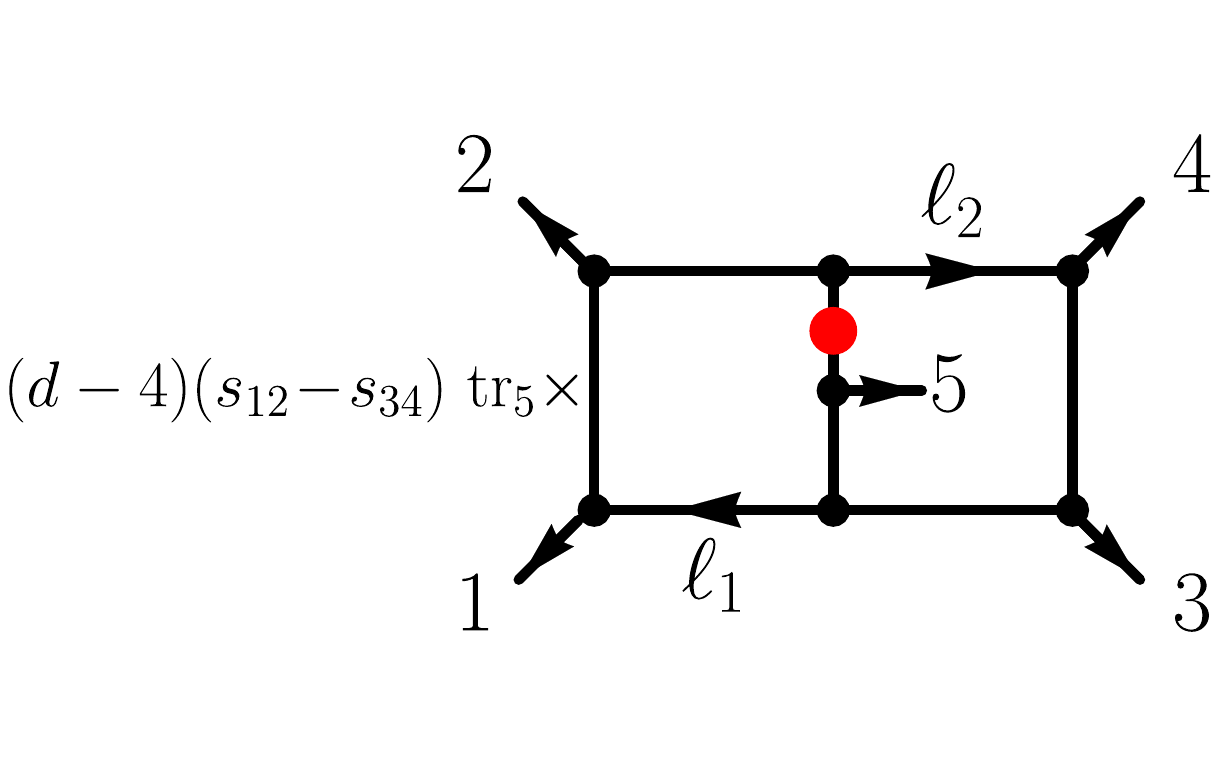}
\vspace{-20pt}
\caption{\label{fig:mastersOdd}The six parity-odd $(6-2\epsilon)$-dimensional master integrals and their normalization factors. The integrals have no numerators. A red dot indicates that the propagator is ``doubled" i.e.\ raised to a squared power.}
\end{figure}

Naively, one could also obtain parity-odd integrals by 
anti-symmetrizing over the spinor-trace expressions of 
ref.~\cite{Bern:2018oao} and their complex conjugates. 
The result
is simply eq.~\eqref{eq:pureNumeratorTraceEven} with $\gamma^5$ 
inserted into both Dirac traces. However, the integral fails to 
be a pure integral in $d$ dimensions (if one tries to use them
as master integrals, the differential equation is not in the
form of eq.~\eqref{eq:diffCanonical}). Instead, our basis of six
parity-odd pure integrals consists of the
$(6-2\epsilon)$-dimensional scalar integrals shown in 
fig.~\ref{fig:mastersOdd}. Each of the integrals has one squared
propagator, denoted by a red dot, as well as a normalization
factor which is written next to each diagram.
These integrals in $(6-2\epsilon)$ dimensions can be
converted to $(4-2\epsilon)$-dimensional integrals by 
dimension-shifting identities~\cite{Bern:1992em, Bern:1993kr, 
Tarasov:1996br, Lee:2009dh}. 
We find it more convenient to use the dimension-shifting
procedure outlined in appendix B of
ref.~\cite{Georgoudis:2016wff}, using the (global) Baikov
representation of Feynman integrals. 
In terms of the Baikov variables $\rho_i$, 
a $(6-2\epsilon)$-dimensional integral with a squared propagator
$1 / \rho_a^2$ is proportional to $1/(d-4)$ times a 
$(4-2\epsilon)$-dimensional integral without any squared
propagator, but with a numerator which is the derivative of the
Baikov polynomial with respect to $\rho_a$.

The purity of the nine nonplanar double-pentagon integrals we
just discussed can be confirmed by evaluating the differential
equations at numerical phase-space points and checking the factorization of
the dimensional regulator $\epsilon$. For this topology, there
are 31 letters ($1\leq\alpha\leq 31$) and 108 master integrals
($1\leq a,b \leq 108$). 
The 31 square matrices of rational numbers $M_\alpha^{ab}$ 
are determined by performing numerical IBP reductions on a
sufficient number of rational phase-space points in a finite
field. Details of the reduction procedure will be discussed in
the next section, as we used the same implementation for
computing the differential equations as we did for reducing the
amplitude to the basis of master integrals.

Once the differential equation has been computed, we obtain
the symbol of the master integrals by evaluating a single 
trivial integral to leading order in $\epsilon$, which fixes the
overall normalization of the functions, and imposing the
first-entry condition~\cite{Gaiotto:2011dt}. Explicit results 
for the master integrals we use can be found in the ancillary
files of ref.~\cite{Abreu:2018aqd}. They satisfy the conjectured
second-entry condition~\cite{Chicherin:2017dob}.

\subsection{Numerical reduction and analytic reconstruction} 
\label{sec:NumericalReduction}

Having discussed the evaluation of the master integrals from their
differential equations, we now describe the
final step in computing the symbol of the $\NeqEight$
amplitude:  the reduction of the representation in 
eq.~\eqref{eq:amp_bcj_rep} to our basis of master integrals. 
Both this step and the calculation of the
differential equation discussed above require performing
numerical IBP reductions. We now discuss our implementation.

\begin{figure}
  \centering
  \includegraphics[width=0.5\textwidth]{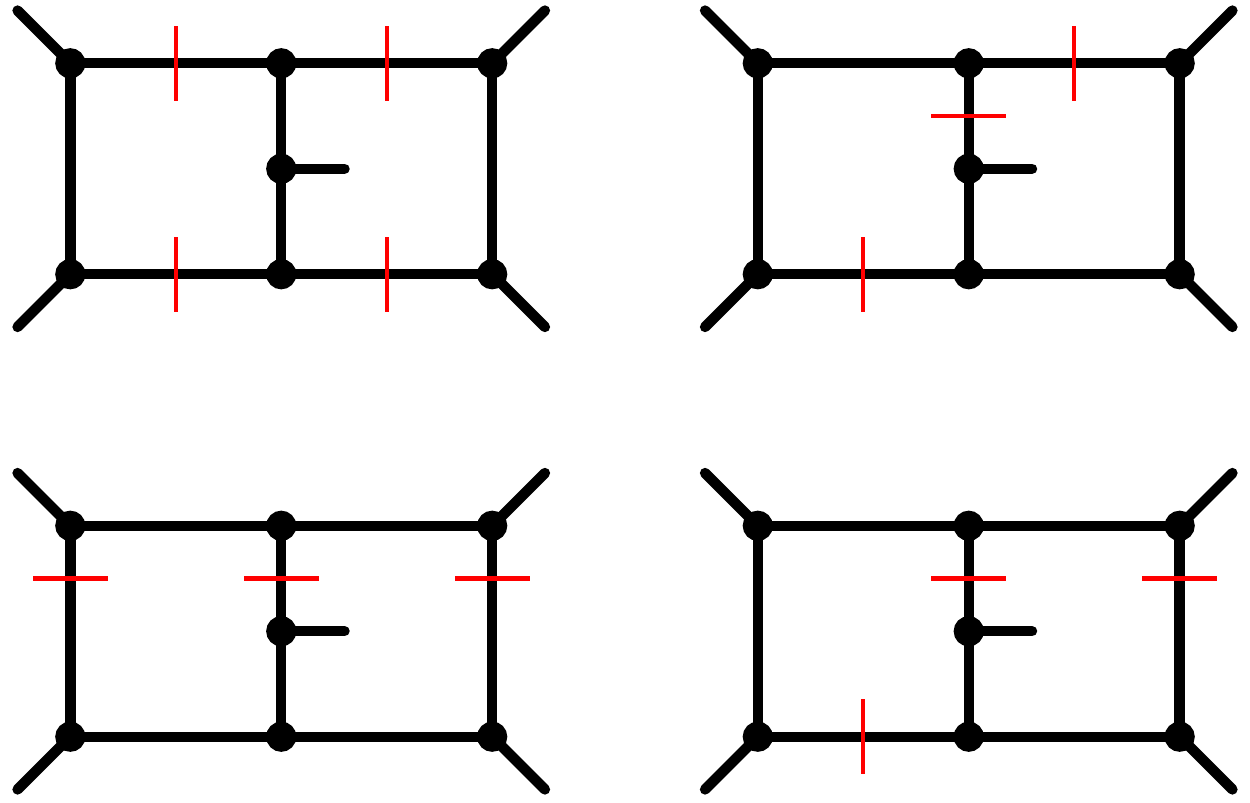}
  \caption{A spanning set of cuts for performing IBP reduction for the nonplanar double pentagon diagram. A cut propagator is indicated by a red line. There are 11 cuts in total, from applying diagram symmetries to the 4 representative cuts shown here.}
  \label{fig:ibpcuts}
\end{figure}

We perform IBP reduction in terms of unitarity cuts and 
computational algebraic geometry~\cite{Gluza:2010ws,
Ita:2015tya, Larsen:2015ped, Abreu:2017hqn, Boehm:2018fpv}.
Once more, we focus on the most challenging topology, 
the nonplanar double-pentagon in diagram $(c)$ of 
fig.~\ref{fig:bcj_integrands}.
The reduction is performed on a set of 11 spanning cuts, which
are the cuts shown in fig.~\ref{fig:ibpcuts} and
their images under the $Z_2\times Z_2$ symmetry of the diagram 
(horizontal and vertical flip).
Merging the reductions on each of the 11 spanning cuts, we
recover the complete IBP reductions for the uncut topology.
(A more detailed description of our implementation can be found in 
ref.~\cite{Abreu:2018rcw}.)

Unitarity cuts are most natural in the absence of doubled 
(squared) propagators. However, doubled propagators are present 
in conventional IBP relations,
\begin{equation}
0 = \int d^d \ell_1 \int d^d \ell_2 \sum_{A=1}^2 \frac{\partial}
{\partial \ell_A^\mu} \frac{v_A^\mu} 
{\rho_1 \rho_2 \dots \rho_N},
\end{equation}
because the derivatives can act on the propagator $1 / \rho_i$. 
This problem is avoided by choosing 
vectors $v_A^\mu$ that satisfy the condition~\cite{Gluza:2010ws}
\begin{equation}
\sum_{A=1}^2 v_A^\mu \, \frac{\partial \rho_i} {\partial
\ell_A^\mu} = f_i \, \rho_i \, ,
\label{eq:GKK}
\end{equation}
where both $v_A^\mu$ and $f_i$ are required to have polynomial
dependence on the components of the loop and external momenta.
Finding a full set of $v_A^\mu$ satisfying eq.~\eqref{eq:GKK} 
is a problem that can be solved by computational algebraic 
geometry. 
State-of-the-art algorithms to solve this equation can be found
in refs.~\cite{Abreu:2017hqn, Boehm:2018fpv, Abreu:2018rcw},
following many earlier devolopments~\cite{Gluza:2010ws, 
Schabinger:2011dz, Ita:2015tya, Larsen:2015ped, Zhang:2016kfo, 
Georgoudis:2016wff, Bern:2017gdk}. 
Avoiding doubled propagators drastically reduces the number of
integrals that are present in the linear system of IBP 
relations, and reduces the computational resources needed for 
solving the linear system via Gaussian elimination. 
Further speed-up is achieved by performing IBP reduction in a 
finite field~\cite{vonManteuffel:2014ixa, Peraro:2016wsq, Maierhoefer:2017hyi, 
Abreu:2017hqn, Smirnov:2019qkx} whose modulus is a 10-digit prime number, at 
numerical rational phase-space points.

We now focus our discussion on the reduction of the amplitude,
but exactly the same strategy applies to the construction of the
differential equation.
IBP reduction is performed separately for each of the top-level 
topologies $(a)$, $(b)$ and $(c)$ in Fig.~\ref{fig:bcj_integrands} 
and the associated ``tower" of sub-topologies.
Each diagram in the representation of the amplitude given 
in eq.~\eqref{eq:amp_bcj_rep} is separately reduced to master
integrals via IBP reduction. 
We add the six diagrams and their permutations \emph{after}
replacing the master integrals by their values in terms of 
symbols. For each of the rational phase-space points where we
perform the reduction of the amplitude, the final result of the
procedure takes the form,
\begin{equation}
    S[M_{5,k}^{(2)}] = \sum^{31}_{\alpha_1=1}\!\! \cdots\!\! \sum^{31}_{\alpha_{k}=1}
    b_{\alpha_1, \ldots, \alpha_{k}}\ \times 
    \left[W_{\alpha_1}, \ldots, W_{\alpha_{k}}\right], \quad
    k=2,3,4 \,,
\end{equation}
where the coefficients $b_{\alpha_1, \ldots, \alpha_{k}}$ 
take numerical values in the finite field. Comparing with
eq.~\eqref{eq:AmplitudeSymbol}, it is clear that the 
$b_{\alpha_1, \ldots, \alpha_{k}}$ are kinematically dependent,
as they depend on the rational functions $r_j$.

To finish our calculation, we must extract the coefficients 
$c^j_{\alpha_1, \ldots, \alpha_{k}}$ in 
eq.~\eqref{eq:AmplitudeSymbol} from IBP reductions at sufficiently many
phase-space points. Generating the numerical data is the most
computationally-intensive
part of the calculation, which is nevertheless much more
efficient than analytic IBP reduction, for the reasons
already highlighted when discussing the construction of
differential equations.
Since the space of rational functions $r_j$ is 45-dimensional,
solving the linear system to determine the
coefficients $c^j_{\alpha_1, \ldots, \alpha_{k}}$ from
numerical evaluations is simple. We first obtain the 
coefficients in the finite field, and since they are very simple
rational numbers this information is sufficient to map them to the field of
rational numbers.

We finish with a comment on the application of this procedure to
compute the differential equation. The equivalent of the 
rational functions $r_j$ are now the $\dlog$-forms 
$\dlog W_\alpha$ in eq.~\eqref{eq:diffCanonical}, which form a
31-dimensional space. The equivalent of the coefficients 
$c^j_{\alpha_1, \ldots, \alpha_{k}}$ are the entries of the
matrices $M_\alpha^{ab}$. They are determined in the same way
and, as for the amplitude, we find they are simple enough that 
only a single finite field is necessary. We note that the IBP 
reductions required for the differential equations are 
harder to obtain than the ones for the $\NeqEight$ amplitude: 
the former require reducing integrals with numerators of at 
least degree 3 in the loop momentum,
while the latter only involve integrals with numerators of degree 2.

\vspace{-0pt}
\section{Validation} 
\label{sec:validation}\vspace{-8pt}

Scattering amplitudes in gauge and gravity theories obey many 
well understood factorization formulae that are given in terms 
of simpler quantities. For example, in special kinematic
configurations such as soft and collinear limits, the analytic
form of the amplitude can be expressed in terms of universal
factors and lower-point amplitudes. Similarly, the
divergence structure of loop amplitudes 
(i.e., the poles in $\e$)
can be written in terms
of lower-loop amplitudes and universal factors. 
These degenerations onto simpler
configurations provide powerful checks for any new calculation.
In the following we shall discuss how our analytic result
satisfies all these conditions.

\subsection{Divergence structure}

On general grounds, the divergences of a scattering amplitude can be
broadly separated into two classes---ultraviolet (UV) and
infrared
(IR). In recent years, understanding the UV structure of supergravity theories has received 
considerable attention and was partially stimulated by the open question about the potential 
UV finiteness of $\NeqEight$ in 4 dimensions, which would clearly impact our
understanding of quantum gravity on a more fundamental level. The critical dimension in which $\NeqEight$ diverges 
has now been explicitly calculated through five 
loops~\cite{Bern:1998ug,GravityThreeLoop,Manifest3,N8FourLoop,ColorKinematics,Bern:2018jmv}. In 4 dimensions, there are various arguments that rule out UV divergences up to at least seven loops~\cite{GreenDuality,BossardHoweStellDuality,BeisertN8,Vanhove:2010nf,Bjornsson:2010wm,Bjornsson:2010wu,VanishingVolume}. The important aspect for our work here is the fact that the two-loop five-point amplitude only has IR divergences.
In comparison to gauge theory, the IR divergence structure of
gravity is rather muted. It has been known for a long time that there
are no virtual collinear divergences in any quantum theory of
gravitation~\cite{Weinberg:1965nx}. Furthermore, it can be shown that
the structure of the soft divergences in gravity is completely
controlled by the one-loop result, which contains a $1/\e$ pole.
Specifically, it can be shown that
the one-loop divergence {\em
exponentiates}~\cite{SchnitzerN8UniformTrans,Weinberg:1965nx,Akhoury:2011kq,Beneke:2012xa,Dunbar:1995ed,Naculich:2011ry,White:2011yy}. In the case of
two-loop four-point amplitudes this was explicitly demonstrated
in ref.~\cite{BoucherVeronneau:2011qv}.

In order to check the divergence structure of the
two-loop five-point
amplitude, we therefore begin by recalling the one-loop result 
\cite{Bern:1998sv},
\begin{align}
\label{eq:1loop_5pt_amp_diag_sum}
 M^{(1)}_5  = \frac{1}{2} \sum_{S_5} \left(\frac{1}{4}\beta_{123(45)} \, I^{d=4\mi 2\eps}_{123(45)}
- \frac{1}{10} \frac{\sqb{12}\sqb{23}\sqb{34}\sqb{45}\sqb{51}}{\ab{12}\ab{23}\ab{34}\ab{45}\ab{51}} (- 2 \epsilon) I^{d=6\mi 2\eps}_{12345}\right) \,,
\end{align}
where the rational factors of $1/4$ and $1/10$ inside the $S_5$
permutation sum remove over-counting, and
\begin{align}
\beta_{123(45)} = - \frac{\sqb{12}^2 \sqb{23}^2 \sqb{45}}
				     {\ab{14}\ab{15}\ab{34}\ab{35}\ab{45}}\,.
\label{beta12345}
\end{align}
$I^{d=4\mi 2\eps}_{123(45)}$ is the one-mass scalar box 
integral in $4-2\epsilon$ dimensions, 
and $I^{d=6\mi 2\eps}_{12345}$
is the massless pentagon integral in
$6-2\epsilon$ dimensions normalized as follows: 
\begin{align}
 I^{d=4\mi 2\eps}_{123(45)} & = e^{\epsilon\gamma_E}
 \int\!\! \frac{d^{4\mi 2\eps}\ell}{i\pi^{2\mi \eps}} 
 	\frac{1}{\ell^2(\ell \mi k_1)^2 (\ell \mi k_1\mi k_2)^2(\ell \pl k_4 \pl k_5)^2}\,, \\
 I^{d=6\mi 2\eps}_{12345} & =  e^{\epsilon\gamma_E}
 \int\!\! \frac{d^{6\mi 2\eps}\ell}{i\pi^{3\mi \eps}} 
 	\frac{1}{\ell^2(\ell \mi k_1)^2 (\ell \mi k_1\mi k_2)^2(\ell \pl k_4)^2(\ell \pl k_4 \pl k_5)^2}\,.
\end{align}
The box integral is known to all orders in
$\epsilon$~\cite{Bern:1993kr} and the symbol of the pentagon integral can be computed to any
order in $\epsilon$ with the techniques 
of~\cite{Abreu:2017mtm,Abreu:2017enx} or by direct
integration with \texttt{HyperInt}~\cite{Panzer:2014caa}.
In ancillary files, we provide symbols for the pure functions (${\cal I}$) obtained by normalizing these integrals by their leading singularities,
\begin{align}\label{eq:pureIntDef}
	\mathcal{I}^{d=4\mi 2\eps}_{123(45)} 
	\equiv s_{12}s_{23}I^{d=4\mi 2\eps}_{123(45)}\,,\qquad
	\mathcal{I}^{d=6\mi 2\eps}_{12345} 
	\equiv-\tr_5\,I^{d=6\mi 2\eps}_{12345}\,.
\end{align}
Despite the presence of $1/\e^2$ soft-collinear divergences in individual box
integrals,
they cancel in the sum to give
\begin{equation} 
 M^{(1)}_5 = \frac{1}{\eps} \left[ \sum_{i<j=1}^5\sab{ij} \log 
 \sab{ij} \right] M^{(0)}_5 + M^{(1),0}_5\, + O(\epsilon),
\end{equation}
where $M_5^{(1),0}$ is the ${\cal O}(\e^0)$ term in the 
one-loop amplitude.
Finally, at two loops, the divergent pieces are dictated by
exponentiation in terms of the square of the one-loop amplitude:
\begin{align}
M^{(2),\text{div}}_5 = \left. \frac{1}{2} \left[ \frac{M^{(1)}_5}{M^{(0)}_5} \right]^2 \!\! \times M^{(0)}_5\right|_{\text{pole-terms}}\,.
\label{OneLoopSquared}
\end{align}
Inserting the symbols of the relevant one-loop integrals and 
comparing against the divergences of our two-loop result we find
perfect agreement.
The factor predicting the pole structure permits many
natural extensions that include different finite pieces.
What the ``ideal'' choice is an interesting question 
which we will discuss in section~\ref{sec:Structure}.

\subsection{Soft factorization}

Gravity amplitudes, similarly to gauge amplitudes, have a
universal factorization
property when a single graviton becomes much softer than the remaining
gravitons.  At tree level, the general factorization when 
the $n^{\rm th}$ graviton becomes soft, $k_n \to 0$,
is~\cite{Weinberg:1965nx,Berends:1988zp},
\begin{align}
M_n^{(0)}(1,\ldots,n\mi1,n^\pm)\ 
\mathop{\Longrightarrow}^{k_n\to0}\
    \mathcal{S}_n^\pm \times
   M_{n\mi1}^{(0)}(1,\ldots,n\mi1)\, ,
\label{TreeSoft}
\end{align}
where the positive-helicity soft factor is
\begin{align}
\mathcal{S}_n^+
= \frac{-1}{\ab{1n} \ab{n\,n\mi1}}
\sum_{i=2}^{n\mi2} \frac{\ab{1i} \ab{i\,n\mi1} \sqb{in}}{\ab{i n}}\,.
\label{TreeSoftFactor}
\end{align}
Naively, the definition of the soft factor $\mathcal{S}_n^+$ seems to pick out two further special legs, $n-1$ and $1$. 
One can however show that this term is independent of that particular choice, which will become important 
momentarily. In ref.~\cite{Bern:1998sv} it was shown that there are no loop corrections
to the leading soft factorization for gravity. That is, 
\bea
&&M_n^{(1)}(1,\ldots,n\mi1,n^\pm)\ \mathop{\Longrightarrow}^{k_n\to0}\
    \mathcal{S}_n^\pm \times M_{n\mi1}^{(1)}(1,\ldots,n\mi1)\,,
\label{OneLoopSoft}\\
&&M_n^{(2)}(1,\ldots,n\mi1,n^\pm)\ \mathop{\Longrightarrow}^{k_n\to0}\
    \mathcal{S}_n^\pm \times M_{n\mi1}^{(2)}(1,\ldots,n\mi1)\,.
\label{TwoLoopSoft}
\eea
We will test our result for the five-point amplitude
against eq.~\eqref{TwoLoopSoft} for $n=5$.

First, we determine the soft behavior of the 31 symbol letters 
in eq.~(\ref{eq:simple_alphabet}).
We parametrize the approach to the $k_5\to 0$ soft limit 
with a parameter $\delta \to 0$.
We then rewrite
the $x_i$ momentum-twistor parametrization of
ref.~\cite{Badger:2013gxa}
(see Appendix \ref{kinematicsappendix} for more details)
as
\begin{align}
\begin{split}
x_1 &= s, \quad x_2 = s x, \quad x_3 = -\frac{sx}{1-z} \,, \\
x_4 &= 1 + \delta \, \frac{x + \bar{z}}{1 - \bar{z}} \,, \quad 
x_5 = 1 + \delta \biggl[ 1 + \frac{x + \bar{z}}{1 - \bar{z}} \biggr] \,,
\label{xisoft5}
\end{split}
\end{align}
where $s=s_{12}$, $x=s_{23}/s_{12}$, 
$z= \ab{14} \ab{35} / (\ab{34} \ab{15})$,
and $\bar{z} = \sqb{14} \sqb{35} / (\sqb{34} \sqb{15})$ at
leading order in the $\delta \to 0$ limit.
The set of 14 letters obtained in this limit are
\begin{align}
\begin{split}
& \{ s, x, 1+x \}\ \cup\
\{ \delta, z, 1-z, x+z, \bar{z}, 1-\bar{z}, x+\bar{z} \} \\
&\cup\ \{ x+z+\bar{z}-z\bar{z},\ x(z+\bar{z}-1)+z\bar{z},\ 
x+z\bar{z},\ z-\bar{z} \} \,.
\label{SoftLetters}
\end{split}
\end{align}
In the soft limit of the two-loop five-point $\mathcal{N}=8$ supergravity
amplitude, it follows from \eqn{TwoLoopSoft} that
only the subset $ \{ s, x, 1+x \}$ should appear, after taking into account
the behavior of the rational prefactors.  In the soft limit of
the two-loop five-point $\mathcal{N}=4$ super-Yang-Mills
amplitude, the second set of letters can also appear in subleading-color
terms, and is consistent with a computation of two-loop soft-gluon
emission using Wilson lines~\cite{LanceToAppear}.

To analyze the soft limit of our five-point amplitude, we perform the
substitution~(\ref{xisoft5}) within the symbol entries, and
refactorize the symbol on the set of letters in \eqref{SoftLetters}.
Then we consider the soft behavior of the rational prefactors.

In the case we are interested in, $n=5$,
the soft factor~(\ref{TreeSoftFactor})
has only two terms,
\be
{\cal S}\ \equiv\ {\cal S}_5^+\ =\ {\cal P}_{14}^2 + {\cal P}_{14}^3 \,, 
\label{SplitUpSoft}
\ee
where
\be
{\cal P}_{jk}^i \equiv - \frac{\ab{ji} \ab{ki} \sqb{i5}}
                             {\ab{j5} \ab{k5} \ab{i5}} \,.
\label{PartSoftFactor}
\ee
In the soft limit, the little group transformation properties 
imply that all the rational
factors $r_j$ in
eq.~\eqref{eq:schematicAmp} are either nonsingular
or become proportional to the four-point amplitude multiplied by
one of these partial soft factors ${\cal P}_{jk}^i$.  Because
\eqref{PartSoftFactor} is symmetric in $j$ and $k$, and $i,j,k \in \{1,2,3,4\}$,
there are 12 such factors.  However, they sum in six pairs to the soft factor,
\be 
{\cal P}_{jk}^{i_1} + {\cal P}_{jk}^{i_2} = {\cal S},
\label{SumIdent}
\ee
where $i_{1,2} \not\in \{j,k,5\}$. Equation~(\ref{SumIdent})
reflects the fact that any two gravitons
can play the role of gravitons 1 and $n-1$ in 
\eqref{TreeSoftFactor}.

There is one more useful identity among the partial soft factors,
\be
s_{13} ({\cal P}_{24}^3 - {\cal P}_{13}^4)
= s_{12} ({\cal P}_{12}^4 - {\cal P}_{34}^2)
  -(\sab{15}+\sab{45})\mathcal{P}^1_{23}
\label{PSoftIdent}
\ee
plus all equations obtained by permuting legs $\{1,2,3,4\}$.
The second term on the right-hand side can be dropped in the soft limit.

Using the identities~(\ref{SumIdent}) and (\ref{PSoftIdent}),
and the symbol substitutions
mentioned above, we find that all letters except $\{s,x,1+x\}$ drop out
of the soft limit.  Furthermore, the limit is proportional to the symbol
of the four-point $\mathcal{N}=8$ supergravity amplitude
given in ref.~\cite{BoucherVeronneau:2011qv}
(see also refs.~\cite{SchnitzerN8UniformTrans,QueenMaryN8UniformTrans}).
That is, the five-point amplitude precisely
satisfies the soft limit~(\ref{TwoLoopSoft}).

\subsection{Collinear factorization}

The behavior of gravity amplitudes as two gravitons $a$ and $b$
become collinear is also universal and well established~\cite{Bern:1998xc},
\be
M_n^{(0)}(1,\ldots,a,b,\ldots,n)\ \mathop{\Longrightarrow}^{k_a||k_b}\
    {\rm Split}^{\rm grav}(\tau,a,b) \times
   M_{n-1}^{(0)}(1,\ldots,P,\ldots,n)\,.
\label{TreeCollinear}
\ee
In eq.~(\ref{TreeCollinear}) we define the common momentum $k_P = k_a + k_b$, 
and write $k_a \approx \tau k_P$, $k_b \approx (1-\tau)k_P$
with the splitting fraction $\tau$ for the longitudinal momentum.
In contrast to the case of
gauge theory, for real collinear kinematics, the amplitude does not diverge
in the limit. Rather, ${\rm Split}^{\rm grav}(\tau,a,b)$
is a pure phase, containing dependence on 
the azimuthal angle as the two nearly-collinear gravitons are rotated
around the axis formed by the sum of their momenta.
This behavior stems from a factor of $[ab]/\langle ab\rangle$
in the amplitude
(or $\langle ab\rangle/[ab]$, depending on the helicity configuration)
as legs $a$ and $b$ become collinear.

At tree level, the form of the gravitational collinear splitting
factor can be understood from the KLT relations~\cite{KLT} to
originate from a product of two singular gauge-theory splitting amplitudes
and a factor of $s_{ab}$ in the numerator~\cite{Bern:1998sv},
\be
{\rm Split}^{\rm grav}_{-2\lambda}
(\tau,a^{2\lambda_a},b^{2\lambda_b}) =  
-s_{ab} \times [ {\rm Split}^{\rm YM}_{-\lambda}(\tau,a^{\lambda_a},b^{\lambda_b}) ]^2 \,.
\label{GravYMSplit}
\ee
Here $\lambda_a$ and $\lambda_b$ are the helicities of the two external
gluons for both of the gauge copies. The sums of their helicities,
$2\lambda_a$ and $2\lambda_b$,
are the external graviton helicities,
and similarly for the intermediate helicities $\lambda$ and $2\lambda$.

For the five-point amplitude in $\mathcal{N}=8$ supergravity,
it is convenient to take all collinear helicities to be positive,
$\lambda_a = \lambda_b = \lambda = +$, and we obtain,
\be
{\rm Split}^{\rm grav}_{-}(\tau,a^{+},b^{+})\ =\ -{ 1 \over \tau
(1-\tau) }
                 { \spb{a}.{b} \over \spa{a}.b }\,.
\label{GravSplitMHV}
\ee

As in the case of soft factorization, there are no loop corrections
to the splitting amplitude~\cite{Bern:1998sv},
so the one- and two-loop amplitudes behave as,
\bea
&&M_n^{(1)}(1,\ldots,a,b,\ldots,n)\ \mathop{\Longrightarrow}^{k_a||k_b}\
    {\rm Split}^{\rm grav}(\tau,a,b) \times
   M_{n-1}^{(1)}(1,\ldots,P,\ldots,n)\,,
\label{OneLoopCollinear}\\
&&M_n^{(2)}(1,\ldots,a,b,\ldots,n)\ \mathop{\Longrightarrow}^{k_a||k_b}\
    {\rm Split}^{\rm grav}(\tau,a,b) \times
   M_{n-1}^{(2)}(1,\ldots,P,\ldots,n)\,.
\label{TwoLoopCollinear}
\eea
We will test the collinear behavior of the two-loop five-graviton
amplitude against~\eqref{TwoLoopCollinear}, with splitting
amplitude~(\ref{GravSplitMHV}).  Since the (super-)amplitude is Bose
symmetric, it does not matter which two legs we take to be parallel.
For convenience we discuss the same limit we studied for the
two-loop five-point $\mathcal{N}=4$ SYM amplitude~\cite{Abreu:2018aqd},
$k_2||k_3$, i.e.~$a=2$ and $b=3$.  The two-loop four-point $\mathcal{N}=8$
supergravity amplitude~\cite{SchnitzerN8UniformTrans,
QueenMaryN8UniformTrans,BoucherVeronneau:2011qv}
should be evaluated with momenta $(k_1,k_P,k_4,k_5)$.

The analysis of the symbol proceeds exactly as in ref.~\cite{Abreu:2018aqd}.
Employing the $x_i$ variables of ref.~\cite{Badger:2013gxa},
we let
\be
x_1 \mapsto s \tau\,, \quad
x_2 \mapsto c s \delta \,, \quad 
x_3 \mapsto r_2 c s \delta\,, \quad
x_4 \mapsto \delta\,, \quad 
x_5 \mapsto -\frac{1}{c\delta}\,,
\label{xicoll}
\ee
where $s = s_{45}$ and $r_2 = s_{15}/s_{45}$ characterize the
four-point kinematics,
$c \sim \sqb{23}/\ab{23}$ corresponds to an azimuthal phase, and
$\delta = \sqrt{s_{23}/(s\ c)}$ vanishes in the collinear limit.
We expand the 31 letter alphabet in the collinear limit
to leading order in $\delta$,
finding 14 multiplicatively independent letters in the collinear limit:
7 physical letters $\{\delta,\ s,\ \tau,\ 1-\tau,\ r_2,\ 1 + r_2,\ c\}$
and 7 spurious letters that are in neither the splitting amplitude nor
the four-point amplitude; hence they must not contribute to the universal limit.

After refactorizing the amplitude on these symbol letters, we choose numerical
kinematics near the collinear limit, and take the difference between
evaluations at two different points corresponding to an azimuthal
rotation of the two collinear gravitons.  Taking this difference removes
non-universal terms that would otherwise be of the same order,
and the results are numerically consistent with the expected
factorization~(\ref{TwoLoopCollinear}). Alternatively, one can use
complexified momenta and perform two non-overlapping BCFW shifts
\cite{BCFW}, e.g. $\lam{2} \to \lam{2} + z \lam{4},\ \lamt{4}\to
\lamt{4}-z\lamt{2}$ and $\lam{5} \to \lam{5} + w \lam{3},\ \lamt{3}\to
\lamt{3}-w\lamt{5}$ and then solve $\ab{23} = \eps_1\,, \sqb{23}=\eps_2$
in terms of $z$ and $w$. Expanding around $\eps_1=0$ then allows one to check that the pole term is proportional to $\eps_2$, which was used as an independent check of the collinear factorization property of our result.

\section{Structure of results}
\label{sec:Structure}

The purpose of this section is to provide some insight into
the structure of the amplitude we have computed.  First we define
a prescription for removing the infrared divergences, which also cleans
up the finite hard remainder.  Then we write the remainder $R_5^{(2)}$
in a manifestly symmetric form, which requires only summing over
permutations of a \emph{single rational structure},
multiplied by a \emph{single weight 4 function} $h$.
We note that the finite quantity $h$ cannot be written only in
terms of the classical polylogarithms $\log$, 
$\operatorname{Li}_2$, $\operatorname{Li}_3$ and
$\operatorname{Li}_4$, but also requires the function
$\operatorname{Li}_{2,2}$
(this can be checked with the procedure described in 
ref.~\cite{Goncharov:2010jf}).
We characterize the properties of $h$ in terms of its final entries
and the weight-3 odd parts of its derivatives.
We go on to characterize the full space of 45 functions in
the (unsubtracted) $\NeqEight$ amplitude, and compare it with
its cousin, the corresponding $\NeqFour$ amplitude, also at the level
of their derivatives (coproducts).  In the course of doing this,
we discovered linear relations between components of
the $\NeqFour$ five-point amplitude at one and two loops.

One interesting ``global'' property of the $\NeqEight$ amplitude
is that the letter $W_{31}$ does not appear at all,
neither in the unsubtracted amplitude
nor in the subtracted hard function to be described shortly.
It does appear in the $\NeqFour$
amplitude~\cite{Abreu:2018aqd,Chicherin:2018yne},
but this appears to be linked solely to 
its contribution to the ${\cal O}(\e^2)$ part of the
$(6-2\epsilon)$-dimensional pentagon integral, 
which is required at two loops for infrared subtractions in
$\NeqFour$, but not in $\NeqEight$ because of the milder IR
divergence structure.

\subsection{A symmetric form of the hard remainder $R^{(2)}_5$}

As mentioned around \eqn{OneLoopSquared}, in order 
to remove the infrared divergences from the two-loop amplitude, 
one could simply subtract the full square of the one-loop amplitude.
That is, one could define
\begin{align}\begin{split}
\tilde{R}_5^{(2)} &\equiv M_5^{(2)}
 - \frac{1}{2} \left[ \frac{M^{(1)}_5}{M^{{(0)}}_5} \right]^2 
\!\! \times M^{{(0)}}_5  \\
&= M_5^{(2)} - \frac{1}{2} \left[ 
\frac{1}{\e} \Bigl( \sum_{i<j} s_{ij} \log s_{ij} \Bigr)
   + \frac{M_5^{(1),0}}{M^{{(0)}}_5}
   + \e \, \frac{M_5^{(1),1}}{M^{{(0)}}_5} \right]^2
\!\! \times M^{{(0)}}_5 \ +\ {\cal O}(\e) \\
&= M_5^{(2)} 
- \Bigl( \sum_{i<j} s_{ij} \log s_{ij} \Bigr)
 \times \left[ 
  \frac{1}{2\e^2} \Bigl( \sum_{i<j} s_{ij} \log s_{ij} \Bigr) M^{{(0)}}_5
 + \frac{1}{\e} M_5^{(1),0} + M_5^{(1),1}  \right] \\
& \hskip1.3cm \null
 -  \frac{\left[M_5^{(1),0}\right]^2}{2 \, M^{{(0)}}_5} \ +\ 
 {\cal O}(\e),
\label{eq:badremainderdef}
\end{split}\end{align}
where $M_5^{(1),0}$ and $M_5^{(1),1}$ are the ${\cal O}(\e^0)$
and ${\cal O}(\e^1)$ terms in the one-loop amplitude.
However, doing this full subtraction would enlarge
the space of rational structures required.  The issue is
with the final term. The tree amplitude
$M^{{(0)}}_5$ is proportional to $\tr_5$, 
see eq.~\eqref{eq:trees}, and appears
in the denominator. 
Within the square of $M_5^{(1),0}$, the products of different permutations
of the box coefficient~(\ref{beta12345}), after dividing by 
$M^{{(0)}}_5$,
cannot be expressed in terms of the 45 rational structures of
eqs.~(\ref{eq:gravity_LS_basic_4d}) and (\ref{eq:gravity_LS_basic_d}).
Instead, we simply omit this last term and define
\be
R_5^{(2)} \equiv M_5^{(2)} 
- \Bigl( \sum_{i<j} s_{ij} \log s_{ij} \Bigr)
 \times \left[ 
  \frac{1}{2\e^2} \Bigl( \sum_{i<j} s_{ij} \log s_{ij} \Bigr) 
  M^{{(0)}}_5
 + \frac{1}{\e} M_5^{(1),0} + M_5^{(1),1}  \right] \,.
\label{eq:remainder_def}
\ee
The ${\cal O}(\e)$ terms in the one-loop amplitude induce a shift
in the finite terms of the two-loop amplitude. In particular, there
is a factor of $\tr_5$ in the denominator of the coefficient of
the $d=6$ pentagon integral, which cancels precisely
against the $1/\tr_5$-containing contributions to the bare
two-loop amplitude.  After this cancellation, there are only 40 
linearly independent rational structures, multiplied by 40
linearly independent weight 4 transcendental functions.

We remark that this cancellation of more complicated structures,
which are associated with $d$-dimensional cuts rather than 4 dimensional
ones, is reminiscent of what was observed for the six-point amplitude
in planar $\NeqFour$~\cite{Bern:2008ap}. In that case, integrals
containing $\mu^2$ factors (extra-dimensional components of the
loop momentum) appeared at two loops and at ${\cal O}(\e)$
in the one-loop amplitude, but cancelled out from the remainder
function.  The physical importance of the finite remainder at
two loops has also been stressed in
the context of constructing finite cross 
sections~\cite{Weinzierl:2011uz}.
The conclusion is that the $1/\tr_5$ rational structures,
originating from $d$-dimensional leading singularities, 
see eq.~\eqref{eq:gravity_LS_basic_d},
should be thought of as unphysical and dependent on the
use of dimensional regularization.

The 40 linearly independent rational structures appearing in the
hard function do not fall into nice orbits under $S_5$.
Consider for example, the structure
\be
r_0 = \frac{\sqb{23} \sqb{34} \sqb{45} \sqb{51}}
{\ab{13}\ab{14}\ab{15}\ab{23}\ab{24}\ab{25}}\,.
\label{r0}
\ee
It is invariant under the $Z_2$ symmetry
that exchanges $1 \leftrightarrow 2$ and $3\leftrightarrow 5$.
So the action of $S_5$ on $r_0$ generates $120/2 = 60$ similar structures,
only 40 of which are linearly independent.

In order to provide a symmetric form for the hard remainder $R^{(2)}_5$,
we find a linear combination $h$ of the 40 weight 4 functions which
is symmetric under the same $Z_2$ as $r_0$, and write $R^{(2)}_5$ as a sum over
the 60 permutations in $S_5/Z_2$,
\be
R^{(2)}_5 = \sum_{\sigma\in S_5/Z_2} r_0(\sigma) \, h(\sigma).
\label{symsumh}
\ee
Requiring that $R^{(2)}_5$ be the same as in the original
linearly-independent
40-term representation leaves a 6 parameter space of solutions.
We pick a particular solution in this space in order to simplify $h$, as we shortly explain.
The symbol of $h$ contains 26,012 terms
and is provided in the ancillary file 
{\tt remainder\char`_h.txt}.

We can characterize $h$ via its derivative, or more technically
the $\{3,1\}$ component of its coproduct, in much the same way
that we characterized the pure function $g_{234}^{\text{DT}}$
appearing in the double-trace coefficient of the $\NeqFour$
amplitude~\cite{Abreu:2018aqd}.  We first remark that the parity odd
part of $h$, like the odd part of $g_{234}^{\text{DT}}$,
has vanishing final entries for
all letters of the form $s_{ij}-s_{kl}$ (letters 6 to 15 and 21
to 25, see appendix \ref{app_subsec:alphabet}).
In addition, the weight 3 odd functions appearing
in the $\{3,1\}$ coproduct component
are all linear combinations of permutations of
the \emph{pure} $d=6$ pentagon integral, the $\I^{d=6}_5$
defined in eq.~\eqref{eq:pureIntDef}, whose
symbol we give in an ancillary file (we use
the same conventions as in ref.~\cite{Abreu:2018aqd}).
However, in contrast to $g_{234}^{\text{DT}}$,
$h$ does not contain letter 31 at all.
By an appropriate choice of solution in the 6 parameter space,
we find that the final entries for letters 17 and 19 vanish as
well,
for the odd part, $h^{\text{odd}}$.
We write the parity-odd part of its derivative as,
\eq{
\partial_{x_i} \left[h^{\text{odd}}\right]\big|_{\text{odd}} 
= \sum^{12}_{j=1}\sum_{\alpha_1} \I^{d=6}_5(\Sigma_j)\ m_{j\alpha_1}
\ \frac{\partial \log W_{\alpha_1}}{\partial x_i}\,,
}
where $j$ labels the 12 inequivalent permutations of the $d=6$
pentagon integral,
\begin{align}
\begin{split}
\Sigma_j &\in  \{ 
\{12543\}, \{12453\}, \{13524\}, \{12534\}, \{13254\}, \{12354\}, \\
&\twhite{.}\hskip0.6cm
\{14325\}, \{13425\}, \{14235\}, \{12435\}, \{13245\}, \{12345\} \},
\label{pentperms}
\end{split}
\end{align}
and $\alpha_1 \in \{1,\!...,\!5\} \cup \{16,\!...,\!20\} \cup \{31\}$
are the nonzero final entries for $g_{234}^{\text{DT}}$.

In these conventions, the matrix $m_{j\alpha_1}$ corresponding to
$h^{\text{odd}}$ is
\begin{small}
\begin{align}
\hskip -.7cm
\label{hmjalpha}
m_{j\alpha_1} = \frac{1}{12} \left(
\begin{array}{rrrrrrrrrrr}
 -3 & -2 &  2 &  2 & -2 &  1 & 0 &  1 & 0 &  1 & 0\\[2pt]
  3 &  1 & -1 & -3 &  0 &  1 & 0 & -1 & 0 &  0 & 0\\[2pt]
 -3 & -2 &  0 &  0 & -2 &  1 & 0 &  5 & 0 &  1 & 0\\[2pt]
  3 &  0 & -3 & -1 &  1 &  0 & 0 & -1 & 0 &  1 & 0\\[2pt]
  3 &  0 &  0 & -2 &  4 & -3 & 0 & -3 & 0 &  1 & 0\\[2pt]
 -3 & -1 &  1 &  3 &  0 & -1 & 0 &  1 & 0 &  0 & 0\\[2pt]
 -3 &  2 & -1 &  3 & -3 &  2 & 0 &  3 & 0 & -3 & 0\\[2pt]
  3 &  4 & -2 &  0 &  0 &  1 & 0 & -3 & 0 & -3 & 0\\[2pt]
  3 & -1 &  0 &  0 & -1 &  2 & 0 & -5 & 0 &  2 & 0\\[2pt]
 -3 &  0 &  3 &  1 & -1 &  0 & 0 &  1 & 0 & -1 & 0\\[2pt]
 -3 & -3 &  3 & -1 &  2 & -3 & 0 &  3 & 0 &  2 & 0\\[2pt]
  3 &  2 & -2 & -2 &  2 & -1 & 0 & -1 & 0 & -1 & 0\\[2pt]
\end{array}
\right) \,.
\end{align}
\end{small}
This matrix has rank 5, so the derivative contains
only five independent combinations of final entries,
and five independent combinations of $d=6$ pentagon permutations.

Similarly, we expand the odd part of the derivative of 
the parity even part of $h$ as,
\begin{align}
\partial_{x_i} \left[h^{\text{even}}\right]\big|_{\text{odd}} 
= \sum^{12}_{j=1}\sum_{\alpha_2} \I^{d=6}_5(\Sigma_j)\ n_{j\alpha_2}
\ \frac{\partial \log W_{\alpha_2}}{\partial x_i}\,,
\end{align}
where $\alpha_2 \in \{26,\ldots,30\}$ runs only over the five
odd letters.
The matrix $n_{j\alpha_2}$ for $h^{\text{even}}$ is given by
\begin{small}
\begin{align}
\hskip -.7cm \label{hnjalpha}
n_{j\alpha_2} = \frac{1}{12} \left(
\begin{array}{rrrrr}
 0 &  1 & 0 &  1 &  1\\[2pt]
 0 & -1 & 0 &  0 &  1\\[2pt]
 0 &  1 & 0 & -1 & -1\\[2pt]
 0 & -1 & 0 &  1 &  0\\[2pt]
 0 & -1 & 0 & -1 &  1\\[2pt]
 0 &  1 & 0 &  0 & -1\\[2pt]
 0 &  1 & 0 &  1 &  0\\[2pt]
 0 & -1 & 0 &  1 & -1\\[2pt]
 0 & -1 & 0 &  0 &  0\\[2pt]
 0 &  1 & 0 & -1 &  0\\[2pt]
 0 &  1 & 0 &  0 &  1\\[2pt]
 0 & -1 & 0 & -1 & -1\\[2pt]
\end{array}
\right) \,.
\end{align}
\end{small}
This matrix has rank 3, corresponding to the vanishing 
of  the final entries 26 and 28.

\subsection{Counting functions for $\NeqEight$ and $\NeqFour$}

It is interesting to compare the spaces of transcendental functions
for $\NeqEight$ and $\NeqFour$.  Before doing so for the two-loop 
five-point case of interest, we review the situation for lower numbers
of loops and/or legs, concentrating on the order ${\cal O}(\e^0)$
terms of weight 2 at one loop and weight 4 at two loops.

For the one-loop four-point amplitudes in both theories~\cite{Green:1982sw}, the space is
three-dimensional, and very simple in terms of the Mandelstam
variables $s,t,u$ (omitting factors of $\log \mu^2$):
\be
\{ \log s \, \log t,\ \log t\, \log u,\ \log u\, \log s \} \,.
\label{fns1l4pt}
\ee
The finite part of the leading-color
one-loop five-point $\NeqFour$ amplitude contains only logarithms~\cite{Bern:1993mq}
\be
V_5\ =\ \sum_{j=1}^5 \Biggl[ -\frac{1}{2} \log^2 s_{j,j+1}
 + \log\left(\frac{s_{j,j+1}}{s_{j+1,j+2}}\right)
   \log\left(\frac{s_{j+2,j-2}}{s_{j-2,j-1}}\right) + \zeta_2 \Biggr] \,.
\label{fns1l5pt}
\ee
The function is invariant under the dihedral $D_5$ symmetry of planar
amplitudes.
In the full-color amplitude, it appears in 12 nontrivial
permutations labeled by $S_5/D_5$.
Subleading-color contributions are also obtained from particular
permutations of this function~\cite{Bern:1990ux}.
The linear span of the 12 permutations of \eqn{fns1l5pt} is
an 11-dimensional space.  Thus there is one linear relation
among the 12 permutations of $V_5$,
\begin{align}\begin{split}
&V_5[12345] + V_5[12453] + V_5[13254] + V_5[13425] + V_5[14235] + V_5[14352]
\\
- &V_5[12435] - V_5[12354] - V_5[13245] - V_5[13452] - V_5[14325] - V_5[14253]
= 0,
\label{V5permsumrelation}
\end{split}\end{align}
corresponding to the totally antisymmetric combination of the twelve
functions.  This relation also holds for the $1/\e$ pole terms as well.
It can be derived by representing $V_5$ as a cyclic sum of one-mass
box integrals, and using the $Z_2\times Z_2$ symmetry of
each such box integral.

What about the one-loop five-point $\NeqEight$ amplitude?
From \eqn{eq:1loop_5pt_amp_diag_sum}, the amplitude contains a sum over
one-mass box integrals, so it might be expected to contain the
dilogarithms
present in the box integral~\cite{Bern:1993kr}.  On the other hand, the same could
be said for the $\NeqFour$ amplitude, where from \eqn{fns1l5pt}
they have long been known to cancel.  We find that the dilogarithms
all cancel from the one-loop five-point $\NeqEight$ amplitude
as well.  (As far as we know, this feature was not recognized before,
even though this amplitude has been available for over 20
years~\cite{Bern:1998sv}.)  In fact, of the 30 permutations of
the box coefficient $\beta_{123(45)}$ in eq.~(\ref{beta12345}), only 10 are linearly independent.
The coefficient of one of these 10 rational structures is,
\be
  \frac{1}{2} \log^2 \left(\frac{s_{41}}{s_{52}}\right)
- \frac{1}{2} \log^2 \left(\frac{s_{51}}{s_{24}}\right)
+ \log s_{12} \, \log\left(\frac{s_{52}s_{41}}{s_{51}s_{24}}\right)
+ \log\left(\frac{s_{34}}{s_{35}}\right)
  \log\left(\frac{s_{51}s_{41}}{s_{52}s_{24}}\right) \,.
\label{gravfns1l5pt}
\ee
Its images under the 30 permutations in $S_5/(Z_2 \times Z_2)$
span a 10 dimensional space, which is entirely contained within
the 11-dimensional space provided by the $V_5$ functions for
$\NeqFour$.\footnote{Note that eq.~(\ref{gravfns1l5pt}) is representative 
of the 10 pure functions, but it does not correspond to a term in a symmetrized 
form like eq.~(\ref{symsumh}).}
So by this measure, $\NeqEight$ is slightly simpler than $\NeqFour$.

Next we turn to the two-loop four-point amplitude.
How many functions should we expect in $\NeqFour$?
For a given color ordering, there is one planar amplitude,
because only a single Parke-Taylor factor appears at leading color.
There are 3 distinct orderings of a single trace, so there are really
3 planar functions.  In the full-color amplitude, group theory implies
that the subleading-color
single-trace coefficients can be traded for the double-trace
coefficients,
or vice versa, and only two of the three of these are
independent~\cite{Bern:2002tk}.  Also, there are two Parke-Taylor structures
in the four-point case, given the one Kleiss-Kuijf ($U(1)$ decoupling)
relation~\cite{KleissKuijf}.
So we expect $2 \times 2 = 4$ nonplanar functions, for a total of $3+4=7$.
Inspecting the actual $\NeqFour$ 
answer~\cite{BRY,SchnitzerN8UniformTrans}, there are 7 independent
functions.  So there are no mysterious relations
like \eqn{V5permsumrelation} at two loops and four points.
There are 3 functions associated with the $\NeqEight$ two-loop four-point
amplitude, with relative prefactors $s^2$, $t^2$, $u^2$ (or $st$, $tu$, $us$).
These three functions are contained within the space of
$\NeqFour$ functions.  This property was anticipated by a relation found in
ref.~\cite{SchnitzerN8UniformTrans} between subleading-color $\mathcal{N}=4$
and $\mathcal{N}=8$ amplitudes, although there are still rational factors
inhabiting this relation.

Finally, we turn to the two-loop five-point amplitudes. 
We need forty-five linearly independent rational 
structures $r_j$
 to describe the full unsubtracted amplitude $M^{(2)}_5$
in $\mathcal{N}=8$ supergravity. 
The weight-4 functions that multiply these 45
structures at ${\cal O}(\e^0)$ are all linearly independent.
As discussed in the previous subsection, 
if we perform the infrared subtraction defined in 
eq.~(\ref{eq:remainder_def}) to remove the pole terms that are
proportional to ``$s\log s$'' times the one-loop amplitude, and if
we also include in this subtraction the ${\cal O}(\e)$ terms
in the one loop amplitude, then the ${\cal O}(\e^0)$ terms in the amplitude
are shifted.  This remainder function has only 40 rational structures,
and the corresponding 40 functions are linearly independent.

We can also compare the functions for $\NeqEight$
with the corresponding number for
$\NeqFour$~\cite{Abreu:2018aqd,Chicherin:2018yne}.
First, we need to understand how many functions there are in the
latter amplitude.
Naively, there are 72 such functions.  The counting is as follows:
The planar (BDS) amplitude has a single pure function $M^{\text{BDS}}$
multiplying a single Parke-Taylor factor.  As the coefficient
of a single trace structure, 
$\tr[T^{a_1} \cdots T^{a_5}]\mi \tr[T^{a_5} \cdots T^{a_1}]$,
$M^{\text{BDS}}$ is invariant under a 10-element dihedral symmetry group, $D_5$.
Thus the sum of $M^{\text{BDS}}$ over $S_5$ permutations
is really over the coset $S_5/D_5$, which gives rise to $120/10 = 12$
planar functions.  In the nonplanar sector, the Edison-Naculich
relations~\cite{Edison:2011ta} show that the subleading-color
terms in the single-trace color structure, $A^{\text{SLST}}$, are linear
combinations of the planar amplitude and the coefficients
of the double-trace structure $A^{\text{DT}}$~\cite{Abreu:2018aqd}.
These latter coefficients can in turn be expanded as Parke-Taylor factors
times pure functions, and in this case all 6 Parke-Taylor factors
(after applying Kleiss-Kuijf identities~\cite{KleissKuijf}) contribute.
Their corresponding pure functions were called
$g_{\sigma(2),\sigma(3),\sigma(4)}^{\text{DT}}$ in ref.~\cite{Abreu:2018aqd}.
The double-trace color structure,
$\tr[T^{a_1}T^{a_5}] \, 
(\tr[T^{a_2}T^{a_3}T^{a_4}]\mi \tr[T^{a_2}T^{a_4}T^{a_3}])$,
is invariant under a 12-element $Z_2 \times S_3$ symmetry group.
Thus there should be $6\times10=60$ nonplanar functions, plus 12
planar functions, for a total of $60+12=72$.

However, the total number of linearly independent $\NeqFour$
functions
at weight 4 is actually 52, not 72.  Therefore there must be 20
separate linear relations between the transcendental functions.
These relations come in two sets of 10.  The first set only involves
permutations of the function $g \equiv g_{2,3,4}^{\text{DT}}$. One such equation
is 
\begin{align}\begin{split}
& g[12345] + g[12453] + g[12534] + g[21345] + g[21453] + g[21534]\\
- &g[12435] - g[12543] - g[12354] - g[21435] - g[21543] - g[21354] = 0.
\label{SYMrelation_g_only}
\end{split}\end{align}
The arguments of $g$ indicate the permutation that is to be applied 
to $g_{2,3,4}^{\text{DT}}$.
The other 9 equations in this set can be found by permuting the labels
further in this equation.
The second representative equation also involves the planar functions
$M \equiv M^{\text{BDS}}$, 
\begin{align}\begin{split}
&6 M[12345] - 6 M[13254]
+ 2 g[23145] + 2 g[25413] - 2 g[32154]
\\\
&
+ g[12354] + g[32451] + g[42531] + g[32541] + g[52314] + g[42513] + g[25431]
\\
&
+ g[41235] + g[31245] - g[12345] - g[52431] - g[52413] - g[42315] -g[21453]
\\
&
- g[21543] - g[23451] - g[24531] - g[23541] - g[51234] - g[31254] = 0.
\label{SYMrelation_Mg}
\end{split}\end{align}
Again the other 9 equations in this set can be found by permuting the
labels further.  These equations all hold, not only at ${\cal O}(\e^0)$
or weight 4, but also for the $1/\e$ pole components, which have lower
weight.

It would be very interesting to understand the origin of
eqs.~(\ref{SYMrelation_g_only}) and (\ref{SYMrelation_Mg}).
They generalize \eqn{V5permsumrelation} to two loops.
Do they reflect some hidden generalization of dual conformal invariance
to the nonplanar sector~\cite{Bern:2014kca,Bern:2015ple,Bern:2018oao,Bern:2017gdk,Chicherin:2018wes}? Could they represent
some integrated version of color-kinematics duality (see e.g.~\cite{Chester:2016ojq, Primo:2016omk})?

In any event, now that we know that there are 52 independent functions
for $\NeqFour$, we can ask, at ${\cal O}(\e^0)$ or weight 4,
how different are the 45 (or 40) $\NeqEight$ functions from them?
To address this question, we take the linear span of the 45 unsubtracted
$\NeqEight$ functions and the 52 $\NeqFour$
functions and find 62 independent functions.  That is, only 10 of the
$\NeqEight$ functions are ``new'', with respect to
those in $\NeqFour$.  (Or to turn it around, only 17 of the 52
$\NeqFour$ functions are ``new'' with respect to $\NeqEight$.)
Thus there is a large overlap between the two sets of functions.

On the other hand, if we take the span of the 40 
{\it subtracted}
$\NeqEight$ functions and the 52 (unsubtracted)
$\NeqFour$ functions, we find 92 independent functions,
i.e.~they are all independent.  The large concordance between
the two sets of unsubtracted functions is lost, when one set is subtracted.

We can also project the sets of (unsubtracted)
functions into the parity-even and parity-odd
sectors and repeat the exercise.  First of all, the number of independent
functions in the even and odd sectors is equal to the number before
projection.  The one exception to this rule is that 5 of the 45 $\NeqEight$
functions, the ones with $1/{\rm tr}_5$ in their rational
function coefficients, are pure parity-odd, so there are only 40 independent
parity-even functions.  For the parity-even part, the 40 $\mathcal{N}=8$
functions and the 52 $\mathcal{N}=4$ functions have a span with dimension 56.
For the parity-odd part, the 45 $\mathcal{N}=8$
functions and the 52 $\mathcal{N}=4$ functions have a span with dimension 62.

\begin{table}[h!]
\centering
\begin{tabular}{|c||c|c|c|c|}
\hline
functions & $\{1,1,1,1\}$ & $\{2,1,1\}$ & $\{3,1\}$ & weight 4 \\[1pt]
\hline
P odd space               &    0    &      9    &     111     &  1191  \\[1pt]
no.~from $\mathcal{N}=8$  &    0    &      9    &      11     &    45  \\[1pt]
no.~from $\mathcal{N}=4$  &    0    &      9    &      12     &    52  \\[1pt]
no.~from both             &    0    &      9    &      12     &    62  \\[1pt]
\hline
P even space              &   10    &     70    &     505     &  3736  \\[1pt]
no.~from $\mathcal{N}=8$  &   10    &     70    &     285     &    40  \\[1pt]
no.~from $\mathcal{N}=4$  &   10    &     70    &     362     &    52  \\[1pt]
no.~from both             &   10    &     70    &     367     &    56  \\[1pt]
\hline
P even with odd letters   &    0    &      0    &     45     & 
711  \\[1pt]
no.~from $\mathcal{N}=8$  &    0    &      0    &      40     &    40  \\[1pt]
no.~from $\mathcal{N}=4$  &    0    &      0    &      40     &    40  \\[1pt]
no.~from both             &    0    &      0    &      40     &    44  \\[1pt]
\hline
\end{tabular}
\caption{\label{tab:coproducts} Table of dimensions of coproducts
of the weight 4 functions for the $\NeqEight$ and $\NeqFour$ twoloop 
five-point amplitudes. By weight 2 they span the full function space
with the second entry condition.}
\end{table}

Because the parity-even overlap involves only 4 additional functions,
and because the parity-even sector has a lot of ``simple'' functions containing
no odd letters, we also ask how many of the even functions require 
odd letters in their symbol (two at a time, of course, by parity
and the first entry condition).  The part of the weight-4 
parity-even
space requiring odd letters is 40 dimensional for both $\NeqFour$
and $\NeqEight$; however the two spaces are not identical
because their span has dimension 44.  In other words, the extra 4 parity-even
functions required by $\NeqEight$ all require odd letters.
It would be interesting to investigate further
the $52-40=12$  functions in $\mathcal{N}=4$ that have no odd
letters, and see just how simple they are.

The right column of table~\ref{tab:coproducts} displays the
dimensions of the weight-4 $\NeqEight$, $\NeqFour$ and combined
spaces,
relative to the full function space proposed in ref.~\cite{Chicherin:2017dob},
which includes the 31 letters, plus an empirical constraint on the first
two entries. This constraint is satisfied by all functions needed
to build both amplitudes, simply because it is satisfied by all
master integrals.

The dimensions in the columns toward the left in table~\ref{tab:coproducts}
correspond to the number of independent functions found by repeated
differentiation of the respective weight 4 functions.
More technically, given a weight $n$ function $F$,
we extract the $\{n-1,1\}$ components $F^\alpha$ for all 31 letters $\alpha$
via the formula, 
\eq{
\partial_{x_i} F 
= \sum_{\alpha=1}^{31} F^{\alpha}\ \frac{\partial \log W_{\alpha}}{\partial x_i}\,.
}
At the level of the symbol, $F^\alpha$ is constructed from $F$
by setting all symbol terms in $F$ to zero unless they have
$W_\alpha$ as their last letter, in which case that letter is clipped off.

Note that parity-even functions can be generated from
odd functions at one weight higher (by clipping off an odd letter),
and vice versa.
At weight two and lower, the amplitudes' coproducts saturate the full space.
However, at weight 3 they occupy a remarkably small fraction
of the nontrivial part of the function space.

In particular, for weight 3 parity odd functions, only 12 of the 111
possible functions are required: the 12 permutations
of the $d=6$ pentagon, $\I^{d=6}_5(\Sigma_j)$.  In the case of $\NeqEight$,
one of the 12 combinations does not appear, and that is the totally 
symmetric sum, $\sum_{j=1}^{12} \I^{d=6}_5(\Sigma_j)$.  We can verify
its absence for the (subtracted) hard function $h$ by observing
that the sum of all column entries vanishes, for all columns
in both the matrix $m_{j\alpha_1}$ in \eqn{hmjalpha}
and $n_{j\alpha_2}$ in \eqn{hnjalpha}.  Of course there are many other
vertical combinations that vanish, since the matrices have ranks 5 and 3
respectively. However, the total sum corresponds to a symmetric combination
that also vanishes for any permutation of $h$.  (The functions
appearing in the subtraction term, coming from
the ${\cal O}(\e)$ term in the one-loop amplitude
also obey this property, and 
so it is true for the unsubtracted amplitude as well.)
Thus in $\NeqEight$, as in $\NeqFour$~\cite{Abreu:2018aqd},
the $d=6$ pentagon integrals provide a key to a lot of the structure
of the final result.

Another key consists of the weight-3 even functions containing
odd letters.
At low weights, most of the even functions do not contain any odd letters.
The bulk of these functions are simply products of logarithms and dilogarithms
whose arguments are rational in the $s_{ij}$ invariants.  More interesting
are the even functions that have two odd letters in some of their
symbol terms.  (They need two odd letters because of parity, and at weight 4,
since an odd letter cannot be in the first entry, 
they cannot have four odd
letters in any term.)  We count these functions in the bottom rows
in the table.  We observe that the $\{3,1\}$ coproducts of both $\NeqEight$
and $\NeqFour$ live in exactly the same 40-dimensional space.

In summary, starting at weight 3, $\NeqEight$
and $\NeqFour$ utilize a remarkably small fraction of the ``interesting''
available pentagon-function space.  Also, there is a surprising
degree of
similarity between the two sets of functions, despite the
fact that the two sets of integrals required for the two-loop five-point
amplitude are different: linear in the loop momentum
for $\NeqFour$ and quadratic for $\NeqEight$ (in the BCJ/double copy
representation).  
It would be interesting to know whether these features have 
further implications for higher loops or other processes.

\section{Outlook} 
\label{sec:outlook}

In this work we have computed the symbol of the two-loop
five-particle scattering amplitude in $\NeqEight$, extending the analytic knowledge of supergravity amplitudes beyond the two-loop four-point examples of 
ref.~\cite{BoucherVeronneau:2011qv}. Our computation relies on
reducing the known supergravity integrand~\cite{Carrasco:2011mn} to the available pure master integrals for all massless
two-loop five-point amplitudes~\cite{Abreu:2018aqd,Chicherin:2018old}.
This step has been significantly simplified and made possible by two key
ideas. First, we employed insights from 
methods based on generalized unitarity
to identify the relevant space of rational
kinematic prefactors which the amplitude spans. All such structures can
be identified by 4- as well as $d$-dimensional leading singularities,
i.e.~maximal codimension residues of the loop integrand that
localize
all internal loop degrees of freedom. Second, we used modern
integration-by-parts methods based on generalized unitarity and
computational algebraic geometry, together with efficient numerical
finite-field methods, to perform the integral reduction. This purely
numerical approach avoids the prohibitive explosion of the size of
intermediate expressions associated with the complexities of the
five-point multi-scale problem. A priori knowledge of the analytic form
of the rational prefactors then allows us to efficiently reconstruct the analytic result from finite-field numerics.

We have verified our result
by checking the universal infrared pole structure as well as matching to known factorization formulae in the soft and collinear limits. We also point out a number of interesting analytic properties of the supergravity symbol and compare it with the recently computed Yang--Mills 
counterpart~\cite{Abreu:2018aqd,Chicherin:2018yne}. 
Like the $\NeqFour$ result, we find that the two-loop five-particle supergravity amplitude is uniformly transcendental. Clearly, $\NeqEight$ must be the ``simplest quantum field
theory''~\cite{ArkaniHamed:2008gz} since its two-loop five-point
amplitude requires 7 fewer functions compared to the other
contender for the title, $\NeqFour$ with full color dependence.
Furthermore, neither its un-subtracted nor subtracted
amplitude requires the letter $W_{31} = \tr_5$.
A further interesting observation is that all pieces related to
the $d$-dimensional leading singularities cancel in a suitably defined IR-subtracted remainder function $R^{(2)}_5$. This observation is reminiscent of earlier observations in the context of planar $\NeqFour$~\cite{Bern:2008ap}. 

Where do we go from here? On a formal level, it would be interesting to
investigate if there is any imprint of BCJ duality on
full amplitudes. This is known to be the case in the different
context of half-maximal supergravity in 5 dimensions, where ``enhanced
cancellations" of two-loop UV divergences can be explained by the
duality~\cite{Bern:2012gh}. As we have discussed in
sec.~\ref{sec:leadingSing}, all supergravity leading
singularities are direct double copies of their super-Yang--Mills
counterparts, but besides rational factors, are there any
indications in
the transcendental functions that originate from the fact that
supergravity integrands are the square of super-Yang--Mills? The two-loop
five-point example presented here seems like an ideal laboratory to
investigate this question, since this is a situation where the BCJ
representation of the integrand involves nontrivial loop-momentum
dependent numerators. 

On a practical level, given the usefulness of the $d$-dimensional
leading singularity method in systematically identifying the rational
functions that appear in the supergravity amplitude from a relatively
simple loop-integrand analysis, it is quite natural to wonder if similar
techniques may help to identify the relevant rational structures of QCD
amplitudes before integration. Just as in the construction of simple forms of loop integrands using generalized unitarity, recyling information from tree-like objects could dramatically simplify otherwise complicated  amplitudes.

\vspace{-12pt}\section*{Acknowledgments}\vspace{-10pt}

L.D.\ and E.H.\ are grateful to Humboldt University, Berlin, for hospitality while this project was completed.
The work of S.A.~is supported by the Fonds de la Recherche Scientifique--FNRS, Belgium.
The work of L.D.\ and E.H.\ is supported by the U.S. Department of Energy (DOE) under contract DE-AC02-76SF00515.  L.D.\ is also supported by a Humboldt Research Award. 
The work of B.P. is supported by the French Agence Nationale pour la Recherche, under grant ANR–17–CE31–0001–01.
The work of M.Z.\ is supported by the Swiss National Science Foundation under contract SNF200021 179016 and the European Commission through the ERC grant pertQCD.

\appendix

\section{Kinematics}
\label{kinematicsappendix}

In this appendix, we summarize various aspects of the
kinematics of massless five-particle scattering for the benefit 
of the reader, without breaking the exposition in the
main text.
More concretely, we first reproduce the pentagon alphabet of 
ref.~\cite{Chicherin:2017dob}. Then we discuss the 
momentum-twistor parametrization of ref.~\cite{Badger:2013gxa},
which rationalizes the symbol alphabet and allows us to use 
powerful finite-field methods in numerical calculations. For 
the validation of the amplitude in section~\ref{sec:validation},
we discuss the
soft- and collinear limits of five-point scattering. These
kinematic limits can also be implemented at the level of the 
twistor parameters in a straightforward manner, as explained in
the main text.

\subsection{Symbol alphabet}
\label{app_subsec:alphabet}
Before discussing details of the five-point kinematics, we 
reproduce the symbol alphabet for five-particle 
scattering first conjectured in ref.~\cite{Chicherin:2017dob},
which is relevant for our discussion 
in section~\ref{sec:amplitude_decomposition}. Here, we first 
follow the kinematic notation of ref.~\cite{Chicherin:2017dob},
and subsequently discuss a few simplifications for the 31 
letters of the alphabet. 

Scattering amplitudes for massless five-point processes depend
on the five massless external momenta $k_i$, involved in the
process, subject to the on-shell constraints $k^2_i = 0$ and 
momentum conservation $\sum^5_{i=1} k_i =0$. The kinematic 
dependence is given in terms of five independent Mandelstam 
invariants $v_i$. In ref.~\cite{Chicherin:2017dob}, the 
following notation is introduced,
\begin{align}
\begin{split}
 v_i & = \sab{i,i\pl1} = 2 k_i \cdot k_{i+1}\,, \qquad \Delta = \tr^2_5 =  \det(2k_i\cdot k_j) \\
 \mathbf{a}_{1,2,3,4}& = \tr[\slashed k_4 \slashed k_5 \slashed k_1 \slashed k_2]
 				  =  v_1v_2-v_2v_3+v_3v_4-v_1v_5-v_4v_5
\end{split}
\end{align}
where $\tr_5 = \tr[\gamma_5\slashed k_1 \slashed k_2 
\slashed k_3\slashed k_4] = \sqb{12}\ab{23}\sqb{34}\ab{41} -  \ab{12}\sqb{23}\ab{34}\sqb{41}$  as defined in 
eq.~\eqref{eq:tr5Def}.
With these definitions, we can now discuss the 31 letters 
$W_{\alpha}$ of the alphabet relevant for the massless 
five-particle scattering problem. These letters can be grouped 
according to spacetime \emph{parity}, which corresponds to
flipping the sign of $\tr_5 \to - \tr_5$, or, equivalently,
conjugating the spinor-bracket expressions 
$\ab{\cdot}\lra \sqb{\cdot}$. The parity-even letters are given 
as the five cyclic images (the index $i$ runs over $1,...,5$) 
of the following basic structures,
\begin{align}
\label{eq:letters_even}
 & W_i = v_i = s_{i,i+1}\,, 		&& W_{5+i} = v_{i+2} + v_{i+3}\,, \nonumber \\
 & W_{10+i} = v_i - v_{i+3}\,, 	&& W_{15+i} = v_i +v_{i+1}-v_{i+3}\,, \\ 
 & W_{20+i} = v_{i+2} + v_{i+3} -v_i -v_{i+1}\,, && W_{31} = \sqrt{\Delta} = \tr_5 \,.\nonumber
\end{align} 
The five parity-odd letters are given by the five cyclic images
of 
\begin{align}
\label{eq:letters_sqrt_parametrization}
 W_{25+i} = \frac{\mathbf{a}_{i,i\pl1,i\pl2,i\pl3} - 
 \sqrt{\Delta}}{\mathbf{a}_{i,i\pl1,i\pl2,i\pl3} + 
 \sqrt{\Delta} }\,.
\end{align}
For real Minkowski momenta $k_i$, complex conjugation is 
realized as $(\sqrt{\Delta})^*=-\sqrt{\Delta}$ so that the odd 
letters invert under complex conjugation $(W_j)^* = W^{-1}_j$ 
for $j\in \{26,...,30\}$.

We can use momentum conservation and spinor-trace identities to
write the alphabet in a more compact way that also eliminates 
the square root in the parity-odd letters, at the cost of having
expressions that are manifestly spinor-helicity valued 
and not written in terms of the five independent Mandelstam
variables $v_i$ alone. 
For concreteness, we write the full list of 31 letters in this 
form:
\begin{align}
\label{eq:simple_alphabet}
\begin{split}
W_\alpha = \Bigg\{
&\sab{12},\sab{23},\sab{34},\sab{45},\sab{15}, \\[-10pt]
&\sab{34}\pl\sab{45},\sab{15}\pl\sab{45},\sab{12}\pl\sab{15},\sab{12}\pl\sab{23},\sab{23}\pl\sab{34}, \\
&\sab{12}\mi\sab{45},\sab{23}\mi\sab{15},\sab{34}\mi\sab{12},\sab{45}\mi\sab{23},\sab{15}\mi\sab{34}, \\
&\mi\sab{13},\mi\sab{24},\mi\sab{35},\mi\sab{14},\mi\sab{25}, \\
&\mi\sab{23}\mi\sab{35},\mi\sab{14}\mi\sab{34},\mi\sab{25}\mi\sab{45},\mi\sab{13}\mi\sab{15},\mi\sab{12}\mi\sab{24}, \\
&\frac{\ab{12} \ab{45} \sqb{15} \sqb{24}}{\ab{15} \ab{24} \sqb{12} \sqb{45}},
\frac{\ab{15} \ab{23} \sqb{12} \sqb{35}}{\ab{12} \ab{35} \sqb{15} \sqb{23}},
\frac{\ab{12} \ab{34} \sqb{14} \sqb{23}}{\ab{14} \ab{23} \sqb{12} \sqb{34}},\\
&\frac{\ab{23} \ab{45} \sqb{25} \sqb{34}}{\ab{25} \ab{34} \sqb{23} \sqb{45}},
\frac{\ab{15} \ab{34} \sqb{13} \sqb{45}}{\ab{13} \ab{45} \sqb{15} \sqb{34}},
\tr_5\Bigg\}\,.
\end{split}
\end{align}
We note that letters $W_6,...,W_{10}$ and $W_{21},...,W_{25}$ 
can also be written in the form $s_{ij}-s_{kl}$. For instance, 
$W_6=s_{34}+s_{45}=s_{12}-s_{35}$.

\subsection{Twistor parametrization and rationalization of the alphabet}
\label{app_subsec:twistor_parametrization}
We have seen in the previous subsection, especially in 
eq.~(\ref{eq:letters_sqrt_parametrization}), that, if one 
chooses five independent Mandelstam variables $\sab{ij}$ as
kinematic variables, the pentagon alphabet contains the square
root of the Gram determinant $\sqrt{\Delta}$. 
From a practical point of view, it is often very desirable to 
rationalize the symbol alphabet. For a recent systematic study 
of rationalizing various roots, see ref.~\cite{Besier:2018jen}. 
Since momentum twistors~\cite{Hodges} give a set of 
unconstrained variables that automatically generate 
momentum-conserving on-shell kinematics, it has been well 
established that choosing such variables is extremely useful in
the context of rationalizing alphabets, 
see e.g.~ref.~\cite{Badger:2013gxa} for the application to
five-point
massless kinematics. 
We now summarize the parametrization established in appendix
A.2 of ref.~\cite{Badger:2013gxa}, which we employ in our 
calculation. For the convenience of the reader, we
also derive the spinor-helicity variables 
that allow us to evaluate all spinor-bracket expressions, such 
as the alphabet in eq.~(\ref{eq:simple_alphabet}). The twistor 
matrix can be parameterized by five independent variables $x_i$ 
in the following way,
\begin{align}
\label{eq:mom_twistor_parametrization}
Z_{(5)} =  \big(Z_1 \, Z_2 \,Z_3 \, Z_4\, Z_5\big) 
	= \begin{pmatrix}
		1 & 0 & \frac{1}{x_1} & \frac{1}{x_1}\pl\frac{1}{x_2} & \frac{1}{x_1}\pl\frac{1}{x_2}\pl\frac{1}{x_3} \\
		0 & 1 & 1 & 1&1 \\
		0&0&0&x_4&1 \\
		0 & 0 & 1 & 1 & \frac{x_5}{x_4}
	   \end{pmatrix}\,.
\end{align}
From the momentum twistor parametrization in 
eq.~(\ref{eq:mom_twistor_parametrization}) there exists a 
straightforward map to the more familiar spinor-helicity 
variables, see e.g. ref.~\cite{Hodges} or sec.~2 of 
ref.~\cite{Bourjaily:2010wh} for more details on this map.
For the five-particle case at hand, this map gives the following
spinor-helicity variables,
\begin{align}
\begin{split}
\lam{(5)} & = \big(\lam{1} \, \lam{2} \, \lam{3} \, \lam{4}\,  \lam{5}\big) 
	      = \begin{pmatrix}
	      		1 & 0 & \frac{1}{x_1} & \frac{1}{x_1}\pl\frac{1}{x_2} & \frac{1}{x_1}\pl\frac{1}{x_2}\pl\frac{1}{x_3} \\
			0 & 1 & 1 & 1&1 \\
	      	 \end{pmatrix}\,, \\
 \lamt{(5)} & = \big(\lamt{1} \, \lamt{2} \, \lamt{3} \, \lamt{4}\,  \lamt{5}\big) 
 		   = \begin{pmatrix}
			 \mi1 & 0 & \mi x_2 x_4 & x_3 \left(x_4\mi1\right)\pl x_2 x_4 & x_3 \left(1\mi x_4\right) \\
 			\mi\frac{x_5}{x_4} & \mi x_1 & x_1 & x_3 \left(1\mi\frac{x_5}{x_4}\right) & x_3 \left(\frac{x_5}{x_4}\mi1\right) \\
		      \end{pmatrix}\,.
 \end{split}
\end{align} 
From these helicity-spinors, all bracket expressions can be
evaluated with
\begin{align}
 	\ab{ij} 	\equiv \det(\{\lam{i},\lam{j}\})\,, \qquad 
	\sqb{ij} 	\equiv \det(\{\lamt{j},\lamt{i}\})\,, \qquad
	\sab{ij} = \ab{ij}\sqb{ji}\,,
\end{align}
where $\det(\{\lam{i},\lam{j}\})$ 
($\det(\{\lamt{j},\lamt{i}\})$) is the instruction to compute 
the $2\times2$ determinant obtained by selecting columns $i$ 
and $j$ from the $2\times 5$ matrix $\lam{(5)}$ ($\lamt{(5)}$).
As an example, we get for instance that $\ab{23}=-1/x_1$. 
Furthermore, the five independent Mandelstam invariants are rationally mapped to the five $x_i$ variables according to
\begin{align}
\begin{split}
 & \sab{12}= x_1\,, \quad 
 \sab{23}=x_2 x_4 \,, \quad 
 \sab{34}=x_1 \left(\!x_4\mi\frac{x_3 \left(1\mi x_4\right)}{x_2} \!\right)\! \pl x_3 \left(x_4\mi x_5\right)\,, \\
 & \sab{45}=x_2 \left(x_4\mi x_5\right)\,, \qquad\quad\ 
 \sab{51}=x_3 \left(1\mi x_5\right)\,.
\end{split}
\end{align}
In the $x_i$ variables, it is clear that the parity-odd letters $W_{26,...30}$ turn into \emph{rational} functions of the $x_i$ as the letters are rational in the spinor brackets and each of the spinors is rationally parameterized, e.g.
\begin{align}
W_{26}= \frac{\ab{12} \ab{45} \sqb{15} \sqb{24}}
            {\ab{15} \ab{24} \sqb{12} \sqb{45}}
 = \frac{x_1\left(x_5\mi1\right)
         \left( x_3 \left(x_4\mi1\right) \pl x_2 x_4\right)}
{x_3 \left(x_1\pl x_2\right) \left(x_4\mi x_5\right)}  \,.
\end{align}
%
%
Similarly, since $\tr_5$ is a rational function of spinor
brackets, it is also a rational function of the $x_i$.
Finally, we would like to emphasize once more that twistor
variables allow us to generate rational kinematics by
choosing rational values for the $x_i$ variables.

\vskip -.3cm
\bibliographystyle{JHEP}
\phantomsection
\bibliography{amp_refs}

\providecommand{\href}[2]{#2}\begingroup\raggedright\begin{thebibliography}{100}

\bibitem{DualConformalMagic}
J.~Drummond, J.~Henn, V.~Smirnov and E.~Sokatchev, \emph{{Magic identities for
  conformal four-point integrals}},
  \href{https://doi.org/10.1088/1126-6708/2007/01/064}{\emph{JHEP} {\bfseries
  0701} (2007) 064} [\href{https://arxiv.org/abs/hep-th/0607160}{{\ttfamily
  hep-th/0607160}}].

\bibitem{Bern:2006ew}
Z.~Bern, M.~Czakon, L.~J. Dixon, D.~A. Kosower and V.~A. Smirnov, \emph{{The
  Four-Loop Planar Amplitude and Cusp Anomalous Dimension in Maximally
  Supersymmetric Yang-Mills Theory}},
  \href{https://doi.org/10.1103/PhysRevD.75.085010}{\emph{Phys.Rev.} {\bfseries
  D75} (2007) 085010} [\href{https://arxiv.org/abs/hep-th/0610248}{{\ttfamily
  hep-th/0610248}}].

\bibitem{Alday:2007hr}
L.~F. Alday and J.~M. Maldacena, \emph{{Gluon scattering amplitudes at strong
  coupling}}, \href{https://doi.org/10.1088/1126-6708/2007/06/064}{\emph{JHEP}
  {\bfseries 0706} (2007) 064}
  [\href{https://arxiv.org/abs/0705.0303}{{\ttfamily 0705.0303}}].

\bibitem{Drummond:2008vq}
J.~Drummond, J.~Henn, G.~Korchemsky and E.~Sokatchev, \emph{{Dual
  superconformal symmetry of scattering amplitudes in N=4 super-Yang-Mills
  theory}},
  \href{https://doi.org/10.1016/j.nuclphysb.2009.11.022}{\emph{Nucl.Phys.}
  {\bfseries B828} (2010) 317}
  [\href{https://arxiv.org/abs/0807.1095}{{\ttfamily 0807.1095}}].

\bibitem{Goncharov:2010jf}
A.~B. Goncharov, M.~Spradlin, C.~Vergu and A.~Volovich, \emph{{Classical
  Polylogarithms for Amplitudes and Wilson Loops}},
  \href{https://doi.org/10.1103/PhysRevLett.105.151605}{\emph{Phys. Rev. Lett.}
  {\bfseries 105} (2010) 151605}
  [\href{https://arxiv.org/abs/1006.5703}{{\ttfamily 1006.5703}}].

\bibitem{Duhr:2011zq}
C.~Duhr, H.~Gangl and J.~R. Rhodes, \emph{{From polygons and symbols to
  polylogarithmic functions}},
  \href{https://doi.org/10.1007/JHEP10(2012)075}{\emph{JHEP} {\bfseries 10}
  (2012) 075} [\href{https://arxiv.org/abs/1110.0458}{{\ttfamily 1110.0458}}].

\bibitem{Duhr:2012fh}
C.~Duhr, \emph{{Hopf algebras, coproducts and symbols: an application to Higgs
  boson amplitudes}},
  \href{https://doi.org/10.1007/JHEP08(2012)043}{\emph{JHEP} {\bfseries 08}
  (2012) 043} [\href{https://arxiv.org/abs/1203.0454}{{\ttfamily 1203.0454}}].

\bibitem{Bern:1994zx}
Z.~Bern, L.~J. Dixon, D.~C. Dunbar and D.~A. Kosower, \emph{{One loop n point
  gauge theory amplitudes, unitarity and collinear limits}},
  \href{https://doi.org/10.1016/0550-3213(94)90179-1}{\emph{Nucl.Phys.}
  {\bfseries B425} (1994) 217}
  [\href{https://arxiv.org/abs/hep-ph/9403226}{{\ttfamily hep-ph/9403226}}].

\bibitem{ArkaniHamed:2010kv}
N.~Arkani-Hamed, J.~L. Bourjaily, F.~Cachazo, S.~Caron-Huot and J.~Trnka,
  \emph{{The All-Loop Integrand For Scattering Amplitudes in Planar N=4 SYM}},
  \href{https://doi.org/10.1007/JHEP01(2011)041}{\emph{JHEP} {\bfseries 1101}
  (2011) 041} [\href{https://arxiv.org/abs/1008.2958}{{\ttfamily 1008.2958}}].

\bibitem{ArkaniHamed:2010gh}
N.~Arkani-Hamed, J.~L. Bourjaily, F.~Cachazo and J.~Trnka, \emph{{Local
  Integrals for Planar Scattering Amplitudes}},
  \href{https://doi.org/10.1007/JHEP06(2012)125}{\emph{JHEP} {\bfseries 06}
  (2012) 125} [\href{https://arxiv.org/abs/1012.6032}{{\ttfamily 1012.6032}}].

\bibitem{Bourjaily:2017wjl}
J.~L. Bourjaily, E.~Herrmann and J.~Trnka, \emph{{Prescriptive Unitarity}},
  \href{https://doi.org/10.1007/JHEP06(2017)059}{\emph{JHEP} {\bfseries 06}
  (2017) 059} [\href{https://arxiv.org/abs/1704.05460}{{\ttfamily
  1704.05460}}].

\bibitem{Bern:1994cg}
Z.~Bern, L.~J. Dixon, D.~C. Dunbar and D.~A. Kosower, \emph{{Fusing gauge
  theory tree amplitudes into loop amplitudes}},
  \href{https://doi.org/10.1016/0550-3213(94)00488-Z}{\emph{Nucl.Phys.}
  {\bfseries B435} (1995) 59}
  [\href{https://arxiv.org/abs/hep-ph/9409265}{{\ttfamily hep-ph/9409265}}].

\bibitem{Britto:2004nc}
R.~Britto, F.~Cachazo and B.~Feng, \emph{{Generalized unitarity and one-loop
  amplitudes in N=4 super-Yang-Mills}},
  \href{https://doi.org/10.1016/j.nuclphysb.2005.07.014}{\emph{Nucl.Phys.}
  {\bfseries B725} (2005) 275}
  [\href{https://arxiv.org/abs/hep-th/0412103}{{\ttfamily hep-th/0412103}}].

\bibitem{MaximalCuts}
Z.~Bern, J.~Carrasco, H.~Johansson and D.~Kosower, \emph{{Maximally
  supersymmetric planar Yang-Mills amplitudes at five loops}},
  \href{https://doi.org/10.1103/PhysRevD.76.125020}{\emph{Phys.Rev.} {\bfseries
  D76} (2007) 125020} [\href{https://arxiv.org/abs/0705.1864}{{\ttfamily
  0705.1864}}].

\bibitem{postnikov}
A.~{Postnikov}, \emph{{Total positivity, Grassmannians, and networks}},
  {\emph{ArXiv Mathematics e-prints} (2006) }
  [\href{https://arxiv.org/abs/math/0609764}{{\ttfamily math/0609764}}].

\bibitem{ArkaniHamed:2012nw}
N.~Arkani-Hamed, J.~L. Bourjaily, F.~Cachazo, A.~B. Goncharov, A.~Postnikov and
  J.~Trnka, \emph{{Grassmannian Geometry of Scattering Amplitudes}}. Cambridge
  University Press, 2016,
  \href{https://doi.org/10.1017/CBO9781316091548}{10.1017/CBO9781316091548},
  [\href{https://arxiv.org/abs/1212.5605}{{\ttfamily 1212.5605}}].

\bibitem{Arkani-Hamed:2013jha}
N.~Arkani-Hamed and J.~Trnka, \emph{{The Amplituhedron}},
  \href{https://doi.org/10.1007/JHEP10(2014)030}{\emph{JHEP} {\bfseries 10}
  (2014) 030} [\href{https://arxiv.org/abs/1312.2007}{{\ttfamily 1312.2007}}].

\bibitem{BCJ}
Z.~Bern, J.~Carrasco and H.~Johansson, \emph{{New Relations for Gauge-Theory
  Amplitudes}},
  \href{https://doi.org/10.1103/PhysRevD.78.085011}{\emph{Phys.Rev.} {\bfseries
  D78} (2008) 085011} [\href{https://arxiv.org/abs/0805.3993}{{\ttfamily
  0805.3993}}].

\bibitem{BjerrumBohr:2009rd}
N.~E.~J. Bjerrum-Bohr, P.~H. Damgaard and P.~Vanhove, \emph{{Minimal Basis for
  Gauge Theory Amplitudes}},
  \href{https://doi.org/10.1103/PhysRevLett.103.161602}{\emph{Phys. Rev. Lett.}
  {\bfseries 103} (2009) 161602}
  [\href{https://arxiv.org/abs/0907.1425}{{\ttfamily 0907.1425}}].

\bibitem{Stieberger:2009hq}
S.~Stieberger, \emph{{Open \& Closed vs. Pure Open String Disk Amplitudes}},
  \href{https://arxiv.org/abs/0907.2211}{{\ttfamily 0907.2211}}.

\bibitem{BCJSquare}
Z.~Bern, T.~Dennen, Y.-t. Huang and M.~Kiermaier, \emph{{Gravity as the Square
  of Gauge Theory}},
  \href{https://doi.org/10.1103/PhysRevD.82.065003}{\emph{Phys.Rev.} {\bfseries
  D82} (2010) 065003} [\href{https://arxiv.org/abs/1004.0693}{{\ttfamily
  1004.0693}}].

\bibitem{BCJLoop}
Z.~Bern, J.~J.~M. Carrasco and H.~Johansson, \emph{{Perturbative Quantum
  Gravity as a Double Copy of Gauge Theory}},
  \href{https://doi.org/10.1103/PhysRevLett.105.061602}{\emph{Phys.Rev.Lett.}
  {\bfseries 105} (2010) 061602}
  [\href{https://arxiv.org/abs/1004.0476}{{\ttfamily 1004.0476}}].

\bibitem{ColorKinematics}
Z.~Bern, J.~Carrasco, L.~Dixon, H.~Johansson and R.~Roiban, \emph{{Simplifying
  Multiloop Integrands and Ultraviolet Divergences of Gauge Theory and Gravity
  Amplitudes}},
  \href{https://doi.org/10.1103/PhysRevD.85.105014}{\emph{Phys.Rev.} {\bfseries
  D85} (2012) 105014} [\href{https://arxiv.org/abs/1201.5366}{{\ttfamily
  1201.5366}}].

\bibitem{Carrasco:2011mn}
J.~J. Carrasco and H.~Johansson, \emph{{Five-Point Amplitudes in N=4
  Super-Yang-Mills Theory and N=8 Supergravity}},
  \href{https://doi.org/10.1103/PhysRevD.85.025006}{\emph{Phys. Rev.}
  {\bfseries D85} (2012) 025006}
  [\href{https://arxiv.org/abs/1106.4711}{{\ttfamily 1106.4711}}].

\bibitem{Mafra:2015mja}
C.~R. Mafra and O.~Schlotterer, \emph{{Two-loop five-point amplitudes of super
  Yang-Mills and supergravity in pure spinor superspace}},
  \href{https://doi.org/10.1007/JHEP10(2015)124}{\emph{JHEP} {\bfseries 10}
  (2015) 124} [\href{https://arxiv.org/abs/1505.02746}{{\ttfamily
  1505.02746}}].

\bibitem{Bern:2017yxu}
Z.~Bern, J.~J. Carrasco, W.-M. Chen, H.~Johansson and R.~Roiban, \emph{{Gravity
  Amplitudes as Generalized Double Copies of Gauge-Theory Amplitudes}},
  \href{https://doi.org/10.1103/PhysRevLett.118.181602}{\emph{Phys. Rev. Lett.}
  {\bfseries 118} (2017) 181602}
  [\href{https://arxiv.org/abs/1701.02519}{{\ttfamily 1701.02519}}].

\bibitem{Bern:2017ucb}
Z.~Bern, J.~J.~M. Carrasco, W.-M. Chen, H.~Johansson, R.~Roiban and M.~Zeng,
  \emph{{Five-loop four-point integrand of $N=8$ supergravity as a generalized
  double copy}}, \href{https://doi.org/10.1103/PhysRevD.96.126012}{\emph{Phys.
  Rev.} {\bfseries D96} (2017) 126012}
  [\href{https://arxiv.org/abs/1708.06807}{{\ttfamily 1708.06807}}].

\bibitem{Bern:2018jmv}
Z.~Bern, J.~J. Carrasco, W.-M. Chen, A.~Edison, H.~Johansson, J.~Parra-Martinez
  et~al., \emph{{Ultraviolet Properties of $\mathcal N = 8$ Supergravity at
  Five Loops}}, \href{https://doi.org/10.1103/PhysRevD.98.086021}{\emph{Phys.
  Rev.} {\bfseries D98} (2018) 086021}
  [\href{https://arxiv.org/abs/1804.09311}{{\ttfamily 1804.09311}}].

\bibitem{BCJreviewToAppear}
Z.~Bern, J.~J. Carrasco, M.~Chiodaroli, H.~Johansson and R.~Roiban, \emph{{The
  Duality Between Color and Kinematics and its Applications}}, {\emph{to
  appear} (2019) }.

\bibitem{Dixon:2011pw}
L.~J. Dixon, J.~M. Drummond and J.~M. Henn, \emph{{Bootstrapping the three-loop
  hexagon}}, \href{https://doi.org/10.1007/JHEP11(2011)023}{\emph{JHEP}
  {\bfseries 1111} (2011) 023}
  [\href{https://arxiv.org/abs/1108.4461}{{\ttfamily 1108.4461}}].

\bibitem{Caron-Huot:2016owq}
S.~Caron-Huot, L.~J. Dixon, A.~McLeod and M.~von Hippel, \emph{{Bootstrapping a
  Five-Loop Amplitude Using Steinmann Relations}},
  \href{https://doi.org/10.1103/PhysRevLett.117.241601}{\emph{Phys. Rev. Lett.}
  {\bfseries 117} (2016) 241601}
  [\href{https://arxiv.org/abs/1609.00669}{{\ttfamily 1609.00669}}].

\bibitem{Dixon:2016nkn}
L.~J. Dixon, J.~Drummond, T.~Harrington, A.~J. McLeod, G.~Papathanasiou and
  M.~Spradlin, \emph{{Heptagons from the Steinmann Cluster Bootstrap}},
  \href{https://doi.org/10.1007/JHEP02(2017)137}{\emph{JHEP} {\bfseries 02}
  (2017) 137} [\href{https://arxiv.org/abs/1612.08976}{{\ttfamily
  1612.08976}}].

\bibitem{Drummond:2018caf}
J.~Drummond, J.~Foster, {\"O}.~G{\"u}rdo{\u{g}}an and G.~Papathanasiou,
  \emph{{Cluster adjacency and the four-loop NMHV heptagon}},
  \href{https://arxiv.org/abs/1812.04640}{{\ttfamily 1812.04640}}.

\bibitem{Bern:1998sv}
Z.~Bern, L.~J. Dixon, M.~Perelstein and J.~S. Rozowsky, \emph{{Multileg one
  loop gravity amplitudes from gauge theory}},
  \href{https://doi.org/10.1016/S0550-3213(99)00029-2}{\emph{Nucl. Phys.}
  {\bfseries B546} (1999) 423}
  [\href{https://arxiv.org/abs/hep-th/9811140}{{\ttfamily hep-th/9811140}}].

\bibitem{Bern:1998ug}
Z.~Bern, L.~J. Dixon, D.~C. Dunbar, M.~Perelstein and J.~S. Rozowsky, \emph{{On
  the relationship between Yang-Mills theory and gravity and its implication
  for ultraviolet divergences}},
  \href{https://doi.org/10.1016/S0550-3213(98)00420-9}{\emph{Nucl. Phys.}
  {\bfseries B530} (1998) 401}
  [\href{https://arxiv.org/abs/hep-th/9802162}{{\ttfamily hep-th/9802162}}].

\bibitem{SchnitzerN8UniformTrans}
S.~G. Naculich, H.~Nastase and H.~J. Schnitzer, \emph{{Two-loop graviton
  scattering relation and IR behavior in N=8 supergravity}},
  \href{https://doi.org/10.1016/j.nuclphysb.2008.07.001}{\emph{Nucl.Phys.}
  {\bfseries B805} (2008) 40}
  [\href{https://arxiv.org/abs/0805.2347}{{\ttfamily 0805.2347}}].

\bibitem{QueenMaryN8UniformTrans}
A.~Brandhuber, P.~Heslop, A.~Nasti, B.~Spence and G.~Travaglini,
  \emph{{Four-point Amplitudes in N=8 Supergravity and Wilson Loops}},
  \href{https://doi.org/10.1016/j.nuclphysb.2008.09.010}{\emph{Nucl.Phys.}
  {\bfseries B807} (2009) 290}
  [\href{https://arxiv.org/abs/0805.2763}{{\ttfamily 0805.2763}}].

\bibitem{BoucherVeronneau:2011qv}
C.~Boucher-Veronneau and L.~J. Dixon, \emph{{$ \mathcal{N}\geq 4$ Supergravity
  Amplitudes from Gauge Theory at Two Loops}},
  \href{https://doi.org/10.1007/JHEP12(2011)046}{\emph{JHEP} {\bfseries 12}
  (2011) 046} [\href{https://arxiv.org/abs/1110.1132}{{\ttfamily 1110.1132}}].

\bibitem{Bern:2015xsa}
Z.~Bern, C.~Cheung, H.-H. Chi, S.~Davies, L.~Dixon and J.~Nohle,
  \emph{{Evanescent Effects Can Alter Ultraviolet Divergences in Quantum
  Gravity without Physical Consequences}},
  \href{https://doi.org/10.1103/PhysRevLett.115.211301}{\emph{Phys. Rev. Lett.}
  {\bfseries 115} (2015) 211301}
  [\href{https://arxiv.org/abs/1507.06118}{{\ttfamily 1507.06118}}].

\bibitem{Bern:2017puu}
Z.~Bern, H.-H. Chi, L.~Dixon and A.~Edison, \emph{{Two-Loop Renormalization of
  Quantum Gravity Simplified}},
  \href{https://doi.org/10.1103/PhysRevD.95.046013}{\emph{Phys. Rev.}
  {\bfseries D95} (2017) 046013}
  [\href{https://arxiv.org/abs/1701.02422}{{\ttfamily 1701.02422}}].

\bibitem{Dunbar:2017qxb}
D.~C. Dunbar, G.~R. Jehu and W.~B. Perkins, \emph{{Two-Loop Gravity amplitudes
  from four dimensional Unitarity}},
  \href{https://doi.org/10.1103/PhysRevD.95.046012}{\emph{Phys. Rev.}
  {\bfseries D95} (2017) 046012}
  [\href{https://arxiv.org/abs/1701.02934}{{\ttfamily 1701.02934}}].

\bibitem{IBP1}
F.~Tkachov, \emph{{A Theorem on Analytical Calculability of Four Loop
  Renormalization Group Functions}},
  \href{https://doi.org/10.1016/0370-2693(81)90288-4}{\emph{Phys.Lett.}
  {\bfseries B100} (1981) 65}.

\bibitem{IBP2}
K.~Chetyrkin and F.~Tkachov, \emph{{Integration by Parts: The Algorithm to
  Calculate Beta Functions in 4 Loops}},
  \href{https://doi.org/10.1016/0550-3213(81)90199-1}{\emph{Nucl.Phys.}
  {\bfseries B192} (1981) 159}.

\bibitem{Gluza:2010ws}
J.~Gluza, K.~Kajda and D.~A. Kosower, \emph{{Towards a Basis for Planar
  Two-Loop Integrals}},
  \href{https://doi.org/10.1103/PhysRevD.83.045012}{\emph{Phys. Rev.}
  {\bfseries D83} (2011) 045012}
  [\href{https://arxiv.org/abs/1009.0472}{{\ttfamily 1009.0472}}].

\bibitem{Ita:2015tya}
H.~Ita, \emph{{Two-loop Integrand Decomposition into Master Integrals and
  Surface Terms}},
  \href{https://doi.org/10.1103/PhysRevD.94.116015}{\emph{Phys. Rev.}
  {\bfseries D94} (2016) 116015}
  [\href{https://arxiv.org/abs/1510.05626}{{\ttfamily 1510.05626}}].

\bibitem{Larsen:2015ped}
K.~J. Larsen and Y.~Zhang, \emph{{Integration-by-parts reductions from
  unitarity cuts and algebraic geometry}},
  \href{https://doi.org/10.1103/PhysRevD.93.041701}{\emph{Phys. Rev.}
  {\bfseries D93} (2016) 041701}
  [\href{https://arxiv.org/abs/1511.01071}{{\ttfamily 1511.01071}}].

\bibitem{Boehm:2018fpv}
J.~B{\"o}hm, A.~Georgoudis, K.~J. Larsen, H.~Sch{\"o}nemann and Y.~Zhang,
  \emph{{Complete integration-by-parts reductions of the non-planar hexagon-box
  via module intersections}},
  \href{https://doi.org/10.1007/JHEP09(2018)024}{\emph{JHEP} {\bfseries 09}
  (2018) 024} [\href{https://arxiv.org/abs/1805.01873}{{\ttfamily
  1805.01873}}].

\bibitem{Abreu:2017hqn}
S.~Abreu, F.~Febres~Cordero, H.~Ita, B.~Page and M.~Zeng, \emph{{Planar
  Two-Loop Five-Gluon Amplitudes from Numerical Unitarity}},
  \href{https://doi.org/10.1103/PhysRevD.97.116014}{\emph{Phys. Rev.}
  {\bfseries D97} (2018) 116014}
  [\href{https://arxiv.org/abs/1712.03946}{{\ttfamily 1712.03946}}].

\bibitem{Kosower:2018obg}
D.~A. Kosower, \emph{{Direct Solution of Integration-by-Parts Systems}},
  \href{https://doi.org/10.1103/PhysRevD.98.025008}{\emph{Phys. Rev.}
  {\bfseries D98} (2018) 025008}
  [\href{https://arxiv.org/abs/1804.00131}{{\ttfamily 1804.00131}}].

\bibitem{vonManteuffel:2014ixa}
A.~von Manteuffel and R.~M. Schabinger, \emph{{A novel approach to integration
  by parts reduction}},
  \href{https://doi.org/10.1016/j.physletb.2015.03.029}{\emph{Phys. Lett.}
  {\bfseries B744} (2015) 101}
  [\href{https://arxiv.org/abs/1406.4513}{{\ttfamily 1406.4513}}].

\bibitem{Peraro:2016wsq}
T.~Peraro, \emph{{Scattering amplitudes over finite fields and multivariate
  functional reconstruction}},
  \href{https://doi.org/10.1007/JHEP12(2016)030}{\emph{JHEP} {\bfseries 12}
  (2016) 030} [\href{https://arxiv.org/abs/1608.01902}{{\ttfamily
  1608.01902}}].

\bibitem{Maierhoefer:2017hyi}
P.~Maierh{\"o}fer, J.~Usovitsch and P.~Uwer, \emph{{Kira --- A Feynman integral
  reduction program}},
  \href{https://doi.org/10.1016/j.cpc.2018.04.012}{\emph{Comput. Phys. Commun.}
  {\bfseries 230} (2018) 99}
  [\href{https://arxiv.org/abs/1705.05610}{{\ttfamily 1705.05610}}].

\bibitem{Smirnov:2019qkx}
A.~V. Smirnov and F.~S. Chukharev, \emph{{FIRE6: Feynman Integral REduction
  with Modular Arithmetic}},
  \href{https://arxiv.org/abs/1901.07808}{{\ttfamily 1901.07808}}.

\bibitem{Kotikov:1990kg}
A.~V. Kotikov, \emph{{Differential equations method: New technique for massive
  Feynman diagrams calculation}},
  \href{https://doi.org/10.1016/0370-2693(91)90413-K}{\emph{Phys. Lett.}
  {\bfseries B254} (1991) 158}.

\bibitem{Bern:1992em}
Z.~Bern, L.~J. Dixon and D.~A. Kosower, \emph{{Dimensionally regulated one loop
  integrals}}, \href{https://doi.org/10.1016/0370-2693(93)90469-X,
  10.1016/0370-2693(93)90400-C}{\emph{Phys. Lett.} {\bfseries B302} (1993) 299}
  [\href{https://arxiv.org/abs/hep-ph/9212308}{{\ttfamily hep-ph/9212308}}].

\bibitem{Gehrmann:1999as}
T.~Gehrmann and E.~Remiddi, \emph{{Differential equations for two loop four
  point functions}},
  \href{https://doi.org/10.1016/S0550-3213(00)00223-6}{\emph{Nucl. Phys.}
  {\bfseries B580} (2000) 485}
  [\href{https://arxiv.org/abs/hep-ph/9912329}{{\ttfamily hep-ph/9912329}}].

\bibitem{Henn:2013pwa}
J.~M. Henn, \emph{{Multiloop integrals in dimensional regularization made
  simple}}, \href{https://doi.org/10.1103/PhysRevLett.110.251601}{\emph{Phys.
  Rev. Lett.} {\bfseries 110} (2013) 251601}
  [\href{https://arxiv.org/abs/1304.1806}{{\ttfamily 1304.1806}}].

\bibitem{Gehrmann:2000zt}
T.~Gehrmann and E.~Remiddi, \emph{{Two loop master integrals for $\gamma^* \,
  \to$ 3 jets: The Planar topologies}},
  \href{https://doi.org/10.1016/S0550-3213(01)00057-8}{\emph{Nucl. Phys.}
  {\bfseries B601} (2001) 248}
  [\href{https://arxiv.org/abs/hep-ph/0008287}{{\ttfamily hep-ph/0008287}}].

\bibitem{Gehrmann:2015bfy}
T.~Gehrmann, J.~M. Henn and N.~A. Lo~Presti, \emph{{Analytic form of the
  two-loop planar five-gluon all-plus-helicity amplitude in QCD}},
  \href{https://doi.org/10.1103/PhysRevLett.116.189903,
  10.1103/PhysRevLett.116.062001}{\emph{Phys. Rev. Lett.} {\bfseries 116}
  (2016) 062001} [\href{https://arxiv.org/abs/1511.05409}{{\ttfamily
  1511.05409}}].

\bibitem{Papadopoulos:2015jft}
C.~G. Papadopoulos, D.~Tommasini and C.~Wever, \emph{{The Pentabox Master
  Integrals with the Simplified Differential Equations approach}},
  \href{https://doi.org/10.1007/JHEP04(2016)078}{\emph{JHEP} {\bfseries 04}
  (2016) 078} [\href{https://arxiv.org/abs/1511.09404}{{\ttfamily
  1511.09404}}].

\bibitem{Gehrmann:2018yef}
T.~Gehrmann, J.~M. Henn and N.~A. Lo~Presti, \emph{{Pentagon functions for
  massless planar scattering amplitudes}},
  \href{https://doi.org/10.1007/JHEP10(2018)103}{\emph{JHEP} {\bfseries 10}
  (2018) 103} [\href{https://arxiv.org/abs/1807.09812}{{\ttfamily
  1807.09812}}].

\bibitem{Gehrmann:2001ck}
T.~Gehrmann and E.~Remiddi, \emph{{Two loop master integrals for
  $\gamma^*\,\to$ 3 jets: The Nonplanar topologies}},
  \href{https://doi.org/10.1016/S0550-3213(01)00074-8}{\emph{Nucl. Phys.}
  {\bfseries B601} (2001) 287}
  [\href{https://arxiv.org/abs/hep-ph/0101124}{{\ttfamily hep-ph/0101124}}].

\bibitem{Chicherin:2017dob}
D.~Chicherin, J.~Henn and V.~Mitev, \emph{{Bootstrapping pentagon functions}},
  \href{https://doi.org/10.1007/JHEP05(2018)164}{\emph{JHEP} {\bfseries 05}
  (2018) 164} [\href{https://arxiv.org/abs/1712.09610}{{\ttfamily
  1712.09610}}].

\bibitem{Abreu:2018rcw}
S.~Abreu, B.~Page and M.~Zeng, \emph{{Differential equations from unitarity
  cuts: nonplanar hexa-box integrals}},
  \href{https://doi.org/10.1007/JHEP01(2019)006}{\emph{JHEP} {\bfseries 01}
  (2019) 006} [\href{https://arxiv.org/abs/1807.11522}{{\ttfamily
  1807.11522}}].

\bibitem{Abreu:2018aqd}
S.~Abreu, L.~J. Dixon, E.~Herrmann, B.~Page and M.~Zeng, \emph{{The two-loop
  five-point amplitude in $\mathcal{N} =4$ super-Yang-Mills theory}},
  \href{https://arxiv.org/abs/1812.08941}{{\ttfamily 1812.08941}}.

\bibitem{Chicherin:2018mue}
D.~Chicherin, T.~Gehrmann, J.~M. Henn, N.~A. Lo~Presti, V.~Mitev and P.~Wasser,
  \emph{{Analytic result for the nonplanar hexa-box integrals}},
  \href{https://doi.org/10.1007/JHEP03(2019)042}{\emph{JHEP} {\bfseries 03}
  (2019) 042} [\href{https://arxiv.org/abs/1809.06240}{{\ttfamily
  1809.06240}}].

\bibitem{Chicherin:2018old}
D.~Chicherin, T.~Gehrmann, J.~M. Henn, P.~Wasser, Y.~Zhang and S.~Zoia,
  \emph{{All master integrals for three-jet production at NNLO}},
  \href{https://arxiv.org/abs/1812.11160}{{\ttfamily 1812.11160}}.

\bibitem{Badger:2017jhb}
S.~Badger, C.~Br{\o}nnum-Hansen, H.~B. Hartanto and T.~Peraro, \emph{{First
  look at two-loop five-gluon scattering in QCD}},
  \href{https://doi.org/10.1103/PhysRevLett.120.092001}{\emph{Phys. Rev. Lett.}
  {\bfseries 120} (2018) 092001}
  [\href{https://arxiv.org/abs/1712.02229}{{\ttfamily 1712.02229}}].

\bibitem{Badger:2018gip}
S.~Badger, C.~Br{\o}nnum-Hansen, T.~Gehrmann, H.~B. Hartanto, J.~Henn, N.~A.
  Lo~Presti et~al., \emph{{Applications of integrand reduction to two-loop
  five-point scattering amplitudes in QCD}},
  \href{https://doi.org/10.22323/1.303.0006}{\emph{PoS} {\bfseries LL2018}
  (2018) 006} [\href{https://arxiv.org/abs/1807.09709}{{\ttfamily
  1807.09709}}].

\bibitem{Abreu:2018jgq}
S.~Abreu, F.~Febres~Cordero, H.~Ita, B.~Page and V.~Sotnikov, \emph{{Planar
  Two-Loop Five-Parton Amplitudes from Numerical Unitarity}},
  \href{https://doi.org/10.1007/JHEP11(2018)116}{\emph{JHEP} {\bfseries 11}
  (2018) 116} [\href{https://arxiv.org/abs/1809.09067}{{\ttfamily
  1809.09067}}].

\bibitem{Badger:2018enw}
S.~Badger, C.~Br{\o}nnum-Hansen, H.~B. Hartanto and T.~Peraro, \emph{{Analytic
  helicity amplitudes for two-loop five-gluon scattering: the single-minus
  case}}, \href{https://doi.org/10.1007/JHEP01(2019)186}{\emph{JHEP} {\bfseries
  01} (2019) 186} [\href{https://arxiv.org/abs/1811.11699}{{\ttfamily
  1811.11699}}].

\bibitem{Abreu:2018zmy}
S.~Abreu, J.~Dormans, F.~Febres~Cordero, H.~Ita and B.~Page, \emph{{Analytic
  Form of Planar Two-Loop Five-Gluon Scattering Amplitudes in QCD}},
  \href{https://doi.org/10.1103/PhysRevLett.122.082002}{\emph{Phys. Rev. Lett.}
  {\bfseries 122} (2019) 082002}
  [\href{https://arxiv.org/abs/1812.04586}{{\ttfamily 1812.04586}}].

\bibitem{Chicherin:2018yne}
D.~Chicherin, J.~M. Henn, P.~Wasser, T.~Gehrmann, Y.~Zhang and S.~Zoia,
  \emph{{Analytic result for a two-loop five-particle amplitude}},
  \href{https://arxiv.org/abs/1812.11057}{{\ttfamily 1812.11057}}.

\bibitem{Cachazo:2008vp}
F.~Cachazo, \emph{{Sharpening The Leading Singularity}},
  \href{https://arxiv.org/abs/0803.1988}{{\ttfamily 0803.1988}}.

\bibitem{Arkani-Hamed:2014bca}
N.~Arkani-Hamed, J.~L. Bourjaily, F.~Cachazo, A.~Postnikov and J.~Trnka,
  \emph{{On-Shell Structures of MHV Amplitudes Beyond the Planar Limit}},
  \href{https://doi.org/10.1007/JHEP06(2015)179}{\emph{JHEP} {\bfseries 06}
  (2015) 179} [\href{https://arxiv.org/abs/1412.8475}{{\ttfamily 1412.8475}}].

\bibitem{Herrmann:2016qea}
E.~Herrmann and J.~Trnka, \emph{{Gravity On-shell Diagrams}},
  \href{https://doi.org/10.1007/JHEP11(2016)136}{\emph{JHEP} {\bfseries 11}
  (2016) 136} [\href{https://arxiv.org/abs/1604.03479}{{\ttfamily
  1604.03479}}].

\bibitem{Heslop:2016plj}
P.~Heslop and A.~E. Lipstein, \emph{{On-shell diagrams for $\mathcal{N}=8$
  supergravity amplitudes}},
  \href{https://doi.org/10.1007/JHEP06(2016)069}{\emph{JHEP} {\bfseries 06}
  (2016) 069} [\href{https://arxiv.org/abs/1604.03046}{{\ttfamily
  1604.03046}}].

\bibitem{Herrmann:2018dja}
E.~Herrmann and J.~Trnka, \emph{{UV cancellations in gravity loop integrands}},
  \href{https://doi.org/10.1007/JHEP02(2019)084}{\emph{JHEP} {\bfseries 02}
  (2019) 084} [\href{https://arxiv.org/abs/1808.10446}{{\ttfamily
  1808.10446}}].

\bibitem{Bourjaily:2018omh}
J.~L. Bourjaily, E.~Herrmann and J.~Trnka, \emph{{Amplitudes at Infinity}},
  \href{https://arxiv.org/abs/1812.11185}{{\ttfamily 1812.11185}}.

\bibitem{Parke:1986gb}
S.~J. Parke and T.~R. Taylor, \emph{{An Amplitude for $n$ Gluon Scattering}},
  \href{https://doi.org/10.1103/PhysRevLett.56.2459}{\emph{Phys. Rev. Lett.}
  {\bfseries 56} (1986) 2459}.

\bibitem{BDS}
Z.~Bern, L.~J. Dixon and V.~A. Smirnov, \emph{{Iteration of planar amplitudes
  in maximally supersymmetric Yang-Mills theory at three loops and beyond}},
  \href{https://doi.org/10.1103/PhysRevD.72.085001}{\emph{Phys.Rev.} {\bfseries
  D72} (2005) 085001} [\href{https://arxiv.org/abs/hep-th/0505205}{{\ttfamily
  hep-th/0505205}}].

\bibitem{LipatovTranscendentality}
A.~Kotikov and L.~Lipatov, \emph{{On the highest transcendentality in N=4
  SUSY}},
  \href{https://doi.org/10.1016/j.nuclphysb.2007.01.020}{\emph{Nucl.Phys.}
  {\bfseries B769} (2007) 217}
  [\href{https://arxiv.org/abs/hep-th/0611204}{{\ttfamily hep-th/0611204}}].

\bibitem{Chicherin:2019xeg}
D.~Chicherin, T.~Gehrmann, J.~M. Henn, P.~Wasser, Y.~Zhang and S.~Zoia,
  \emph{{The two-loop five-particle amplitude in $\mathcal{N}=8$
  supergravity}},  \href{https://arxiv.org/abs/1901.05932}{{\ttfamily
  1901.05932}}.

\bibitem{Berends:1988zp}
F.~A. Berends, W.~T. Giele and H.~Kuijf, \emph{{On relations between multi -
  gluon and multigraviton scattering}},
  \href{https://doi.org/10.1016/0370-2693(88)90813-1}{\emph{Phys. Lett.}
  {\bfseries B211} (1988) 91}.

\bibitem{Bern:2014kca}
Z.~Bern, E.~Herrmann, S.~Litsey, J.~Stankowicz and J.~Trnka, \emph{{Logarithmic
  Singularities and Maximally Supersymmetric Amplitudes}},
  \href{https://doi.org/10.1007/JHEP06(2015)202}{\emph{JHEP} {\bfseries 06}
  (2015) 202} [\href{https://arxiv.org/abs/1412.8584}{{\ttfamily 1412.8584}}].

\bibitem{Log}
N.~Arkani-Hamed, J.~L. Bourjaily, F.~Cachazo and J.~Trnka, \emph{{Singularity
  Structure of Maximally Supersymmetric Scattering Amplitudes}},
  \href{https://doi.org/10.1103/PhysRevLett.113.261603}{\emph{Phys. Rev. Lett.}
  {\bfseries 113} (2014) 261603}
  [\href{https://arxiv.org/abs/1410.0354}{{\ttfamily 1410.0354}}].

\bibitem{N5FourLoop}
Z.~Bern, S.~Davies and T.~Dennen, \emph{{Enhanced ultraviolet cancellations in
  $\mathcal N=5$ supergravity at four loops}},
  \href{https://doi.org/10.1103/PhysRevD.90.105011}{\emph{Phys. Rev.}
  {\bfseries D90} (2014) 105011}
  [\href{https://arxiv.org/abs/1409.3089}{{\ttfamily 1409.3089}}].

\bibitem{Weinberg:1965nx}
S.~Weinberg, \emph{{Infrared photons and gravitons}},
  \href{https://doi.org/10.1103/PhysRev.140.B516}{\emph{Phys. Rev.} {\bfseries
  140} (1965) B516}.

\bibitem{Akhoury:2011kq}
R.~Akhoury, R.~Saotome and G.~Sterman, \emph{{Collinear and Soft Divergences in
  Perturbative Quantum Gravity}},
  \href{https://doi.org/10.1103/PhysRevD.84.104040}{\emph{Phys. Rev.}
  {\bfseries D84} (2011) 104040}
  [\href{https://arxiv.org/abs/1109.0270}{{\ttfamily 1109.0270}}].

\bibitem{Hodges}
A.~Hodges, \emph{{Eliminating spurious poles from gauge-theoretic amplitudes}},
  \href{https://doi.org/10.1007/JHEP05(2013)135}{\emph{JHEP} {\bfseries 1305}
  (2013) 135} [\href{https://arxiv.org/abs/0905.1473}{{\ttfamily 0905.1473}}].

\bibitem{Badger:2013gxa}
S.~Badger, H.~Frellesvig and Y.~Zhang, \emph{{A Two-Loop Five-Gluon Helicity
  Amplitude in QCD}},
  \href{https://doi.org/10.1007/JHEP12(2013)045}{\emph{JHEP} {\bfseries 12}
  (2013) 045} [\href{https://arxiv.org/abs/1310.1051}{{\ttfamily 1310.1051}}].

\bibitem{Drummond:2013nda}
J.~Drummond, C.~Duhr, B.~Eden, P.~Heslop, J.~Pennington and V.~A. Smirnov,
  \emph{{Leading singularities and off-shell conformal integrals}},
  \href{https://doi.org/10.1007/JHEP08(2013)133}{\emph{JHEP} {\bfseries 08}
  (2013) 133} [\href{https://arxiv.org/abs/1303.6909}{{\ttfamily 1303.6909}}].

\bibitem{Abreu:2017ptx}
S.~Abreu, R.~Britto, C.~Duhr and E.~Gardi, \emph{{Cuts from residues: the
  one-loop case}}, \href{https://doi.org/10.1007/JHEP06(2017)114}{\emph{JHEP}
  {\bfseries 06} (2017) 114}
  [\href{https://arxiv.org/abs/1702.03163}{{\ttfamily 1702.03163}}].

\bibitem{Baikov:1996rk}
P.~A. Baikov, \emph{{Explicit solutions of the three loop vacuum integral
  recurrence relations}},
  \href{https://doi.org/10.1016/0370-2693(96)00835-0}{\emph{Phys. Lett.}
  {\bfseries B385} (1996) 404}
  [\href{https://arxiv.org/abs/hep-ph/9603267}{{\ttfamily hep-ph/9603267}}].

\bibitem{Baikov:1996iu}
P.~A. Baikov, \emph{{Explicit solutions of the multiloop integral recurrence
  relations and its application}},
  \href{https://doi.org/10.1016/S0168-9002(97)00126-5}{\emph{Nucl. Instrum.
  Meth.} {\bfseries A389} (1997) 347}
  [\href{https://arxiv.org/abs/hep-ph/9611449}{{\ttfamily hep-ph/9611449}}].

\bibitem{Grozin:2011mt}
A.~G. Grozin, \emph{{Integration by parts: An Introduction}},
  \href{https://doi.org/10.1142/S0217751X11053687}{\emph{Int. J. Mod. Phys.}
  {\bfseries A26} (2011) 2807}
  [\href{https://arxiv.org/abs/1104.3993}{{\ttfamily 1104.3993}}].

\bibitem{Frellesvig:2017aai}
H.~Frellesvig and C.~G. Papadopoulos, \emph{{Cuts of Feynman Integrals in
  Baikov representation}},
  \href{https://doi.org/10.1007/JHEP04(2017)083}{\emph{JHEP} {\bfseries 04}
  (2017) 083} [\href{https://arxiv.org/abs/1701.07356}{{\ttfamily
  1701.07356}}].

\bibitem{Kosower:2011ty}
D.~A. Kosower and K.~J. Larsen, \emph{{Maximal Unitarity at Two Loops}},
  \href{https://doi.org/10.1103/PhysRevD.85.045017}{\emph{Phys. Rev.}
  {\bfseries D85} (2012) 045017}
  [\href{https://arxiv.org/abs/1108.1180}{{\ttfamily 1108.1180}}].

\bibitem{Bern:2015ple}
Z.~Bern, E.~Herrmann, S.~Litsey, J.~Stankowicz and J.~Trnka, \emph{{Evidence
  for a Nonplanar Amplituhedron}},
  \href{https://doi.org/10.1007/JHEP06(2016)098}{\emph{JHEP} {\bfseries 06}
  (2016) 098} [\href{https://arxiv.org/abs/1512.08591}{{\ttfamily
  1512.08591}}].

\bibitem{Bern:2018oao}
Z.~Bern, M.~Enciso, C.-H. Shen and M.~Zeng, \emph{{Dual Conformal Structure
  Beyond the Planar Limit}},
  \href{https://doi.org/10.1103/PhysRevLett.121.121603}{\emph{Phys. Rev. Lett.}
  {\bfseries 121} (2018) 121603}
  [\href{https://arxiv.org/abs/1806.06509}{{\ttfamily 1806.06509}}].

\bibitem{Bern:2017gdk}
Z.~Bern, M.~Enciso, H.~Ita and M.~Zeng, \emph{{Dual Conformal Symmetry,
  Integration-by-Parts Reduction, Differential Equations and the Nonplanar
  Sector}}, \href{https://doi.org/10.1103/PhysRevD.96.096017}{\emph{Phys. Rev.}
  {\bfseries D96} (2017) 096017}
  [\href{https://arxiv.org/abs/1709.06055}{{\ttfamily 1709.06055}}].

\bibitem{Chicherin:2018wes}
D.~Chicherin, J.~M. Henn and E.~Sokatchev, \emph{{Implications of nonplanar
  dual conformal symmetry}},
  \href{https://doi.org/10.1007/JHEP09(2018)012}{\emph{JHEP} {\bfseries 09}
  (2018) 012} [\href{https://arxiv.org/abs/1807.06321}{{\ttfamily
  1807.06321}}].

\bibitem{Bern:1993kr}
Z.~Bern, L.~J. Dixon and D.~A. Kosower, \emph{{Dimensionally regulated pentagon
  integrals}},
  \href{https://doi.org/10.1016/0550-3213(94)90398-0}{\emph{Nucl.Phys.}
  {\bfseries B412} (1994) 751}
  [\href{https://arxiv.org/abs/hep-ph/9306240}{{\ttfamily hep-ph/9306240}}].

\bibitem{Tarasov:1996br}
O.~V. Tarasov, \emph{{Connection between Feynman integrals having different
  values of the space-time dimension}},
  \href{https://doi.org/10.1103/PhysRevD.54.6479}{\emph{Phys. Rev.} {\bfseries
  D54} (1996) 6479} [\href{https://arxiv.org/abs/hep-th/9606018}{{\ttfamily
  hep-th/9606018}}].

\bibitem{Lee:2009dh}
R.~N. Lee, \emph{{Space-time dimensionality D as complex variable: Calculating
  loop integrals using dimensional recurrence relation and analytical
  properties with respect to D}},
  \href{https://doi.org/10.1016/j.nuclphysb.2009.12.025}{\emph{Nucl. Phys.}
  {\bfseries B830} (2010) 474}
  [\href{https://arxiv.org/abs/0911.0252}{{\ttfamily 0911.0252}}].

\bibitem{Georgoudis:2016wff}
A.~Georgoudis, K.~J. Larsen and Y.~Zhang, \emph{{Azurite: An algebraic geometry
  based package for finding bases of loop integrals}},
  \href{https://doi.org/10.1016/j.cpc.2017.08.013}{\emph{Comput. Phys. Commun.}
  {\bfseries 221} (2017) 203}
  [\href{https://arxiv.org/abs/1612.04252}{{\ttfamily 1612.04252}}].

\bibitem{Gaiotto:2011dt}
D.~Gaiotto, J.~Maldacena, A.~Sever and P.~Vieira, \emph{{Pulling the straps of
  polygons}}, \href{https://doi.org/10.1007/JHEP12(2011)011}{\emph{JHEP}
  {\bfseries 12} (2011) 011} [\href{https://arxiv.org/abs/1102.0062}{{\ttfamily
  1102.0062}}].

\bibitem{Schabinger:2011dz}
R.~M. Schabinger, \emph{{A New Algorithm For The Generation Of
  Unitarity-Compatible Integration By Parts Relations}},
  \href{https://doi.org/10.1007/JHEP01(2012)077}{\emph{JHEP} {\bfseries 01}
  (2012) 077} [\href{https://arxiv.org/abs/1111.4220}{{\ttfamily 1111.4220}}].

\bibitem{Zhang:2016kfo}
Y.~Zhang, \emph{{Lecture Notes on Multi-loop Integral Reduction and Applied
  Algebraic Geometry}},  2016,
  \href{https://arxiv.org/abs/1612.02249}{{\ttfamily 1612.02249}}.

\bibitem{GravityThreeLoop}
Z.~Bern, J.~Carrasco, L.~J. Dixon, H.~Johansson, D.~Kosower and R.~Roiban,
  \emph{{Three-Loop Superfiniteness of N=8 Supergravity}},
  \href{https://doi.org/10.1103/PhysRevLett.98.161303}{\emph{Phys.Rev.Lett.}
  {\bfseries 98} (2007) 161303}
  [\href{https://arxiv.org/abs/hep-th/0702112}{{\ttfamily hep-th/0702112}}].

\bibitem{Manifest3}
Z.~Bern, J.~Carrasco, L.~J. Dixon, H.~Johansson and R.~Roiban, \emph{{Manifest
  Ultraviolet Behavior for the Three-Loop Four-Point Amplitude of N=8
  Supergravity}},
  \href{https://doi.org/10.1103/PhysRevD.78.105019}{\emph{Phys.Rev.} {\bfseries
  D78} (2008) 105019} [\href{https://arxiv.org/abs/0808.4112}{{\ttfamily
  0808.4112}}].

\bibitem{N8FourLoop}
Z.~Bern, J.~Carrasco, L.~J. Dixon, H.~Johansson and R.~Roiban, \emph{{The
  Ultraviolet Behavior of N=8 Supergravity at Four Loops}},
  \href{https://doi.org/10.1103/PhysRevLett.103.081301}{\emph{Phys.Rev.Lett.}
  {\bfseries 103} (2009) 081301}
  [\href{https://arxiv.org/abs/0905.2326}{{\ttfamily 0905.2326}}].

\bibitem{GreenDuality}
M.~B. Green, J.~G. Russo and P.~Vanhove, \emph{{String theory dualities and
  supergravity divergences}},
  \href{https://doi.org/10.1007/JHEP06(2010)075}{\emph{JHEP} {\bfseries 1006}
  (2010) 075} [\href{https://arxiv.org/abs/1002.3805}{{\ttfamily 1002.3805}}].

\bibitem{BossardHoweStellDuality}
G.~Bossard, P.~Howe and K.~Stelle, \emph{{On duality symmetries of supergravity
  invariants}}, \href{https://doi.org/10.1007/JHEP01(2011)020}{\emph{JHEP}
  {\bfseries 1101} (2011) 020}
  [\href{https://arxiv.org/abs/1009.0743}{{\ttfamily 1009.0743}}].

\bibitem{BeisertN8}
N.~Beisert, H.~Elvang, D.~Z. Freedman, M.~Kiermaier, A.~Morales and
  S.~Stieberger, \emph{{$E_{7(7)}$ constraints on counterterms in $N=8$
  supergravity}},
  \href{https://doi.org/10.1016/j.physletb.2010.09.069}{\emph{Phys.Lett.}
  {\bfseries B694} (2010) 265}
  [\href{https://arxiv.org/abs/1009.1643}{{\ttfamily 1009.1643}}].

\bibitem{Vanhove:2010nf}
P.~Vanhove, \emph{{The Critical ultraviolet behaviour of N=8 supergravity
  amplitudes}},  \href{https://arxiv.org/abs/1004.1392}{{\ttfamily 1004.1392}}.

\bibitem{Bjornsson:2010wm}
J.~Bj{\"o}rnsson and M.~B. Green, \emph{{5 loops in 24/5 dimensions}},
  \href{https://doi.org/10.1007/JHEP08(2010)132}{\emph{JHEP} {\bfseries 08}
  (2010) 132} [\href{https://arxiv.org/abs/1004.2692}{{\ttfamily 1004.2692}}].

\bibitem{Bjornsson:2010wu}
J.~Bj{\"o}rnsson, \emph{{Multi-loop amplitudes in maximally supersymmetric pure
  spinor field theory}},
  \href{https://doi.org/10.1007/JHEP01(2011)002}{\emph{JHEP} {\bfseries 01}
  (2011) 002} [\href{https://arxiv.org/abs/1009.5906}{{\ttfamily 1009.5906}}].

\bibitem{VanishingVolume}
G.~Bossard, P.~Howe, K.~Stelle and P.~Vanhove, \emph{{The vanishing volume of
  D=4 superspace}},
  \href{https://doi.org/10.1088/0264-9381/28/21/215005}{\emph{Class.Quant.Grav.}
  {\bfseries 28} (2011) 215005}
  [\href{https://arxiv.org/abs/1105.6087}{{\ttfamily 1105.6087}}].

\bibitem{Beneke:2012xa}
M.~Beneke and G.~Kirilin, \emph{{Soft-collinear gravity}},
  \href{https://doi.org/10.1007/JHEP09(2012)066}{\emph{JHEP} {\bfseries 09}
  (2012) 066} [\href{https://arxiv.org/abs/1207.4926}{{\ttfamily 1207.4926}}].

\bibitem{Dunbar:1995ed}
D.~C. Dunbar and P.~S. Norridge, \emph{{Infinities within graviton scattering
  amplitudes}}, \href{https://doi.org/10.1088/0264-9381/14/2/009}{\emph{Class.
  Quant. Grav.} {\bfseries 14} (1997) 351}
  [\href{https://arxiv.org/abs/hep-th/9512084}{{\ttfamily hep-th/9512084}}].

\bibitem{Naculich:2011ry}
S.~G. Naculich and H.~J. Schnitzer, \emph{{Eikonal methods applied to
  gravitational scattering amplitudes}},
  \href{https://doi.org/10.1007/JHEP05(2011)087}{\emph{JHEP} {\bfseries 05}
  (2011) 087} [\href{https://arxiv.org/abs/1101.1524}{{\ttfamily 1101.1524}}].

\bibitem{White:2011yy}
C.~D. White, \emph{{Factorization Properties of Soft Graviton Amplitudes}},
  \href{https://doi.org/10.1007/JHEP05(2011)060}{\emph{JHEP} {\bfseries 05}
  (2011) 060} [\href{https://arxiv.org/abs/1103.2981}{{\ttfamily 1103.2981}}].

\bibitem{Abreu:2017mtm}
S.~Abreu, R.~Britto, C.~Duhr and E.~Gardi, \emph{{Diagrammatic Hopf algebra of
  cut Feynman integrals: the one-loop case}},
  \href{https://doi.org/10.1007/JHEP12(2017)090}{\emph{JHEP} {\bfseries 12}
  (2017) 090} [\href{https://arxiv.org/abs/1704.07931}{{\ttfamily
  1704.07931}}].

\bibitem{Abreu:2017enx}
S.~Abreu, R.~Britto, C.~Duhr and E.~Gardi, \emph{{Algebraic Structure of Cut
  Feynman Integrals and the Diagrammatic Coaction}},
  \href{https://doi.org/10.1103/PhysRevLett.119.051601}{\emph{Phys. Rev. Lett.}
  {\bfseries 119} (2017) 051601}
  [\href{https://arxiv.org/abs/1703.05064}{{\ttfamily 1703.05064}}].

\bibitem{Panzer:2014caa}
E.~Panzer, \emph{{Algorithms for the symbolic integration of hyperlogarithms
  with applications to Feynman integrals}},
  \href{https://doi.org/10.1016/j.cpc.2014.10.019}{\emph{Comput. Phys. Commun.}
  {\bfseries 188} (2015) 148}
  [\href{https://arxiv.org/abs/1403.3385}{{\ttfamily 1403.3385}}].

\bibitem{LanceToAppear}
L.~J. Dixon, E.~Herrmann, K.~Yan and H.~X. Zhu, \emph{{Soft emission function
  at two loops}}, {\emph{to appear} (2019) }.

\bibitem{Bern:1998xc}
Z.~Bern, L.~J. Dixon, M.~Perelstein and J.~S. Rozowsky, \emph{{One loop n point
  helicity amplitudes in (selfdual) gravity}},
  \href{https://doi.org/10.1016/S0370-2693(98)01397-5}{\emph{Phys. Lett.}
  {\bfseries B444} (1998) 273}
  [\href{https://arxiv.org/abs/hep-th/9809160}{{\ttfamily hep-th/9809160}}].

\bibitem{KLT}
H.~Kawai, D.~Lewellen and S.~H.~H. Tye, \emph{{A Relation Between Tree
  Amplitudes of Closed and Open Strings}},
  \href{https://doi.org/10.1016/0550-3213(86)90362-7}{\emph{Nucl.Phys.}
  {\bfseries B269} (1986) 1}.

\bibitem{BCFW}
R.~Britto, F.~Cachazo, B.~Feng and E.~Witten, \emph{{Direct proof of tree-level
  recursion relation in Yang-Mills theory}},
  \href{https://doi.org/10.1103/PhysRevLett.94.181602}{\emph{Phys.Rev.Lett.}
  {\bfseries 94} (2005) 181602}
  [\href{https://arxiv.org/abs/hep-th/0501052}{{\ttfamily hep-th/0501052}}].

\bibitem{Bern:2008ap}
Z.~Bern, L.~Dixon, D.~Kosower, R.~Roiban, M.~Spradlin et~al., \emph{{The
  Two-Loop Six-Gluon MHV Amplitude in Maximally Supersymmetric Yang-Mills
  Theory}}, \href{https://doi.org/10.1103/PhysRevD.78.045007}{\emph{Phys.Rev.}
  {\bfseries D78} (2008) 045007}
  [\href{https://arxiv.org/abs/0803.1465}{{\ttfamily 0803.1465}}].

\bibitem{Weinzierl:2011uz}
S.~Weinzierl, \emph{{Does one need the ${\cal O}(\epsilon)$- and ${\cal
  O}(\epsilon^2)$-terms of one-loop amplitudes in an NNLO calculation?}},
  \href{https://doi.org/10.1103/PhysRevD.84.074007}{\emph{Phys. Rev.}
  {\bfseries D84} (2011) 074007}
  [\href{https://arxiv.org/abs/1107.5131}{{\ttfamily 1107.5131}}].

\bibitem{Green:1982sw}
M.~B. Green, J.~H. Schwarz and L.~Brink, \emph{{N=4 Yang-Mills and N=8
  Supergravity as Limits of String Theories}},
  \href{https://doi.org/10.1016/0550-3213(82)90336-4}{\emph{Nucl. Phys.}
  {\bfseries B198} (1982) 474}.

\bibitem{Bern:1993mq}
Z.~Bern, L.~J. Dixon and D.~A. Kosower, \emph{{One loop corrections to five
  gluon amplitudes}},
  \href{https://doi.org/10.1103/PhysRevLett.70.2677}{\emph{Phys. Rev. Lett.}
  {\bfseries 70} (1993) 2677}
  [\href{https://arxiv.org/abs/hep-ph/9302280}{{\ttfamily hep-ph/9302280}}].

\bibitem{Bern:1990ux}
Z.~Bern and D.~A. Kosower, \emph{{Color decomposition of one loop amplitudes in
  gauge theories}},
  \href{https://doi.org/10.1016/0550-3213(91)90567-H}{\emph{Nucl. Phys.}
  {\bfseries B362} (1991) 389}.

\bibitem{Bern:2002tk}
Z.~Bern, A.~De~Freitas and L.~J. Dixon, \emph{{Two loop helicity amplitudes for
  gluon-gluon scattering in QCD and supersymmetric Yang-Mills theory}},
  \href{https://doi.org/10.1088/1126-6708/2002/03/018}{\emph{JHEP} {\bfseries
  03} (2002) 018} [\href{https://arxiv.org/abs/hep-ph/0201161}{{\ttfamily
  hep-ph/0201161}}].

\bibitem{KleissKuijf}
R.~Kleiss and H.~Kuijf, \emph{{Multi - Gluon Cross-sections and Five Jet
  Production at Hadron Colliders}},
  \href{https://doi.org/10.1016/0550-3213(89)90574-9}{\emph{Nucl. Phys.}
  {\bfseries B312} (1989) 616}.

\bibitem{BRY}
Z.~Bern, J.~Rozowsky and B.~Yan, \emph{{Two loop four gluon amplitudes in N=4
  superYang-Mills}},
  \href{https://doi.org/10.1016/S0370-2693(97)00413-9}{\emph{Phys.Lett.}
  {\bfseries B401} (1997) 273}
  [\href{https://arxiv.org/abs/hep-ph/9702424}{{\ttfamily hep-ph/9702424}}].

\bibitem{Edison:2011ta}
A.~C. Edison and S.~G. Naculich, \emph{{SU(N) group-theory constraints on
  color-ordered five-point amplitudes at all loop orders}},
  \href{https://doi.org/10.1016/j.nuclphysb.2012.01.019}{\emph{Nucl. Phys.}
  {\bfseries B858} (2012) 488}
  [\href{https://arxiv.org/abs/1111.3821}{{\ttfamily 1111.3821}}].

\bibitem{Chester:2016ojq}
D.~Chester, \emph{{Bern-Carrasco-Johansson relations for one-loop QCD integral
  coefficients}}, \href{https://doi.org/10.1103/PhysRevD.93.065047}{\emph{Phys.
  Rev.} {\bfseries D93} (2016) 065047}
  [\href{https://arxiv.org/abs/1601.00235}{{\ttfamily 1601.00235}}].

\bibitem{Primo:2016omk}
A.~Primo and W.~J. Torres~Bobadilla, \emph{{BCJ Identities and $d$-Dimensional
  Generalized Unitarity}},
  \href{https://doi.org/10.1007/JHEP04(2016)125}{\emph{JHEP} {\bfseries 04}
  (2016) 125} [\href{https://arxiv.org/abs/1602.03161}{{\ttfamily
  1602.03161}}].

\bibitem{ArkaniHamed:2008gz}
N.~Arkani-Hamed, F.~Cachazo and J.~Kaplan, \emph{{What is the Simplest Quantum
  Field Theory?}}, \href{https://doi.org/10.1007/JHEP09(2010)016}{\emph{JHEP}
  {\bfseries 1009} (2010) 016}
  [\href{https://arxiv.org/abs/0808.1446}{{\ttfamily 0808.1446}}].

\bibitem{Bern:2012gh}
Z.~Bern, S.~Davies, T.~Dennen and Y.-t. Huang, \emph{{Ultraviolet Cancellations
  in Half-Maximal Supergravity as a Consequence of the Double-Copy Structure}},
  \href{https://doi.org/10.1103/PhysRevD.86.105014}{\emph{Phys. Rev.}
  {\bfseries D86} (2012) 105014}
  [\href{https://arxiv.org/abs/1209.2472}{{\ttfamily 1209.2472}}].

\bibitem{Besier:2018jen}
M.~Besier, D.~Van~Straten and S.~Weinzierl, \emph{{Rationalizing roots: an
  algorithmic approach}},  \href{https://arxiv.org/abs/1809.10983}{{\ttfamily
  1809.10983}}.

\bibitem{Bourjaily:2010wh}
J.~L. Bourjaily, \emph{{Efficient Tree-Amplitudes in N=4: Automatic BCFW
  Recursion in Mathematica}},
  \href{https://arxiv.org/abs/1011.2447}{{\ttfamily 1011.2447}}.

\end{thebibliography}\endgroup
\clearpage

\end{document}